\documentclass{egpubl}
\usepackage{eg2025}

\ConferencePaper        
\CGFccby

\usepackage[T1]{fontenc}
\usepackage{dfadobe}  

\BibtexOrBiblatex
\electronicVersion
\PrintedOrElectronic
\ifpdf \usepackage[pdftex]{graphicx} \pdfcompresslevel=9
\else \usepackage[dvips]{graphicx} \fi

\usepackage{cite}  
\usepackage{tcolorbox}
\usepackage{colortbl} 
\usepackage{xcolor}   
\usepackage{tabularx} 
\usepackage{egweblnk} 
\usepackage{amsmath}
\usepackage{amssymb}
\usepackage{booktabs}
\usepackage{caption}
\usepackage{subcaption}
\usepackage{gensymb}
\usepackage{overpic}
\usepackage{longtable}
\usepackage{array}
\usepackage{tikz}
\usetikzlibrary{arrows.meta}
\usepackage{graphicx}
\usepackage{bm}
\usepackage[normalem]{ulem}

\newcommand{\revision}[1]{#1}

\newcommand{\Dim}{\ensuremath{\mathcal{D}}}
\newcommand{\vr}[1]{\ensuremath{\bm{#1}}}
\newcommand{\mat}[1]{\ensuremath{\bm{#1}}}
\newcommand*\diff{\mathop{}\!\mathrm{d}}

\newcommand{\xvec}[0]{\ensuremath{\vr{x}}}

\newcommand{\ray}[0]{\ensuremath{\mathcal{R}}}
\newcommand{\pixel}[0]{\ensuremath{\vr{p}}}

\title[Does 3D Gaussian Splatting Need Accurate Volumetric Rendering?]{Does 3D Gaussian Splatting Need Accurate Volumetric Rendering?}

\author[A. Celarek, G. Kopanas, G. Drettakis, M. Wimmer, B. Kerbl]
{\parbox{\textwidth}{\centering A. Celarek$^{1}$, G. Kopanas$^{2, 3, 4}$\orcid{0009-0002-5829-2192}, G. Drettakis$^{3, 4}$\orcid{0000-0002-9254-4819}, M. Wimmer$^{1}$\orcid{0000-0002-9370-2663} and B. Kerbl$^{1}$\orcid{0000-0002-5168-8648}
			}
		\\
		{\parbox{\textwidth}{\centering $^1$TU Wien, Austria, $^2$Google, United Kingdom, $^3$Inria, France, $^4$Université Côte d'Azur, France
					}
			}
	}

\begin{document}
	
\teaser{
\begin{overpic}[width=\linewidth]{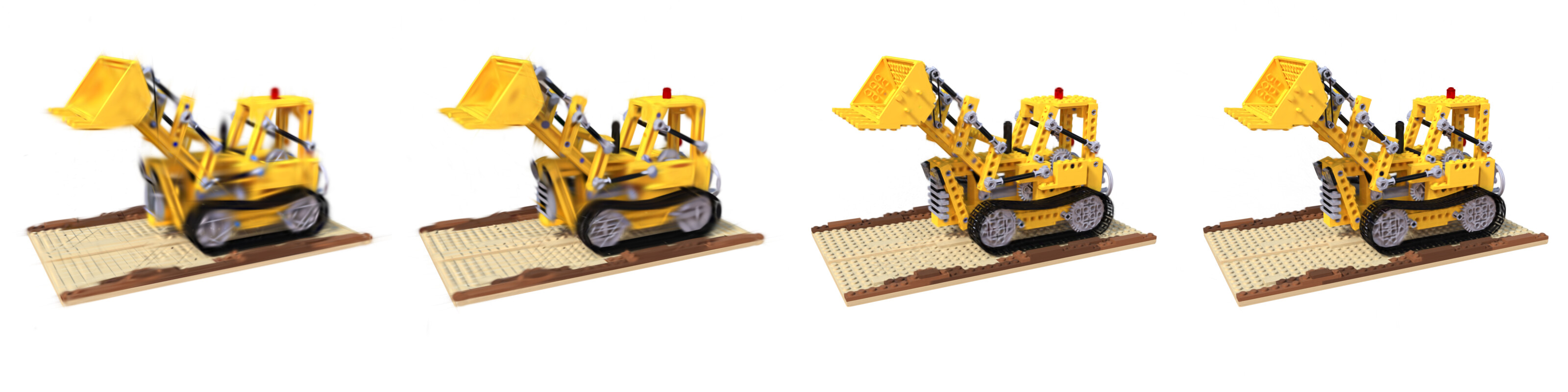}
\put(18,24){\Large \textbf{4k Gaussians}}
\put(4,2){\large 3D Gaussian Splatting}
\put(33,2){\large Ray Marching}
\put(66,24){\Large \textbf{100k Gaussians}}
\put(54,2){\large 3D Gaussian Splatting}
\put(83,2){\large Ray Marching}
\end{overpic}
  \caption{\textsc{Lego} scene from the NeRF-synthetic dataset. 
  \textbf{Left}: 4k Gaussians, rendered with 3D Gaussian Splatting and with extinction-based volume ray marching. \textbf{Right}: 100k Gaussians with the same two techniques. While the more principled ray-marching technique yields superior quality for fewer Gaussians, this benefit vanishes in qualitative and quantitative assessment when increasing their number. }
\label{fig:teaser}
}
	
\maketitle
\begin{abstract}
Since its introduction, 3D Gaussian Splatting (3DGS) has become an important reference method for learning 3D representations of a captured scene, allowing real-time novel-view synthesis with high visual quality and fast training times.  Neural Radiance Fields (NeRFs), which preceded 3DGS, are based on a principled ray-marching approach for volumetric rendering. In contrast, while sharing a similar image formation model with NeRF, 3DGS uses a hybrid rendering solution that builds on the strengths of volume rendering and primitive rasterization. A crucial benefit of 3DGS is its performance, achieved through a set of approximations, in many cases with respect to volumetric rendering theory. A naturally arising question is whether replacing these approximations with more principled volumetric rendering solutions can improve the quality of 3DGS.
In this paper, we present an in-depth analysis of the various approximations and assumptions used by the original 3DGS solution. We demonstrate that, while more accurate volumetric rendering can help for low numbers of primitives, the power of efficient optimization and the large number of Gaussians allows 3DGS to outperform volumetric rendering despite its approximations. 

\ccsdesc[500]{Computing methodologies~Image-based rendering}
\ccsdesc[400]{Computing methodologies~Volumetric models}
\ccsdesc[300]{Computing methodologies~Rasterization}
\ccsdesc[300]{Computing methodologies~Ray tracing}

\printccsdesc   
\end{abstract}  
\section{Introduction}

Neural Radiance Fields (NeRFs)~\cite{mildenhall2020nerf,barron2022mipnerf360} revolutionized novel-view synthesis by introducing a differentiable rendering optimization based on \emph{principled} volumetric ray marching. \revision{This common volume rendering formulation is inspired by physical principles (although it uses a simplified model as an approximation of real, physics-based behavior)}. 
Key to \revision{NeRF's} success was the representation of scene geometry as a continuous, differentiable volume of \emph{optical density}.
However, the volumetric ray marcher and neural network architecture made rendering and training excessively slow.
Several methods have been proposed to enhance performance \cite{yu2021plenoctrees, mueller2022instant, MERF, SMERF}, but these improvements often involve trade-offs, such as reduced visual quality.

3D Gaussian Splatting (3DGS)~\cite{kerbl3Dgaussians} uses a \emph{hybrid, primitive-based} rendering method.
It leverages volumetric rendering principles to preserve the differentiable component and aligns its image formation model with that of NeRF, which aids in optimizing 3D reconstruction.
Furthermore, it employs point-based rasterization methods for both training and rendering, significantly reducing the computational demands relative to NeRF.
However, in contrast to NeRF, \revision{3DGS learns an \emph{opacity} value \revision{for each Gaussian primitive} that it uses for fast alpha blending.
In fact, 3DGS significantly deviates from volume rendering theory}, making many simplifying assumptions and approximations compared to more principled approaches to achieve its fast computation.
Thus, a natural question to ask is: Can the visual fidelity of 3DGS improve if we remove these approximations and use principled volume rendering solutions instead?
In this paper, we perform an in-depth analysis of 3DGS to answer this question.

We explicitly clarify the difference between the learned opacity value in 3DGS and the extinction function used in volumetric rendering (extinction is referred to as ``density'' in the NeRF~\cite{mildenhall2020nerf} literature).
To illustrate this difference and to allow further analysis, we introduce an extinction-based splatting solution.
Our experiments show that for small numbers of primitives, the extinction-based solution performs better than opacity-based splatting.
However, this is reversed with more primitives, where opacity splatting performs best.
As we will analyze further, as Gaussians grow in number, rendering them with 3DGS becomes equally expressive to volumetric rendering.

One of the reasons 3DGS is so fast is its approach to resolving visibility, which is done via a single global sorting step.
This again is an approximation, since Gaussians are sorted based only on their centers.
This results in problematic popping, which has been identified in previous work~\cite{radl2024stopthepop}.
In addition, spatial overlap of Gaussians is not handled, essentially ignoring parts of the conventional volumetric rendering integral.
In order to study the impact of this approximation, we implemented a ray-marching algorithm on 3D Gaussians.
We show that these approximations have negligible impact on still images, especially with a large number of Gaussians.

Several other approximations are made by 3DGS: in particular, self-attenuation is not treated correctly, and a Gaussian's projected shape in screen space is approximate~\cite{huang2024erroranalysis3dgaussian}. We show that these also have little impact on the effectiveness of 3DGS.

\noindent
To summarize, in this paper, we systematically analyze and assess the approximations made in 3DGS. Specifically:

\begin{itemize}
	\item We introduce a mathematical framework that allows us to clarify the differences between 3DGS and accurate volumetric rendering (i.e., opacity \emph{vs.\ }extinction-based rendering) 
   \item We introduce extinction-based splatting and ray-marching algorithms for Gaussian primitives to analyze this difference. In this context, we also provide a closed-form solution for splatting self-attenuated Gaussians.
	\item Our analysis shows that opacity-based splatting results in lower error compared to extinction-based methods given a sufficiently high number of primitives.
    \item We also show that for a low number of Gaussians, correct overlap resolution and extinction-based rendering improves image quality, while correct sorting does not affect results as much.
\end{itemize}

Overall, this work fills an important gap in the understanding of 3DGS as a rendering technique 
and provides an in-depth analysis of how the method can achieve quality rivaling the state of the art, despite several simplifying assumptions and approximations.

\section{Related Work}
Volume rendering equations and theory~\cite{Max95, Novak18Course} were developed to explain and model how light scatters in participating media, e.g. clouds, smoke, but also wax and skin/flesh. Such media can emit, absorb, and scatter photons. This is described by the volumetric rendering integral, and volumetric path tracers are capable of computing the solution to such integrals.
This theory proved useful for modern image-based 3D reconstruction methods. 

In the context of novel-view synthesis, NeRF~\cite{mildenhall2020nerf} uses a simplified volume rendering model that only considers emission and absorption, ignoring scattering. 
For every position, density and directional outgoing radiance are stored in an MLP. This allows recovering
a 3D volumetric scene representation for a set of unstructured images using Stochastic Gradient Descent (SGD). For every iteration of SGD, NeRF renders an image via ray marching, sampling an MLP along the ray, and computing the volumetric integral using quadrature. 
Since this process is differentiable, the weights of the MLP receive gradients, continuously pushing image formation closer to the ground truth. 

Ray marching paired with quadrature provides an accurate solution of the volume rendering integral, but historically, rasterization is faster, albeit less accurate. Elliptical Weighted Average (EWA) splatting~\cite{ewa_splatting} is an efficient algorithm for rendering volumes through rasterization. It is based on Heckbert's thesis~\cite{heckbert1989fundamentals} and traditional signal processing. EWA volume splatting considers discrete samples of density and color and reconstructs the underlying continuous signal in screen space. In practice, the volume is represented with Gaussian primitives that are splatted and alpha-blended on screen. To allow for fast rasterization, a series of assumptions and simplifications are made: 1) Gaussian primitives do not overlap in 3D, 2) self-attenuation---i.e., the reduction of light intensity along a ray inside a primitive---is ignored, 3) the perspective projection is linearized and approximated by a first-order Taylor expansion, and 4) $e^{-x}$ is approximated by the faster $1-x$.

3D Gaussian Splatting (3DGS)~\cite{kerbl3Dgaussians} is a fast and high-quality alternative for NeRF.
3DGS builds on EWA Volume Splatting and several other works on point-based rendering~\cite{lassner2021pulsar,wiles2020synsin,zhang2022differentiable,kopanas2021point, Yifan:DSS:2019}, while simplifying the formulation even further.
Most notably, 3DGS significantly simplifies the handling of transparency (respectively its inverse, opacity). \revision{Instead of computing transparency by integrating extinction along a finite segment of a ray according to the volume rendering equation, 3DGS just assigns one single opacity value per primitive. This value is not related to an integrated extinction, as we will show in more detail in Sec.~\ref{sec:anal_opa_vs_dens}. }
The 3DGS opacity value therefore has no \revision{direct equivalent in volume rendering theory}, but it allows the splatted primitives to have a constant transparency, no matter the viewing direction, and allows for optimizing a well-behaved and bounded scalar.
The subtle difference between opacity and extinction is often overlooked in subsequent work, or the quantities are even conflated~\cite{guedon2023sugar,bolanos2024gaussian,lee2024gscore,huang2023photo}.
In this paper, we will analyze the differences and the impact of these approximations and seek to find their importance in terms of final quality for forward rendering, as well as the full optimization process.

A wealth of literature has appeared following up on the original 3DGS paper:
In line with our own work, Mip-Splatting~\cite{yu2023mip} identifies an approximation for convolution used in 3DGS and introduces a principled pre-filtering of the underlying signal to provide a correct solution for antialiasing, allowing consistent multi-scale and multi-resolution training.
Concurrent work has explored various methods to leverage benefits of ray-Gaussian intersection.
These approaches typically involve creating shells around Gaussian functions and employing conventional ray-tracing techniques to query the Gaussians.
Condor et al.\ and Zhou et al.\ concentrate on principled volumetric path tracing; however, the first does not demonstrate competitive reconstruction quality \cite{condor2024dontsplatgaussiansvolumetric}, while the latter addresses geometry reconstruction rather than novel-view synthesis\cite{zhou2024unified}.
Moenne-Loccoz et al. implement a range of ray-tracing effects, including motion blur and refraction, and they utilize ray tracing for scene composition \cite{3dgrt2024}.
Their method identifies the point of highest density, but does not incorporate volumetric marching.
Their reconstruction performance is comparable to that of 3DGS.
Finally, Blanc et al.\ use a ray-marching method that traverses the scene with a fixed step size.
However, rather than enabling direct comparison and analysis, their work focuses on improving 3DGS reconstruction quality, which they achieve by complementing the view-dependent representation with spherical Gaussians \cite{blanc2024raygaussvolumetricgaussianbasedray}.
For further analysis of the research that succeeds 3DGS, we redirect the readers to the following surveys\cite{chen2024survey,fei20243d,wu2024recent,dalal2024gaussian}.

\revision{
\section{Overview}

Our goal is to analyze the various approximations of 3DGS and their effect on visual quality. The first step in this analysis is a detailed mathematical framework (Sec.~\ref{sec:math}), starting from first principles of volumetric rendering, and then presenting a consistent notation for image formation methods based on Gaussian primitives. This presentation clarifies the mathematical properties of 3DGS and EWA (on which it is based), clearly presenting some of the approximations these techniques use. We then provide our in-depth assessment of 3DGS representations and approximations (Sec.~\ref{sec:analysis}). We first perform a detailed mathematical analysis. Importantly, we then introduce a set of algorithms, progressively replacing each of the identified approximations, allowing us to experimentally evaluate their respective effect on the visual quality of Gaussian-based rendering. Their respective implementations (Sec.~\ref{sec:impl}) enable us to run an extensive experimental analysis (Sec.~\ref{sec:experiments}), leading to several significant insights into which approximations matter, and under which conditions.
}
\section{Mathematical Framework}
\label{sec:math}

In the following, we revisit the necessary theory and provide the mathematical tools required for our in-depth analysis. 
We first recall the classic volumetric rendering integral.
With this as a basis, we show how Gaussian functions can be used as a representation of extinction within this integral.
Finally, we make the connection to elliptical weighted average (EWA) splatting and 3DGS, which is required for the subsequent theoretical and experimental analysis.
It is important to note that the former models volumetric extinction,  which is also the case for NeRF methods~\cite{mildenhall2020nerf}. Our analysis is thus also relevant in understanding how 3DGS relates to NeRF.
\revision{A list of recurring symbols in our analysis is provided in Table \ref{tab:symbols}).}

\begin{table}[h]
\revision{
    \centering
    \renewcommand{\arraystretch}{1.2}
        \caption{\revision{List of commonly used symbols in this paper.}}
    \begin{tabularx}{\linewidth}{c X}
        \hline
        Symbol & Meaning \\
        \hline
        $I$ & Image function, parameterized by pixel $\vr{p}$\\
        $f$ & Extinction coefficient function\\
        $o$ & Opacity function (used in 3DGS)\\
        $\ray$ & Viewing ray, parameterized by distance $t$\\
        $c$ & Constant or view-dependent color\\
        $\mathcal{G}^n_\mathcal{D}$ & Normalized $\mathcal{D}$-dimensional Gaussian function \\
        $\mathcal{G}^u_\mathcal{D}$ & Unnormalized $\mathcal{D}$-dimensional Gaussian function \\
        $\mu, \mu'$ & Mean and projected mean \\
        $\Sigma, \Sigma'$ & Covariance matrix and projected covariance matrix \\
        $a, a'$ & Amplitude and projected amplitude of unnormalized Gaussian function\\$\mathcal{I}_\mathcal{D}$ & Normalization factor for the exponential part of a $\mathcal{D}$-dimensional normalized Gaussian function\\
        $w$ & Weight (integral) of normalized Gaussian function\\
        $_i,_j$ & Subscripts to refer to the attributes of the $i$-th or $j$-th Gaussian in a mixture or along a viewing ray\\
        $\theta$ & Data term modeling a Gaussian's appearance\\
        \hline
    \end{tabularx}

    \label{tab:symbols}
    }
\end{table}

\subsection{The Volumetric Rendering Integral}
The integral for direct volume rendering with attenuation and source terms \revision{(Eq.\ 1 in \cite{max_et_al, Max95})} is given by:
\begin{align}
	\label{eq:volumetric_integration}
	I(\pixel) = \int_0^\infty c(\ray,t) f(\ray(t)) e^{-\int_0^t f(\ray(\tau)) \diff{\tau}} \diff{t},
\end{align}
where $\ray$ is the viewing ray going through pixel $\pixel$, parameterized by \revision{distance} $t$ \revision{along the ray} to obtain a three dimensional point on $\ray$ at that distance. \revision{Further, }$c$ is the function returning the (wavelength- and view-dependent) radiance at $\ray(t)$ in the direction of $\ray$, and $f$ is the function returning the extinction coefficient (or, for matter, particle density) at $\ray(t)$.
In this integral, $c(\ray, t) f(\ray(t))$ models the entire source (emission and in-scattering) and $e^{-\int_0^t f(\ray(\tau)) \diff{\tau}}$ the entire attenuation (absorption and out-scattering) of light.
This simplified model is also employed for the derivation of NeRF methodology \cite{mildenhall2020nerf}, where the extinction coefficient is referred to as \emph{density}.
It does not explicitly consider the scattering of light, which means that light refraction (e.g., through glass) is not modeled.
Notwithstanding the lack of this nuance, the formulation is powerful enough to describe such scattering implicitly, that is, by attenuating all light and sourcing appropriate new light through emission.

\subsection{3D Gaussian Data Model}

\label{sec:gaussian_data_model}
Eq.~\ref{eq:volumetric_integration} does not assume a specific representation for extinction in the domain of integration.
The following specialize it for a \emph{Gaussian representation} of the extinction function, as is used by EWA splatting \cite{ewa_splatting}, 3D Gaussian splatting \cite{kerbl3Dgaussians}, and others. 
We first define $\Dim$-dimensional Gaussian functions. A familiar example is their use in the Gaussian (or \emph{normal}) \revision{distribution's PDF:
\begin{align}\label{eq:normal_dist}
	\mathcal{N}_\Dim(\vr{x};\vr{\vr{\mu}}, \mat{\Sigma}) = \frac{1}{\mathcal{I}_\Dim(\Sigma)} e^{-\frac{1}{2}(\vr{x}-\vr{\vr{\mu}})^T\mat{\Sigma}^{-1}(\vr{x}-\vr{\vr{\mu}})},
\end{align}
}
where 
\vr{x} is a point in $\mathbb{R}^{\Dim}$,
$\vr{\vr{\mu}}$ is the ${\Dim}$-dimensional position (mean), $\mat{\Sigma}$ the shape (covariance matrix), and \revision{$\mathcal{I}_\Dim(\mat{\Sigma})$} is the unbounded integral of the exponential part, ensuring that $\mathcal{N}_\Dim$ integrates to one:
\begin{align}
	\label{eq:exponential_integral}
    \revision{
	\mathcal{I}_\Dim(\mat{\Sigma}) = \int_{\mathbb{R}^\Dim} e^{-\frac{1}{2}(\vr{x}-\vr{\vr{\mu}})^T\mat{\Sigma}^{-1}(\vr{x}-\vr{\vr{\mu}})} \diff \xvec = \sqrt{(2\pi)^\Dim det(\mat{\Sigma})}.
    }
\end{align}

\paragraph*{Normalised Gaussian functions} are the basic building block of EWA splatting.
There, they are paired with an additional weight parameter $w$, which effectively models the \emph{total} capacity of the corresponding Gaussian to block and emit light:
\begin{align}\label{eq:gaussian_definition_normalised}
    \revision{
	\mathcal{G}^n_\Dim(\vr{x}, w, \vr{\mu}, \mat{\Sigma}) = w \mathcal{N}_\Dim(\vr{x}; \vr{\mu}, \mat{\Sigma}),
    }
\end{align}
Since $\mathcal{N}_\Dim$ integrates to 1, $w$ is equivalent to the unbounded integral of the full normalized Gaussian function:
\begin{align}
    \label{eq:total_gaussian_integral}
	\int_{\mathbb{R}^\Dim} \mathcal{G}^n_\Dim(\xvec, w, \vr{\mu}, \mat{\Sigma}) \diff \xvec = w.
\end{align}

\paragraph*{Unnormalised Gaussian functions} are defined by
\begin{align}\label{eq:gaussian_definition_unnormalised}
\revision{\mathcal{G}^u_\Dim(\vr{x}, a, \vr{\mu}, \mat{\Sigma}) = a \mathcal{I}_\Dim(\mat{\Sigma})  \mathcal{N}_\Dim(\vr{x}; \vr{\mu}, \mat{\Sigma}),}
\end{align}
where $a$ is the amplitude, or value, at $\vr{\mu}$, and all other variables are the same as in the normalised case.
This representation is used in 3DGS, where $a$ equates to ``opacity'', i.e., the \emph{peak} capacity of a Gaussian to block/emit light \emph{at its center}. Its evaluation is easier compared to that of normalized Gaussians, since \revision{$\mathcal{I}_\Dim(\mat{\Sigma})$} and the normalisation factor in $\mathcal{N}_\Dim$ cancel each other out. 
Note that it is always possible to convert between the normalised and unnormalised parameterization of a given Gaussian using
\begin{align}\label{eq:gaussian_definition_conversion}
\revision{
	a = \frac{w}{\mathcal{I}_\Dim(\mat{\Sigma})}.
    }
\end{align}

\paragraph*{Image formation:} 
To use Gaussian functions for volume rendering, the extinction function is modeled by a mixture of Gaussians:
\begin{align}
    \label{eq:gmm}
    \revision{
	f(\xvec) = \sum_{i=0}^N \mathcal{G}^n_3(\vr{x}, w_i, \vr{\mu}_i, \mat{\Sigma}_i).
    }
\end{align}
\revision{
Each Gaussian can be supplemented by a (learnable) function $c_i(\ray)$, which returns the radiance value for a given view ray $\ray$}.
In the case of 3DGS, the directional radiance is modeled with spherical harmonics, but other representations are possible as well. 
Gaussians can serve as the basis for the radiance function \revision{on ray $\ray$ at $t$}:
\revision{
\begin{align}
	c(\ray, t) = \sum_{i=0}^N c_i(\ray) \mathcal{G}^n_3(\vr{x}, w_i, \vr{\mu}_i, \mat{\Sigma}_i) .
\end{align}
}
\revision{Using the shortand 
$\mathcal{G}_3^n(\vr{x}, i) = \mathcal{G}^n_3(\vr{x}, w_i, \vr{\mu}_i, \mat{\Sigma}_i)$ for compactness, we can substitute extinction and radiance (color) functions into Eq.~\ref{eq:volumetric_integration} to arrive at a 3D Gaussian-based solution for volume rendering:
\begin{align}
    \label{eq:volumetric_integration_c}
	I(\pixel) = &\int_0^\infty \sum_{i=0}^N c_i(\ray)\mathcal{G}_3^n(\ray(t), i) e^{-\int_0^t \sum_{i=0}^N \mathcal{G}_3^n(\ray(t), i) \diff{\tau}} \diff{t} 
\end{align}
}

\subsection{EWA and 3D Gaussian Splatting}
\label{sec:bg_3dgs}
\begin{figure}
    \centering
    \includegraphics[width=\linewidth]{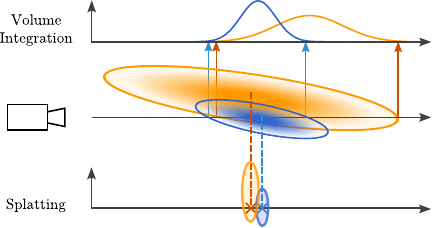}
    \caption{\revision{A setup with two Gaussian primitives (center). For each viewing ray, volumetric integration (top) considers the intersection with all primitives and evaluates their joint extinction at each point along the ray. In splatting (bottom), Gaussians are projected to flat, camera-facing disks. Each 2D disk provides an isolated contribution to the ray, which 3DGS blends in the order of projected means.}}
    \label{fig:splatting}
\end{figure}
Eq.~\ref{eq:volumetric_integration_c} represents a basic solution for volumetric integration of Gaussian mixtures along a ray.
However, the image formation models used in EWA and 3DGS are, at best, an approximation of this integral. 
The focus of our work is to clarify how EWA and 3DGS relate to volumetric integration and consider the \revision{implications of} algorithmic choices for approximations in the above integral.

To avoid the high cost of volume integration, both EWA and 3DGS simplify the rendering of 3D Gaussians by reducing them to 2D Gaussians that can be easily ``splatted''.
Intuitively, it seems desirable that the appearance of splatted 2D Gaussians matches the result of volumetrically rendering their 3D counterparts. 
With this perspective, we can consider how splatting-based approaches \revision{deviate from} Eq.~\ref{eq:volumetric_integration_c}:
\revision{First, Gaussians are assumed \emph{not to overlap}} and are sorted based on their means, front to back, by view-space depth of the 3D Gaussian position.
\emph{Within} a Gaussian, the color $c_i$ is uniform; view-dependent $c_i$ of 3DGS are evaluated w.r.t.\ the mean only. 
Also \emph{within} a Gaussian, continuous accumulation of extinction and the corresponding effect on its appearance is not considered. 
\revision{Further ignoring perspective distortion, EWA exploits these simplifications to find the 2D extinction contribution function $f_i$ of Gaussian $i$ from its 3D definition \revision{(see Fig.~\ref{fig:splatting})}. This disk-like 2D function is found by projecting a 3D Gaussian along a view axis:}
\begin{align}
	\label{eq:gaussian2d_ewa}
    \revision{f_i(\vr{p}) = \mathcal{G}^n_2(\pixel, w_i, \vr{\mu}'_i, \mat{\Sigma}'_i) = \int_{-\infty}^\infty \mathcal{G}^n_3(\ray(t), w_i, \vr{\mu}_i, \mat{\Sigma}_i) \diff t,}
\end{align}
where $\vr{\mu}'$ and $\mat{\Sigma}'$ are projected 2D mean and covariance matrix, and normalised Gaussians are used.
This corresponds to the marginal distribution of a normal distribution, which is again a normal distribution, thus we have
\begin{align}
\label{eq:integra_of_normalised_2d3dgaussians_is_wl}
     \int_{\mathbb{R}^3} w\mathcal{N}_3 (\xvec, \vr{\mu}, \mat{\Sigma}) \diff \xvec &= \int_{\mathbb{R}^3} \mathcal{G}^n_3 (\xvec, w, \vr{\mu}, \mat{\Sigma}) \diff \xvec
     \\ \nonumber = \int_{\mathbb{R}^2} w\mathcal{N}_2(\pixel, \vr{\mu}', \mat{\Sigma}') \diff \pixel &= \int_{\mathbb{R}^2} \mathcal{G}^n_2(\pixel, w, \vr{\mu}', \mat{\Sigma}') \diff \pixel = w,
\end{align}
i.e., $w$ is preserved across all projections of a 3D Gaussian.
Recall that the mixture of Gaussians for EWA defines an \emph{extinction function} that is the same quantity as the density learned by NeRF. 

\begin{figure}
    \centering
    \includegraphics[width=\linewidth]{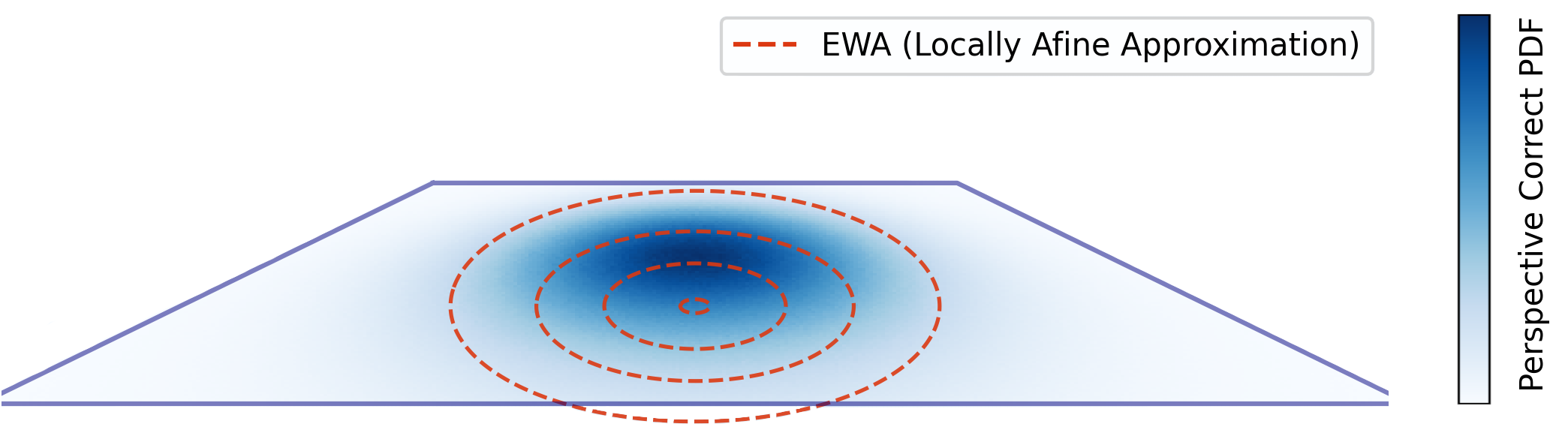}
    \caption{\revision{A flat 3D Gaussian (one scale dimension is 0), positioned in a slanted angle to the camera. Blue shows exact projection;
    notice the skewness between the top and bottom.
    Red iso-curves show the affine approximation first used in EWA, a camera-facing disk.}}
    \label{fig:ewa-approximation}
\end{figure}

\revision{In contrast, 3DGS uses unnormalized Gaussians and preserves 2D \emph{amplitude} $a'$  (which they call "opacity") across all projections:}
\begin{align}
    \label{eq:gaussian2d_3dgs}
    \revision{
	o_i(\pixel) = \mathcal{G}^u_2(\pixel, a_i', \vr{\mu}'_i, \mat{\Sigma}'_i)
    }
\end{align}
We will shortly see that due to this convention, 3DGS does not allow for defining a corresponding 3D extinction (or density) function $\mathcal{G}_3$ in the established volume integration framework; in particular, we must \emph{add view-dependent factors} to consolidate the two.

\revision{The computation of $\Sigma'$ involves the transformation of the Gaussian from world-space coordinates to screen space. Gaussian distributions are closed under affine transformations, however, perspective projection is usually non-linear. To circumvent this, EWA and 3DGS approximate perspective transformation with its locally-affine counterpart (see Fig.~\ref{fig:ewa-approximation}), using first-order Taylor expansion:}
\begin{align}
	\mat{J} &= \begin{pmatrix}
		1 / \vr{\mu}_2 & 0 & -\vr{\mu}_{0}/\vr{\mu}^2_{2} \\
		0 & 1 / \vr{\mu}_{2} & -\vr{\mu}_{1}/\vr{\mu}^2_{2} \\
		\vr{\mu}_{0}/l & \vr{\mu}_{1}/l & \vr{\mu}_{2}/l \\
	\end{pmatrix}\nonumber\\
	\label{eq:splatting_transformation_of_covariance}
	\mat{\Sigma}' &= \mat{JW}\mat{\Sigma} \mat{W}^T\mat{J}^T,
\end{align}
where $\vr{\mu}$ is the Gaussian's position in camera space, $l=\lVert \vr{\mu} \rVert_2$ and $\mat{W}$ the transformation to camera space \revision{(Eq.\ 29 and 31 in \cite{ewa_splatting})}.

Finally, the attenuation term is approximated by the first-order Taylor expansion of $e^x$, which is $\approx 1-x$.
Applying all these approximations to Eq. \ref{eq:volumetric_integration_c} yields \revision{the image function (Eq.\ 3 in \cite{kerbl3Dgaussians})}:
\begin{equation}
	\label{eq:alphablend}
    \revision{
	I(\pixel) = \sum_{i=0}^N c_i(\ray) g_i(\pixel)\prod_{j=0}^{i - 1}(1-g_j(\pixel)) + c_b \prod_{i=0}^N(1-g_i(\pixel)),
    }
\end{equation}
where 
\revision{$c_b$ is the background color, $i$ iterates over Gaussians on ray $\ray$ going through $\vr{p}$, and $g_i$ is the $i$-th Gaussian's partial contribution, either extinction (Eq.~\ref{eq:gaussian2d_ewa}) for EWA, or opacity (Eq.~\ref{eq:gaussian2d_3dgs}) for 3DGS.}


\section{Analysis of 3DGS Representation and Approximations}
\label{sec:analysis}

Given the above mathematical framework, 
we now present a detailed analysis of splatting algorithms compared to accurate volumetric rendering and assess the representation and approximations.
In Sec.~\ref{sec:anal_opa_vs_dens}, we expand on the key difference between the EWA and 3DGS splatting image formation method, i.e., the use of 2D opacity instead of extinction-based values \revision{(see Fig.~\ref{fig:rendering_methods})}.

\revision{We also aim} to experimentally evaluate the \revision{\emph{effect}} of the different approximations in 3DGS \revision{and EWA}. To do this, we define a set of algorithms that replace each specific approximation with the more principled volumetric rendering solution. We start this in
Sec.~\ref{sec:ewabasedsplat}, where we adapt the 3DGS method to implement an extinction-based algorithm that uses splatting, making it more similar to EWA.
In Sec.~\ref{sec:self_attenuation_and_exp}, we propose an extended version of extinction-based splatting that correctly accounts for light attenuation \emph{within} a Gaussian (self-attenuation).
Sec.~\ref{sec:overlap_proj} then examines issues arising from 3DGS taking substantial liberties in resolving visibility during rendering, entirely ignoring Gaussian overlap, and using only a single depth value per Gaussian for ordering.
To \revision{evaluate their impact in practice}, we introduce a differentiable ray marcher for 3D Gaussians that correctly handles per-pixel order and primitive overlap.

\subsection{A Unified Framework for Gaussian Extinction Functions}
\label{sec:anal_opa_vs_dens}
Our goal is to analyze the effect of the different choices made by traditional volumetric extinction-based EWA and ``opacity''-based 3DGS. To this end, we
provide a unified framework for computing Gaussian-based extinction functions across EWA and 3DGS. 
We introduce an abstract data term $\theta$, stored (and potentially learned) per Gaussian, from which its appearance is derived.
To prepare the necessary foundation for evaluating \revision{the different approaches} in 3D (e.g., for ray marching) and 2D (e.g., for 2D splatting), we will focus on solutions for computing the 3D and 2D amplitude parameters $a$ and $a'$ of \revision{unnormalised} Gaussian functions $\mathcal{G}^u_3$ and $\mathcal{G}^u_2$\revision{, since they are more convenient for rendering}. 

\begin{figure}[t]
	\centering
        \hspace*{\fill}
	\begin{subfigure}[t]{0.32\columnwidth}
		\includegraphics[width=\linewidth]{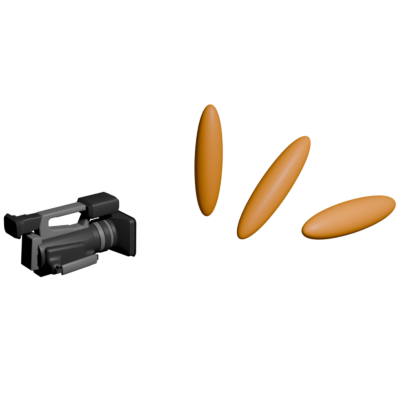}
		\caption{Viewing Scenario}
	\end{subfigure}%
    	\hfill
	\begin{subfigure}[t]{0.32\columnwidth}
		\includegraphics[width=\linewidth]{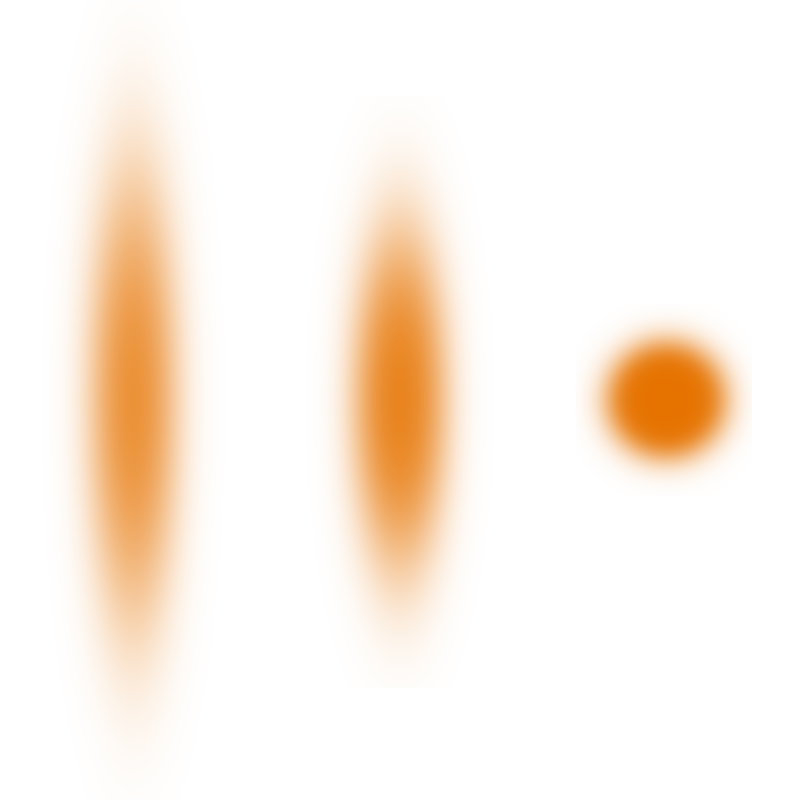}
		\caption{EWA-based}
	\end{subfigure}%
	\hfill
	\begin{subfigure}[t]{0.32\columnwidth}
		\includegraphics[width=\linewidth]{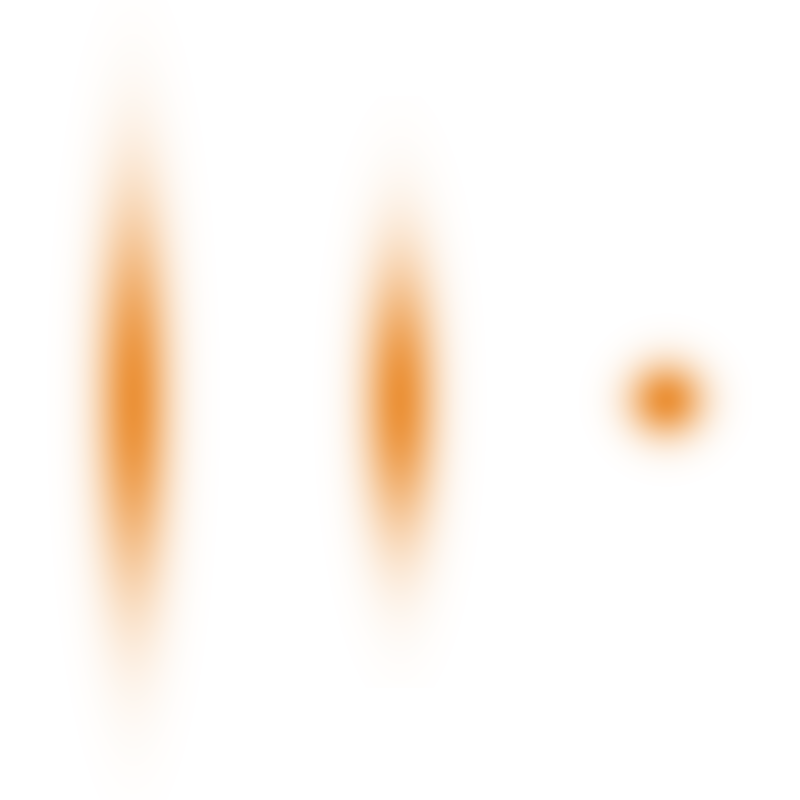}
		\caption{3DGS-based}
	\end{subfigure}%

  \hspace*{\fill}
	\caption{\revision{Illustration of rendering methods for the non-overlapping, anisotropic Gaussians in (a). The left-hand Gaussian presents its thinnest side, the one in the middle is at a 45\degree\ angle and the right-hand one shows its thickest side. (b) EWA-based splatting preserves the integral of Gaussian extinction and thus the three Gaussians appear progressively darker. (c) With 3DGS-based splatting, all Gaussians have the same opacity (and thus intensity) at their center.}}  
	\label{fig:rendering_methods}
\end{figure}


\paragraph*{EWA Splatting} uses normalised Gaussians and the stored per-Gaussian data term $\theta$ corresponds to $w$, the total integral of each normalised Gaussian function.

From the conversion between parameterizations in Eq.~\ref{eq:gaussian_definition_conversion}, \revision{we get unnormalised Gaussian amplitudes
\begin{align}
\label{eq:a_mass_formulation}
	a &=\frac{\theta}{\mathcal{I}_3(\mat{\Sigma})},
\end{align}
and
\begin{align}
	\label{eq:aprime_mass_formulation}
	a' &=\frac{\theta}{\mathcal{I}_2(\mat{\Sigma'})},
\end{align}
where $\mathcal{I}_3$ and $\mathcal{I}_2$ are the normalization factors for the exponential part of the 3D Gaussian function and its 2D projection in the current view, respectively} (Eq.~\ref{eq:exponential_integral}).
The behavior of $a'$ can be observed in Fig.~\ref{fig:rendering_methods}b, where we clearly see that the peak intensity increases with distance travelled through the Gaussian. This is the expected behavior when Gaussians model volumetric extinction.

\begin{figure}
	\centering
	\includegraphics[width=\linewidth]{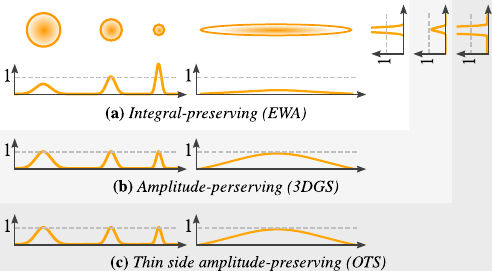}
	\caption{\revision{Behaviors for the different formulations when changing covariance and fixing $\theta=1$ for a Gaussian primitive. For illustration, we use 2D Gaussian functions: Plotted curves indicate their contribution after projecting/marginalizing them along the vertical and horizontal "viewing axis". The left side varies the Gaussian covariance uniformly, while the right stretches it to produce anisotropy.
    (a) In EWA, $\theta$ represents the Gaussian's \emph{total} extinction, i.e., its integral (area under the curve): as we change the covariance, we must thus adapt its amplitude. (b) In contrast, 3DGS preserves the amplitude of the Gaussian's projection: all curves peak at the same value, regardless of covariance shift or projection axis. For anisotropic Gaussians (right), area under the curve \emph{varies} with the viewing axis. (c) OTS also uses $\theta$ to store the projected Gaussian amplitude, but through its thinnest side specifically, preserving total extinction across different projection axes (i.e., view directions).}}
	\label{fig:extingi}
\end{figure}

Assuming the Gaussians describe density of matter, $\theta$ in this formulation would represent the total \emph{mass} of a single Gaussian ``particle cloud''.
Intuitively, marginalizing a 3D representation of density $\mathcal{G}_3$ along an axis should turn it into a 2D Gaussian with the same total mass, and this is indeed the case for EWA.
Eq.~\ref{eq:aprime_mass_formulation} \revision{and Fig.~\ref{fig:extingi}a illustrate} this: the total mass $\theta$ does not change with the view, but the projected shape $\mat{\Sigma}'$ does, and therefore \revision{$\mathcal{I}_2(\Sigma')$} and $a'$ do too.

\paragraph*{3D Gaussian Splatting,} on the other hand, stores an ``opacity'' term on the 3D primitives, which is a constant, view-independent quantity for amplitude $a'$ of \revision{projected} Gaussians in 2D directly:
\begin{align}
	\label{eq:aprime_opacity_formulation}
	a' = \theta.
\end{align}
Note that compared to Eq.~\ref{eq:aprime_mass_formulation}, the view-dependent factor $\mathcal{I}_2^{-1}$ is missing.
While EWA preserves the \emph{integral} of Gaussian extinction functions, 3DGS preserves the \emph{amplitude} of projected Gaussians in 2D, \emph{regardless of viewing direction}. Intuitively, this means that the distance travelled along the ray within each Gaussian primitive does not affect the peak opacity of the primitive; this can be clearly seen in Fig.~\ref{fig:rendering_methods}c, where the resulting image-space splats have the same intensity at the center.
An alternative interpretation is that 3DGS uses a volumetric rendering model with \emph{view-dependent} extinction, implying that the density (and mass) of a supposed \revision{Gaussian particle} cloud varies with the viewpoint---contradicting the physically\revision{-inspired} volume integration framework \revision{(see Fig.~\ref{fig:extingi}b)}.

As a consequence, there is no globally valid parameterisation of $\mathcal{G}_3$ from the Gaussian shape and the data term $\theta$ in 3DGS.
However, we can recover the view-dependent solution for $a$ in 3D from $\theta$.
We again use Eq.~\ref{eq:gaussian_definition_conversion} to convert the 2D opacity $\theta$ to $w$ and back to $a$:
\begin{align}
\revision{
\label{eq:a_opacity_formulation}
    w = \mathcal{I}_2(\mat{\Sigma'})\theta, \quad 
	a =\frac{w}{\mathcal{I}_3(\mat{\Sigma})} = \frac{\mathcal{I}_2(\mat{\Sigma'})}{\mathcal{I}_3(\mat{\Sigma})}\theta,
    }
\end{align}
%
\revision{While this derivation does not benefit splatting-based techniques, it is still useful since it} permits the use of the 3DGS image formation model in the 3D domain (e.g., for ray marching).


\revision{We note that 3DGS' use of} view-independent opacity is not only an approximation for the sake of performance, but also a 
design decision that clearly works well in practice. However, this fundamental distinction between 3DGS and volumetric extinction-based solutions such as EWA splatting and NeRFs is sometimes overlooked in literature, and in some cases, ``extinction'', ``density'', and ``opacity'' are even conflated~\cite{guedon2023sugar,bolanos2024gaussian,lee2024gscore,huang2023photo}.

\subsection{Optimization with EWA-based Extinction}
\label{sec:ewabasedsplat}



Having analyzed the deviation of 3DGS from volume rendering theory, we shift our focus to \revision{the more principled} \emph{EWA-based} splatting.
Our aim is to identify the necessary modifications required to adapt EWA-based splatting for gradient-descent-based optimization \revision{like 3DGS}.
Directly learning \revision{$\theta$ as the weight $w$ of normalized Gaussians (Eq.~\ref{eq:a_mass_formulation} and \ref{eq:aprime_mass_formulation}) impedes training speed and robustness}: 
Consider the case of a single camera capturing a Gaussian that has already reached the appropriate opacity for the object it represents. If the Gaussian changes in size (e.g., to better fit the object's shape), $\theta$ is not affected. However, $\mathcal{I}_3$ and $\mathcal{I}_2$ will change, and $a'$ inversely so. Hence, expanding the Gaussian implicitly renders it more transparent \revision{and shrinking makes it more opaque (see Fig.~\ref{fig:extingi}a)}.

We want to find a parameterization that adheres to Eq.~\ref{eq:gaussian2d_ewa} and has robust behavior.
Our initial idea is to learn peak extinction:
\revision{\begin{align}
	a = \theta \nonumber, \quad
	a' = \theta \frac{\mathcal{I}_3(\mat{\Sigma})}{\mathcal{I}_2(\mat{\Sigma'})},
\end{align}
}
where $a'$ was derived using Eq.~\ref{eq:gaussian_definition_conversion} again.
This somewhat offsets the initial issue, but only to the degree to which changes in $\mathcal{I}_3$ and $\mathcal{I}_2$ are correlated. 
Unfortunately, this is not generally the case. Consider a Gaussian trying to model a flat solid surface, frontally facing the current view: to appear opaque, an almost-flat Gaussian requires a much higher extinction than a thicker Gaussian \revision{(Fig.~\ref{fig:extingi}a, right)}.

To ensure robustness under optimization as well as the ability to model thin, solid objects,
we therefore arrive at a scheme we call \emph{opacity-thin-side} (OTS). 
OTS dynamically scales the learned weight $\theta$ such that $a'= \theta$ when looking at the Gaussian facing its thinnest side (i.e., viewing along its shortest axis).
Under rotation it behaves like other EWA-based Gaussians, i.e., the 3D unbounded integral is view-independent, and the 2D projection is an ordinary integral \revision{(see Fig.~\ref{fig:extingi}c).
This is achieved by
\begin{align}
	\label{eq:gaussian_weight_ots}
	a = \theta \frac{\mathcal{I}^*_2(\mat{\Sigma})}{\mathcal{I}_3(\mat{\Sigma})}, \quad
	a' = \theta \frac{\mathcal{I}^*_2(\mat{\Sigma})}{\mathcal{I}_2(\mat{\Sigma'})},
\end{align}
with $\mathcal{I}^*_2$ being the largest possible $\mathcal{I}_2$, which can be computed as:
\begin{align}
	\mathcal{I}^*_2(\mat{\Sigma}) = 2 \pi \sqrt{\lambda_1\lambda_2},\nonumber
\end{align}
}
where $\lambda_1$ is the largest and $\lambda_2$ the second-largest eigenvalue of the 3D covariance matrix.
\revision{With this, OTS contains all necessary definitions for robust, optimizable splatting with EWA-based extinction.}

\subsection{Attenuation and Self-Attenuation}
\revision{
\label{sec:self_attenuation_and_exp}

Both EWA splatting and 3DGS ignore how a Gaussian's extinction affects its own appearance. We call this \emph{self-attenuation}: it affects the visual result, even when Gaussians do not overlap (see Figure \ref{fig:selfattn}). For the attenuation of one Gaussian by another, both EWA and 3DGS approximate $e^{-g(\pixel)}$ with the first two terms of its Taylor series expansion, $1-g(\pixel)$ (Eq.~\ref{eq:alphablend}).
Note that this solution \emph{cannot} be used directly if $g(\pixel)$ can exceed $1$; this, however, is the case in our OTS solution, where an unrestricted $\theta$, associated with a Gaussian's \emph{mass}, is learned. A na\"{i}ve solution is to simply clamp $g(\pixel)$ in Eq.~\ref{eq:alphablend} to $[0, 1)$, but this results in sharp, non-Gaussian falloffs, as well as the elimination of gradients, which are essential for training. 

\begin{figure}
\includegraphics[width=\columnwidth]{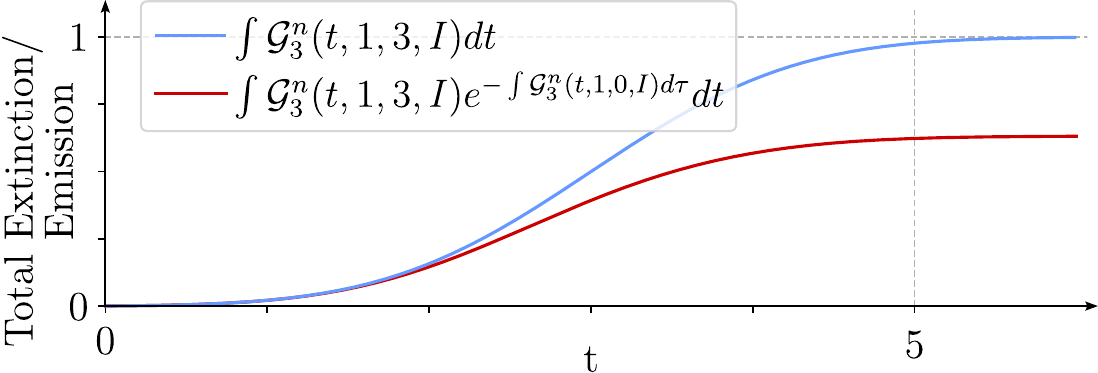}
\caption{\revision{Impact of self-attenuation. For a ray intersecting a 3D Gaussian, ignoring (blue) and accounting for (red/lower) self-attenuation show similar initial behavior in Gaussians with identical parameters. However, as the ray goes on, self-attenuation gradually reduces the Gaussian's ability to emit/extinguish additional light, due to the extinction already accumulated since entering it.}}
\label{fig:selfattn}
\end{figure}

To address attenuation in a principled manner, we revisit Eq.~\ref{eq:volumetric_integration_c}.
We first consider the case of a Gaussian mixture with just one Gaussian (index 0) intersecting a viewing ray. We move the ray's starting point to $-\infty$ to ensure that it traverses the entire Gaussian:
\begin{align}
	\label{eq:volumetric_equation_single_gaussian}
	I(\pixel) = c_0(\ray) \int_{-\infty}^\infty \mathcal{G}^n_3(\ray(t), 0) e^{-\int_{-\infty}^t \mathcal{G}^n_3(\ray(\tau), 0) \diff{\tau}} \diff{t},
\end{align}
For simplicity, EWA and 3DGS set the exponential term---the self-attenuation---to $1$. However, it is possible to derive a closed-form solution if we use the 2D Gaussian extinction $f_i$ defined in Eq.~\ref{eq:gaussian2d_ewa}:
\begin{align}
	\label{eq:volumetric_equation_single_gaussian_closed_form}
	I(\pixel) = c_0(\ray) \left(1 - e^{-f_0(\vr{p})} \right).
\end{align}
The proof for this solution is provided in the supplemental material.
When splatting a single Gaussian, this approach accurately accounts for self-attenuation and exponential extinction, and has been used for producing the rendering in Figure \ref{fig:rendering_methods}c.}
\revision{Exploiting the assumption of non-overlapping Gaussians that splatting inherently depends on, this solution easily extends to multiple Gaussians:
\begin{equation}
	\label{eq:self_attenuation_blend}
	I(\pixel) = \sum_{i=0}^N c_i(\ray) \left(1 - e^{-f_i(\pixel)} \right) \prod_{j=0}^{i - 1}(e^{-f_j(\pixel)}) + c_b \prod_{i=0}^N(e^{-f_i(\pixel)}),
\end{equation}
Note that it also removes the need for clamping, since any $f_i(\pixel) \in [0, \infty)$ yields coefficients $\in [0, 1)$, making it compatible with OTS.

}

\subsection{Visibility}
\label{sec:overlap_proj}
\begin{figure*}
	\centering
        \hspace*{\fill}
	\begin{subfigure}[t]{0.24\textwidth}
		\includegraphics[width=0.8\linewidth,trim=0 2cm 0cm 0, clip]{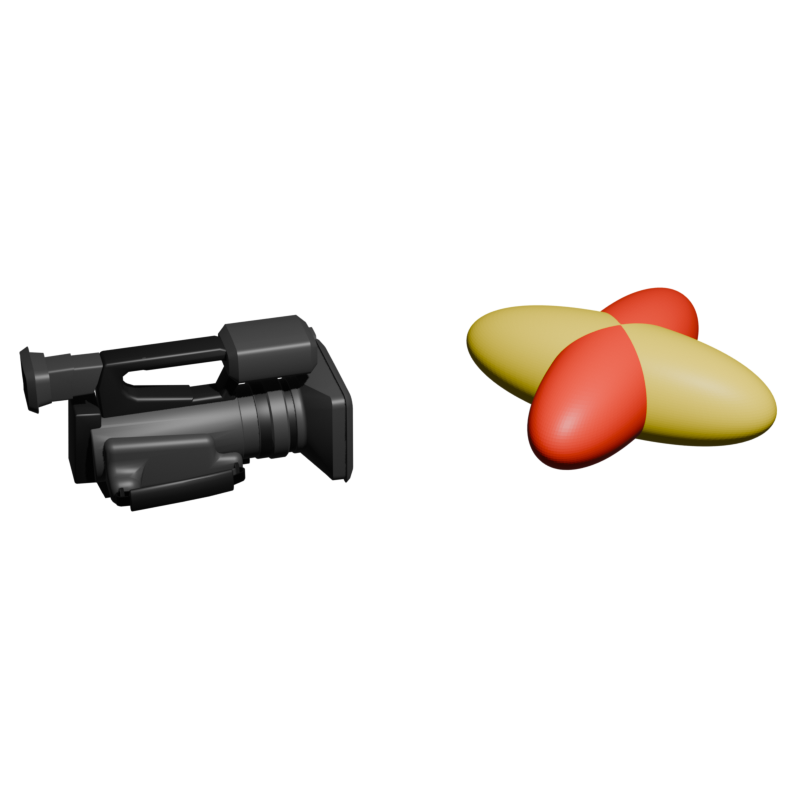}
		\caption{Viewing Scenario}
	\end{subfigure}%
	\hspace{1mm}
	\begin{subfigure}[t]{0.24\textwidth}
		\scalebox{-1}[1]{\includegraphics[width=0.8\linewidth,trim=0 2cm 0cm 0, clip]{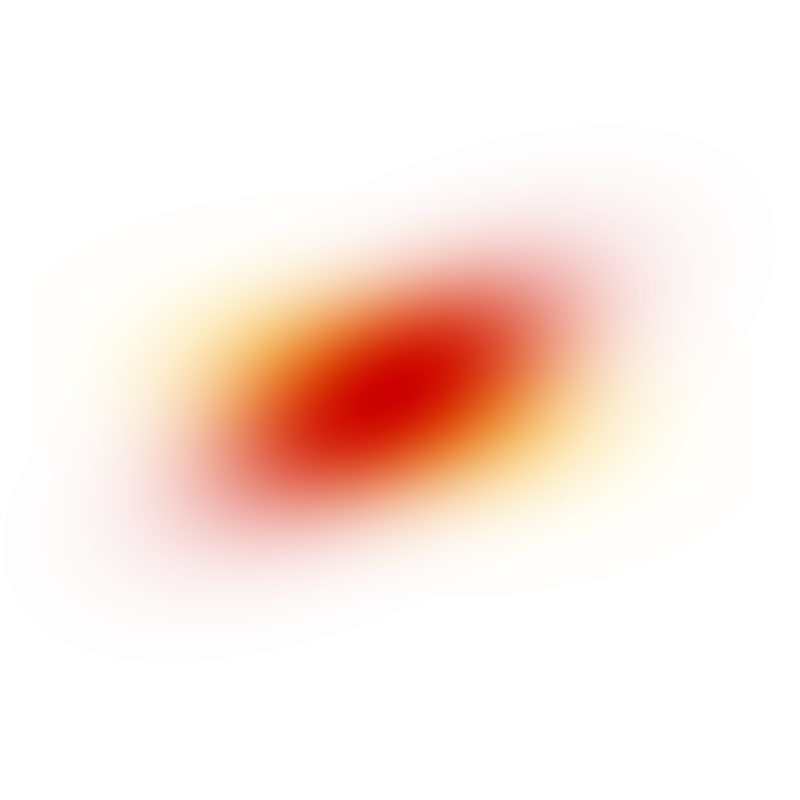}}
		\caption{Per-Gaussian sorting (3DGS)}
	\end{subfigure}%
	\hspace{1mm}
	\begin{subfigure}[t]{0.24\textwidth}
		\scalebox{-1}[1]{\includegraphics[width=0.8\linewidth,trim=0 2cm 0cm 0, clip]{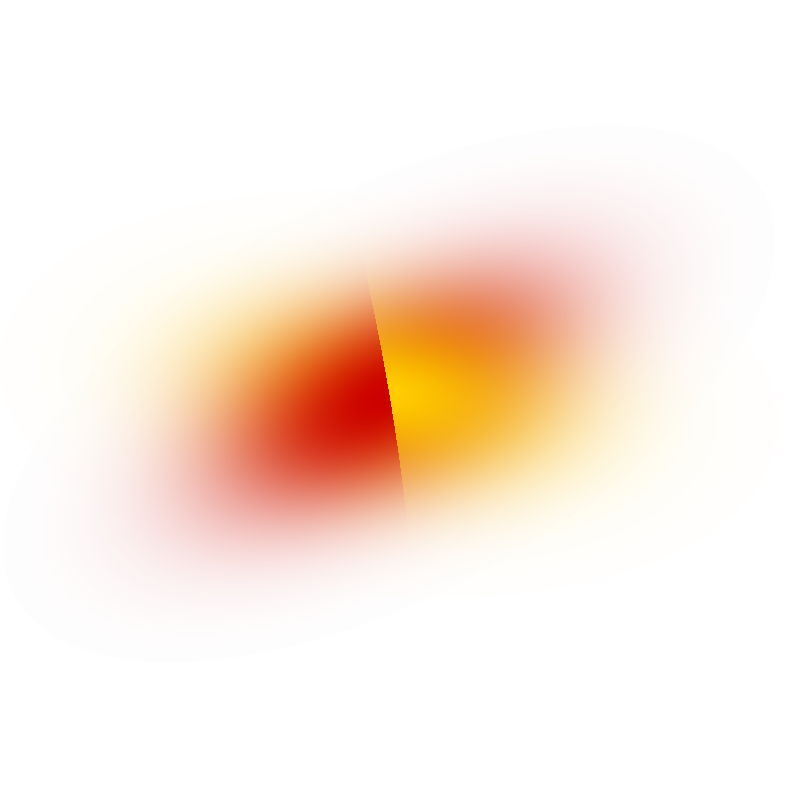}}
		\caption{Per-pixel sorting (STP)}
	\end{subfigure}%
	\hspace{1mm}
	\begin{subfigure}[t]{0.24\textwidth}
		\scalebox{-1}[1]{\includegraphics[width=0.8\linewidth,trim=0 2cm 0cm 0, clip]{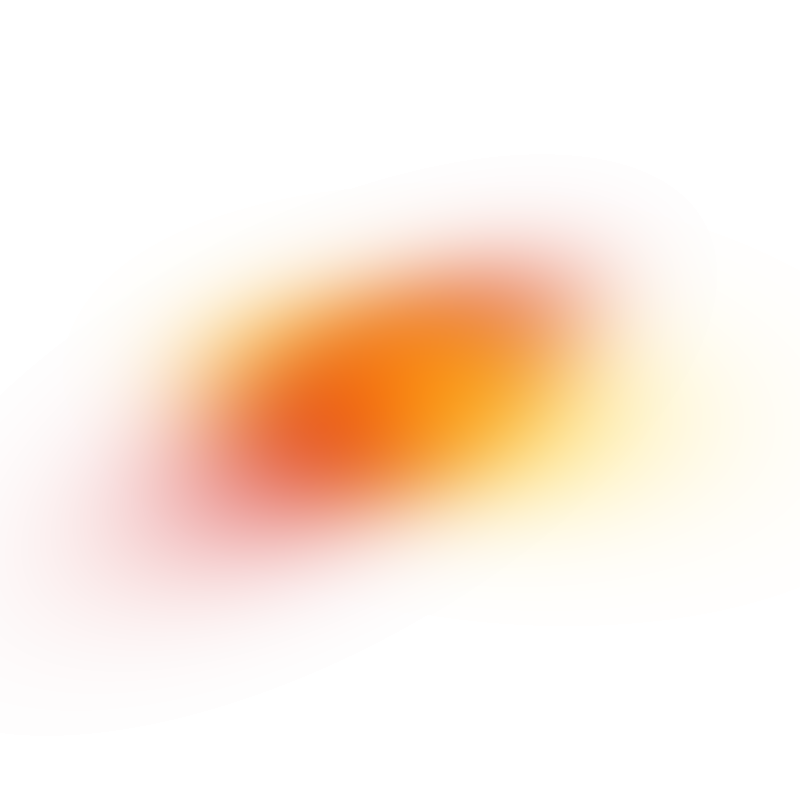}}
		\caption{Exact integration (ray marcher)}
	\end{subfigure}%
         \hspace*{\fill}
	\caption{Visibility in Gaussian Splatting. Illustration of the effect of sorting and overlap when rendering the 2-Gaussians setup shown in (a). The two Gaussians have the same position and are at 90\degree\ to each other. (b) shows the behaviour of the original 3DGS paper: One of the Gaussians ends up in front as a global sort is used. (c) shows the effect of a per-pixel sort (StopThePop). Only (d) correctly mixes the colours.}
	\label{fig:overlap_and_sorting}
\end{figure*}

So far, our analysis has neglected the issues of Gaussian overlap and depth sorting (see Fig.~\ref{fig:splatting}); yet both impact the rendering of Gaussian models.
3DGS renders Gaussians sequentially, based on their mean view space depth.
When Gaussians overlap in 3D, this causes popping artifacts due to abrupt changes in visibility, especially with large primitives (Fig.~\ref{fig:overlap_and_sorting}b).
StopThePop (STP) \cite{radl2024stopthepop} mitigates this issue by using per-pixel sorting. However, it does not resolve overlap during splatting (Fig.~\ref{fig:overlap_and_sorting}c).
Proper visibility resolution in the spirit of volumetric rendering requires modifications for integration along a ray, impacting performance.
To assess the importance of exact visibility for scene reconstruction and novel-view synthesis, we have designed a principled ray marching-based renderer for 3D Gaussians (Sec.~\ref{sec:raymarcher}): It evaluates the full Gaussian mixture in $f$ and accurately handles overlap (see Fig.~\ref{fig:overlap_and_sorting}d).

			
			
			
			
			

Additionally, projection distortions arise from the non-linear nature of perspective transformation, which is only approximated by the Jacobian at the Gaussians’ position (Equation \ref{eq:splatting_transformation_of_covariance}).
While reasonable for small Gaussians near the focal point, distortion becomes significant for larger Gaussians at peripheral regions.
Previous work partially addresses this projection inconsistency \cite{huang2024erroranalysis3dgaussian}, but cannot fully eliminate the approximations.
Again, our ray-marching implementation provides a testbed to examine the importance of precise perspective projection for 3D Gaussians.

\section{Implementation}
\label{sec:impl}
This section discusses the implementation details for the rendering algorithms.
We build these algorithms on top of the original 3DGS code base; this enables meaningful comparisons since the alterations for replacing each specific approximation are kept to a minimum.
First, we describe the modifications made to 3DGS for a consistent comparison, and to implement the EWA-based (OTS) variant and self-attenuation.
We continue by describing the details of our 3D Gaussian ray-marching solution.
All newly developed gradient computations for these modifications were rigorously validated through unit testing against numerical approximations.
Finally, we discuss the necessary changes to initialization.
All source code is available on \url{github.com/cg-tuwien/does_3d_gaussian_splatting_need_accurate_volumetric_rendering}.

\subsection{Modifying 3DGS-Based Splatting (3DGS \& 3DGS+STP)}
In 3DGS, the rendering process is split into several passes:
$\vr{\mu}'$ and $\mat{\Sigma}'$ are computed as described \revision{in Sec.~\ref{sec:bg_3dgs}} in a per-primitive pass.
Also in this pass, the RGB color is computed by evaluating spherical harmonics for the direction from camera to position $\vr{\mu}_i$.
Next, Gaussians are sorted first by affected screen-space tiles and then by depth for rendering.
Finally, in the rendering pass, for each pixel in a tile, its portion of the sorted list is traversed.
For each entry, the Gaussian's contribution is computed by evaluating the \revision{projected} 2D Gaussian, multiplied with the RGB color, and blended following Equation \ref{eq:alphablend}.
In practice, the traversal can stop early if the remaining transmittance \revision{$\prod_{j=0}^{i-1}(1 -f_i)$} drops below a given threshold.

In order to be consistent with the EWA-based variant and the ray marcher, we used a convolution instead of a dilation for anti-aliasing as described by \cite{yu2023mip}.
For StopThePop (3DGS+STP) \cite{radl2024stopthepop}, we used a configuration enabling hierarchical sorting with culling and their convolution \revision{correction} for anti-aliasing.

\subsection{EWA-Based Splatting (OTS)}
In the per-primitive pass (Sec.~\ref{sec:bg_3dgs}) we changed the computation of $a'$ to follow Eq.~\ref{eq:gaussian_weight_ots}.
$\vr{\mu}'$ and $\mat{\Sigma}'$ are computed as is.
In addition, we note that the $\mat{\Sigma}'$ is affected by perspective projection and can change for a given Gaussian, depending on its screen-space position.
Consequently, to preserve the integral of the extinction coefficient, the weight $a'$ must be multiplied by the determinant of the Jacobian, as done by Zwicker et al.\ \cite{ewa_splatting}.
In the rendering kernel, it is necessary to clamp $f_i(x)$, as 2D Gaussians can now produce values greater than 1 (see Sec.~\ref{sec:self_attenuation_and_exp}).
Incidentally, this clamp was already present in the 3DGS codebase for numerical stability \cite{kerbl3Dgaussians}.

\subsection{EWA-Based Splatting with Self-Attenuation (OTS+SAtn)}
\label{sec:impl_extinction_and_self_shadowing}
Implementing Eq.~\ref{eq:self_attenuation_blend} requires changing the render- or rasterisation kernel.
Doing so in forward rendering mode is trivial.
However, computing the gradient in the backward pass required switching from back-to-front traversal to front-to-back, requiring an extensive rewrite (see supplemental material).
The self-attenuation version does not use the Taylor approximation for $\texttt{exp}$ in the attenuation factor (Eq. \ref{eq:volumetric_integration_c}) and therefore a ray integral of $1$ would not result in an opacity value of $1$.
Hence, we cannot use the sigmoid activation, because Gaussians could not become fully opaque from all directions.
Instead, we use the softplus activation with $\beta = 2$.

\subsection{Ray Marching (OTS Marcher \& 3DGS Marcher)}
\label{sec:raymarcher}
Our ray marcher performs adaptive sampling along each ray to accurately resolve Gaussian overlap to compute the volumetric integral with a precision governed by the sample count.
Two variants were implemented, one follows the 3DGS model of opacity (3DGS Marcher), and the other the \revision{extinction-based} model with the OTS scheme (OTS Marcher), with differences being explained in Sec.~\ref{sec:anal_opa_vs_dens}.
Consequently, we can reuse the data structures (tensors with weights, positions, scales, and rotations), activation functions, and learning rates.
It is possible to ray march scenes learned with splatting, and vice-versa, but the reconstruction is less accurate.
The rendering pipeline shares similarities with 3DGS, including a per-primitive pass, identical tiling and sorting kernels, and a per-pixel rendering kernel.
However, the key distinction is the traversal and integration of 3D Gaussian primitives.

Ray marching fundamentally involves making incremental steps along a ray and evaluating the Gaussian mixture.
At each step, all Gaussians are intersected with the viewing ray, creating 1D Gaussians.
These are integrated analytically between step positions.
Effectively this creates bins of variable extent, each holding ray-mixture-integrals for opacity and colours.
For this to be accurate, it is essential to adapt bin borders to the Gaussian data.
We do this by iterating over the mixture twice, first establishing bin borders and then for analytical integration.
Finally, the bins are blended together following the volume integral in Eq.~\ref{eq:volumetric_integration}.
Further details are in the supplemental material. 
While performance is not the main focus of this paper, we note that our ray marcher, implemented in CUDA, is considerably slower than splatting (between one and two orders of magnitude, depending on the scene).
This is in spite of us following the guidelines for efficient GPU programming, highlighting the implementation challenges of principled rendering variants.
\revision{Although recent research demonstrates high performance for \emph{ray-tracing} Gaussians~\cite{3dgrt2024}, in contrast to \emph{ray marching}, their solution does not account for primitive overlap in the integration.}

\subsection{Initialization}

The original 3DGS solution initializes trainings of the NeRF-synthetic dataset~\cite{mildenhall2020nerf} by placing Gaussians randomly within the unit cube and setting a constant opacity.
Using this approach, the overall opacity increases with the number of Gaussians, preventing convergence for very large counts.
As a solution, we fit a power function that returns an appropriate opacity for the number of Gaussians used (\revision{see supplemental material for  details}).
This approach was repeated for the \revision{OTS and OTS+SAtn} variants.

\subsection{Learning Rates}
The learning rates for 3DGS were adjusted with densification in mind.
Therefore, we performed a grid search to find optimal learning rates for all algorithms on 5 of our test scenes with 60,000 Gaussians.
Results show that the OTS, and OTS+SAtn variants (splatters and marchers) all need a slightly lower learning rate for the weight.
All other learning rates are the same for all algorithms.
The exact numbers can be found in our code.

\section{Evaluation}
\label{sec:experiments}
We conducted several numerical experiments to determine which approximations have a significant impact on the reconstruction quality of 3D Gaussian splatting.
To avoid having too many different variables that would complicate analysis and to isolate each method's ability to optimize Gaussian primitives, we have removed the 3DGS densification logic from the optimization process and instead list the achieved metrics for different numbers of randomly initialized Gaussians.

\begin{figure}[h]
    \centering
	\begin{subfigure}[t]{0.24\linewidth}
		\includegraphics[width=\linewidth]{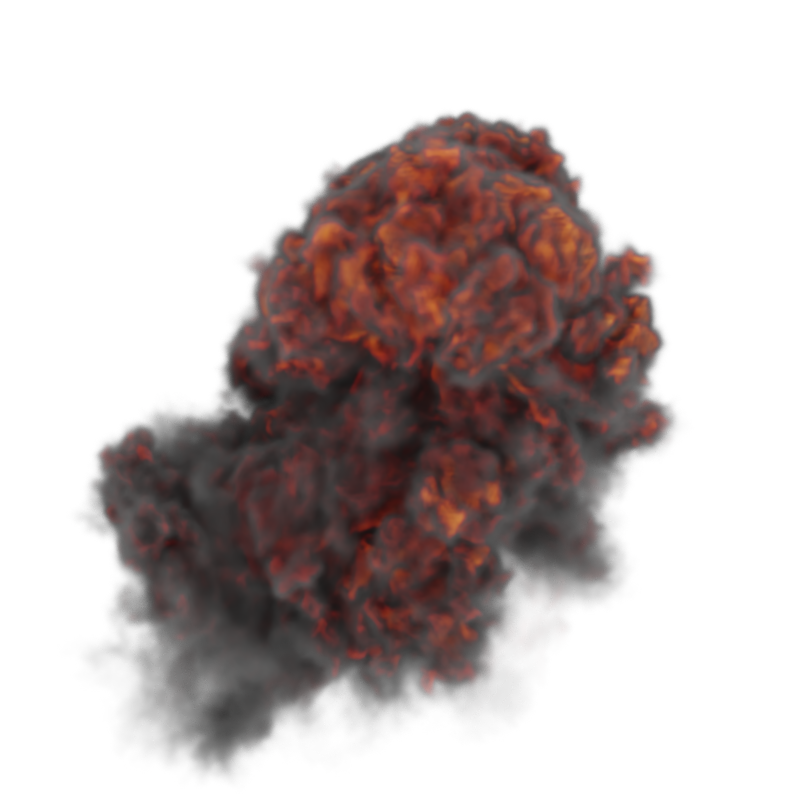}
		\caption*{\textsc{Explosion}}
	\end{subfigure}%
	\hfill
	\begin{subfigure}[t]{0.25\linewidth}
		\includegraphics[width=\linewidth]{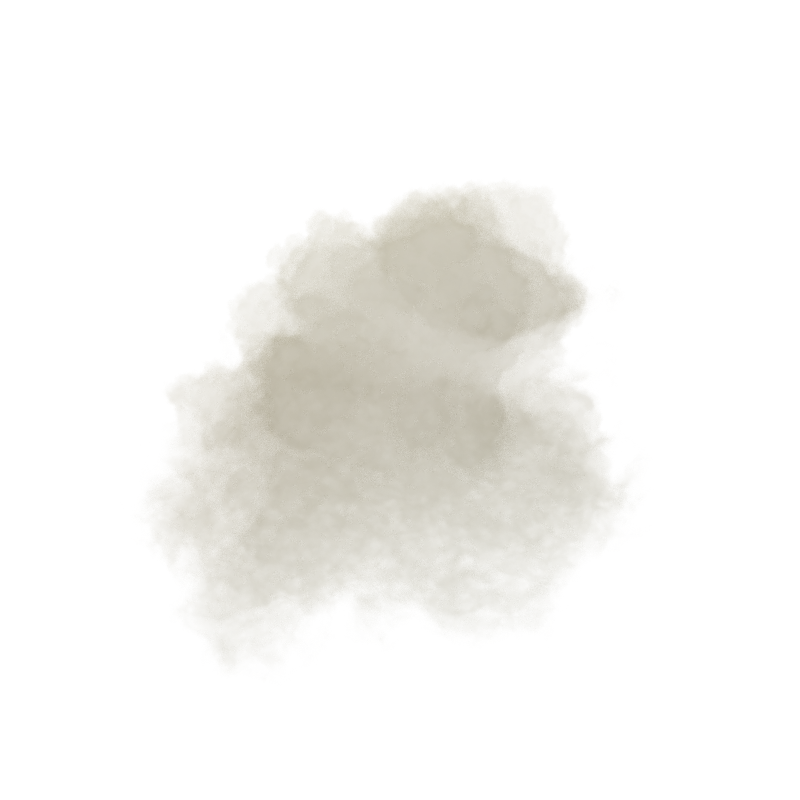}
		\caption*{\textsc{Cloud 2}}
	\end{subfigure}%
	\hfill
	\begin{subfigure}[t]{0.25\linewidth}
		\includegraphics[width=\linewidth]{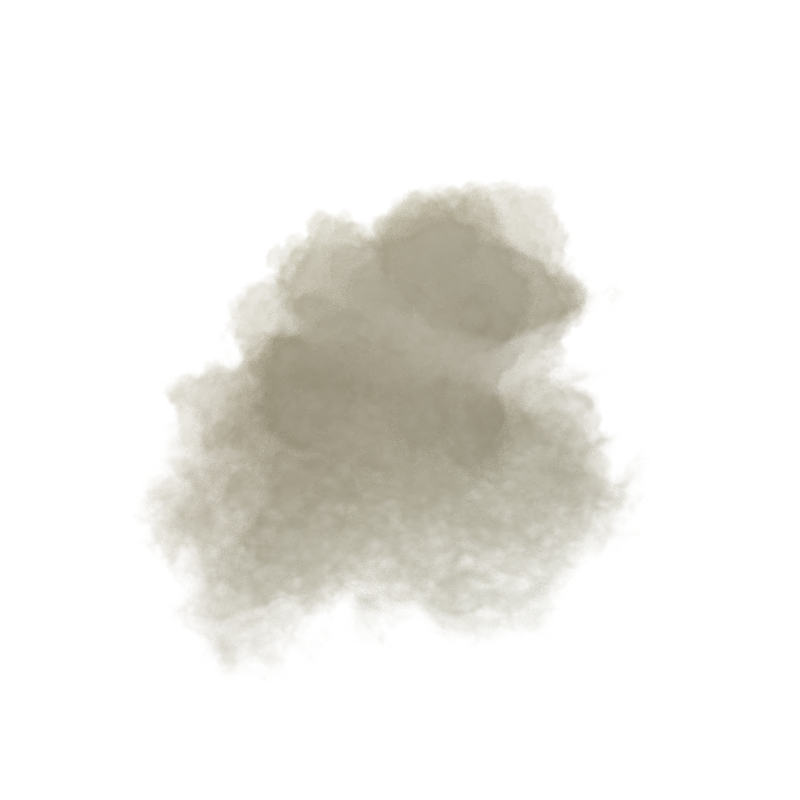}
		\caption*{\textsc{Cloud 3}}
	\end{subfigure}%
	\hfill
	\begin{subfigure}[t]{0.26\linewidth}
		\includegraphics[width=\linewidth]{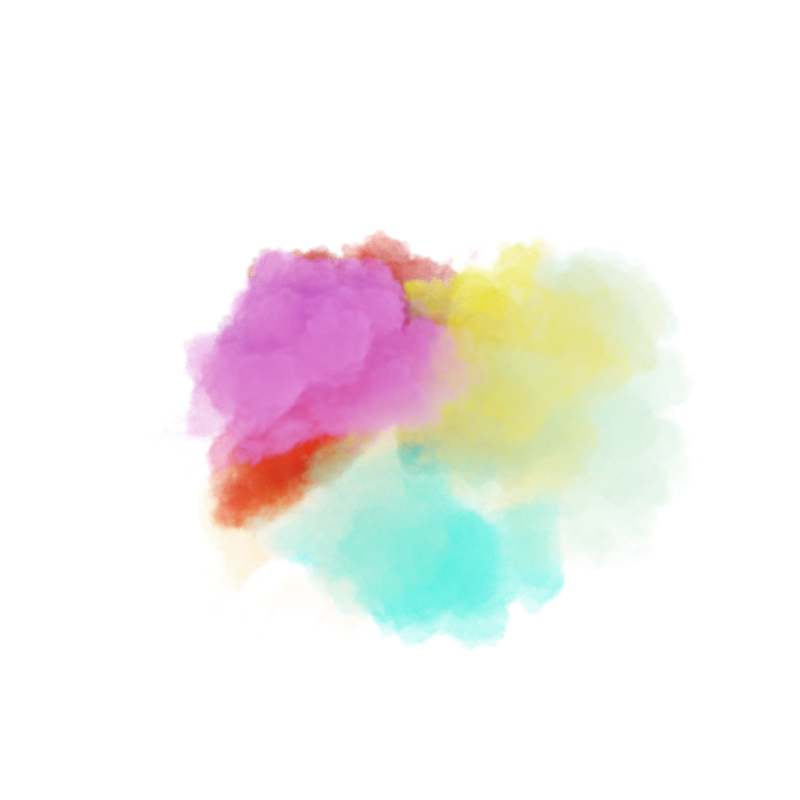}
		\caption*{\textsc{Colored Cloud}}
	\end{subfigure}%
    \caption{Four volumetric scenes, from left to right: \textsc{Explosion}, \textsc{WDAS Cloud} 2 and 3, \textsc{Colored WDAS Cloud}.}
    \label{fig:somefigs}
\end{figure}

In a similar spirit, we perform our evaluation on the established NeRF synthetic dataset~\cite{mildenhall2020nerf}, which provides a well-defined environment with exhaustive camera coverage, such that scenes can be optimized from randomly initialized Gaussians. However, these scenes focus mostly on solid objects:
to further assess the impact of correct volumetric rendering in detail, we created additional volumetric datasets with varying parameters for transparency, frequency, scattering coefficients, and color.
These volumetric scenes are shown in Fig.~\ref{fig:somefigs}, \ref{fig:example_burning_ficus}, and \ref{fig:example_materials}. 

\begin{table*}[htbp]
\centering
\caption{Averages over all scenes (individual results in the supplemental material). Red is best, orange second and yellow is third.}
\newcommand{\first}[1]{\cellcolor{red!30}#1}

\newcommand{\second}[1]{\cellcolor{orange!30}#1}

\newcommand{\third}[1]{\cellcolor{yellow!30}#1}

\centering
\scalebox{0.93}
{
\setlength{\tabcolsep}{3pt} 
\begin{tabular}{lc|c|c|c|c|c|c|c|c|c|c|c|c|c|c|c|c|c}
	& \multicolumn{16}{c}{Number of Gaussians
 } \\
Algorithm & \multicolumn{3}{c|}{4\,000} & \multicolumn{3}{c|}{12\,000} & \multicolumn{3}{c|}{36\,000} & \multicolumn{3}{c|}{100\,000} & \multicolumn{3}{c|}{330\,000} & \multicolumn{3}{c}{1\,000\,000} \\
\hline
& \tiny SSIM & \tiny PSNR & \tiny LPIPS & \tiny SSIM & \tiny PSNR & \tiny LPIPS & \tiny SSIM & \tiny PSNR & \tiny LPIPS & \tiny SSIM & \tiny PSNR & \tiny LPIPS & \tiny SSIM & \tiny PSNR & \tiny LPIPS & \tiny SSIM & \tiny PSNR & \tiny LPIPS\\
\midrule
3DGS           &         .9320  &         31.63  &         .1231         &         .9471  &         33.22  &         .1007  &                .9596  &         34.73  &         .0800  &                .9684  &  \third{36.05} &         .0632  &         \third{.9736} & \second{37.02} &         .0523  &         \first{.9760} & \second{37.58} &  \first{.0467} \\
3DGS+STP       &         .9322  &         31.69  &  \third{.1222}        &         .9476  &         33.30  &         .0995  &                .9601  &         34.81  &         .0790  &         \third{.9688} &  \first{36.15} &         .0625  &        \second{.9738} &  \first{37.10} &  \third{.0520} &         \first{.9760} &  \first{37.62} &  \first{.0467} \\
OTS            &  \third{.9394} &         32.17  &  \first{.1115}        &  \third{.9524} &  \third{33.66} &  \first{.0906} &                .9619  &         34.83  & \second{.0742} &                .9686  &         35.78  &  \third{.0611} &                .9729  &         36.52  &  \first{.0517} &                .9747  &         36.93  & \second{.0471} \\
OTS+SAtn       &  \third{.9394} & \second{32.24} &  \third{.1122}        & \second{.9526} & \second{33.72} &  \third{.0913} &         \third{.9621} &  \third{34.90} &         .0744  &         \third{.9688} &         35.81  &  \third{.0611} &                .9730  &         36.51  & \second{.0518} &                .9748  &         36.93  & \second{.0471} \\
3DGS Marcher   & \second{.9401} &  \third{32.23} &  \first{.1115}        &  \first{.9536} &  \first{33.77} & \second{.0908} &         \first{.9635} &  \first{35.08} &  \first{.0736} &         \first{.9701} & \second{36.11} &  \first{.0606} &         \first{.9739} &  \third{36.79} &         .0521  &        \second{.9756} &  \third{37.14} &  \third{.0477} \\
OTS Marcher    &  \first{.9402} &  \first{32.28} & \second{.1116}        &  \first{.9536} &  \first{33.77} & \second{.0908} &        \second{.9629} & \second{34.94} &  \third{.0743} &        \second{.9694} &         35.85  & \second{.0610} &                .9732  &         36.45  &         .0523  &         \third{.9749} &         36.77  &         .0481  \\

	\bottomrule
\end{tabular}
}

\label{tab:results_psnr}
\end{table*}

\subsection{Results}

For each scene and rendering method described so far, we evaluate three established image-quality metrics (SSIM, PSNR, LPIPS).
For each rendering method, the optimization described by 3DGS is performed for 30k iterations. Our supplemental video shows several scenes rendered with the different algorithms at 4k and 1 million Gaussians. The videos help to convey the popping issue with 3DGS, as well as illustrate the quality trends discussed below.

\begin{figure}
	\centering
	\includegraphics[width=\linewidth]{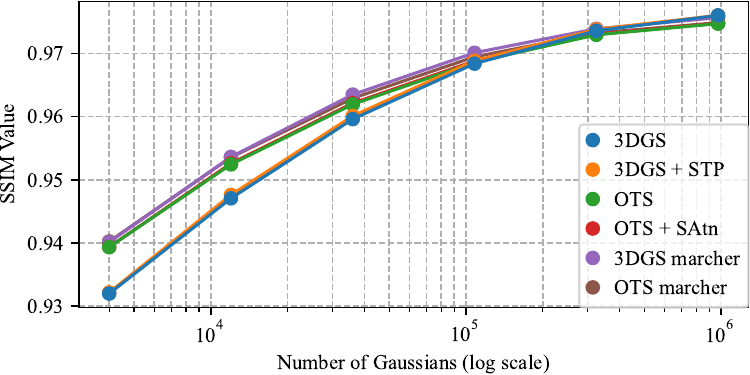}
	\caption{SSIM averages over all scenes for each model size.}
	\label{fig:results_ssim_avg_all}
\end{figure}

\begin{figure}
\centering
\includegraphics[width=\linewidth]{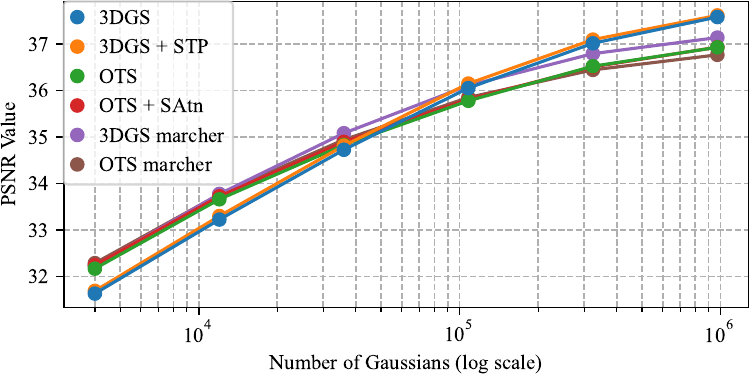}
\caption{PSNR averages over all scenes for each model size.}
\label{fig:results_psnr_avg_all}
\end{figure}

\begin{figure}
\centering
\includegraphics[width=\linewidth]{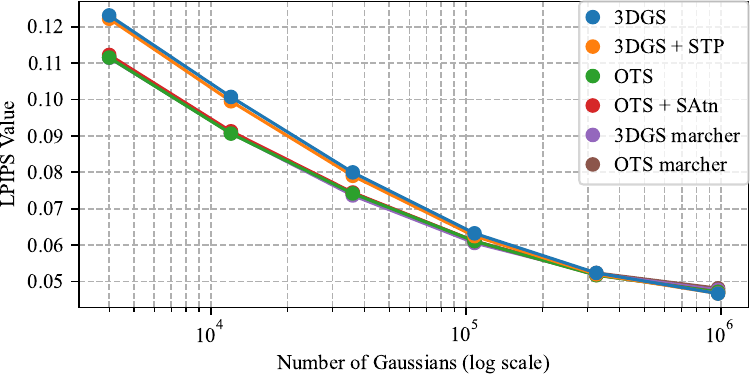}
\caption{LPIPS averages over all scenes for each model size.}
\label{fig:results_lips_avg_all}
\end{figure}

\begin{figure*}
	\centering
        \hspace*{\fill}
	\begin{subfigure}[t]{0.16\linewidth}
		\includegraphics[width=\linewidth]{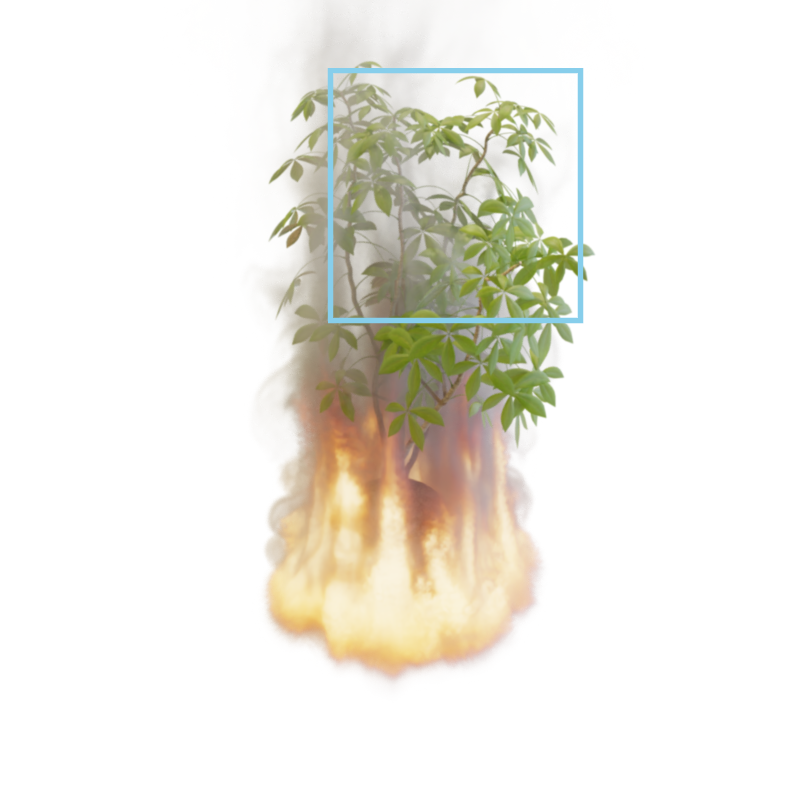}
		\caption{Overview}
	\end{subfigure}%
	\hfill
	\begin{subfigure}[t]{0.16\linewidth}
		\includegraphics[width=\linewidth]{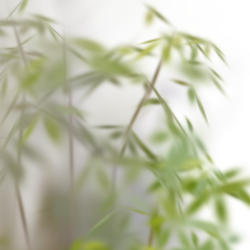}
		\caption{3DGS 4k}
	\end{subfigure}%
	\hfill
	\begin{subfigure}[t]{0.16\linewidth}
	\includegraphics[width=\linewidth]{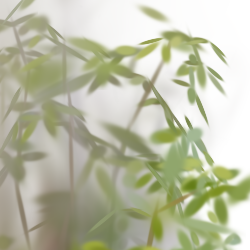}
	\caption{OTS + SAtn 4k}
	\end{subfigure}%
	\hfill
	\begin{subfigure}[t]{0.16\linewidth}
	\includegraphics[width=\linewidth]{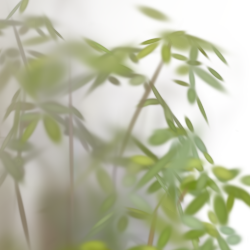}
	\caption{3DGS Marcher 4k}
	\end{subfigure}%
	\hfill
	\begin{subfigure}[t]{0.16\linewidth}
	\includegraphics[width=\linewidth]{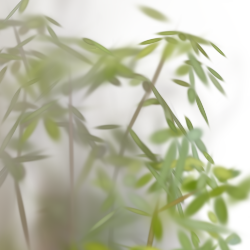}
	\caption{OTS Marcher 4k}
	\end{subfigure}%
	\hspace*{\fill}

 \hspace*{\fill}
	\begin{subfigure}[t]{0.16\linewidth}
	\includegraphics[width=\linewidth]{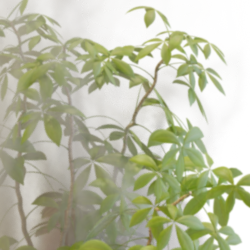}
	\caption{Ground Truth}
	\end{subfigure}%
	\hfill
	\begin{subfigure}[t]{0.16\linewidth}
	\includegraphics[width=\linewidth]{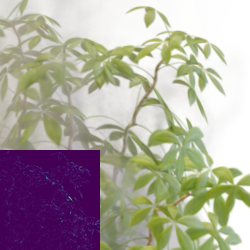}
	\caption{3DGS 1M}
	\end{subfigure}%
	\hfill
	\begin{subfigure}[t]{0.16\linewidth}
	\includegraphics[width=\linewidth]{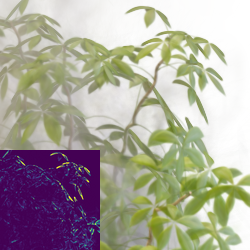}
	\caption{OTS + SAtn 1M}
	\end{subfigure}%
	\hfill
	\begin{subfigure}[t]{0.16\linewidth}
	\includegraphics[width=\linewidth]{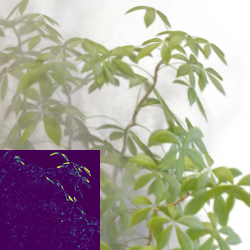}
	\caption{3DGS Marcher 1M}
	\end{subfigure}%
	\hfill
	\begin{subfigure}[t]{0.16\linewidth}
	\includegraphics[width=\linewidth]{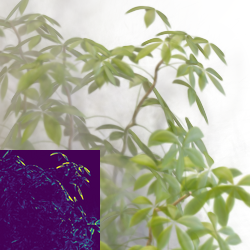}
	\caption{OTS Marcher 1M}
	\end{subfigure}%
 \hspace*{\fill}
	\caption{\textsc{Burning Ficus} with 4k and 1M Gaussians. The OTS renderers and the 3DGS Marcher have an advantage when using few Gaussians. In these cases, they manage to achieve a crispier reconstruction of delicate structures. Volumetric effects like smoke are unaffected. For the second row, squared error w.r.t.\ ground truth is shown in bottom left corner.}
	\label{fig:example_burning_ficus}
\end{figure*}

Tab.~\ref{tab:results_psnr} lists the results obtained over all tested scenes, and Fig.~\ref{fig:results_ssim_avg_all} through \ref{fig:results_lips_avg_all} illustrate them visually. Per-scene results are provided in our supplemental material. With the exception of special cases, which we discuss below, the full results establish several insightful trends. The most striking is the dependence on model size: while ray marchers with correct volumetric integration are superior for lower Gaussian counts, they are eventually matched or surpassed by 3DGS as the number of Gaussians increases.
A similar trend can be observed for EWA-based splatters (OTS and OTS+SAtn): they also perform better than 3DGS for small model sizes but are outperformed for larger ones. Concrete examples of this behavior are given in Figure~\ref{fig:teaser} and \ref{fig:example_burning_ficus}. In the following, we discuss these trends for individual variants in detail and explore their causes.

\paragraph*{3DGS-Based Splatting vs. Ray Marching}
Compared to 3DGS, the presented 3DGS ray-marching variant is more faithful to volumetric integration. In addition to exact projection, it avoids discontinuities in the optimization due to the strict per-primitive sorting, which affects both 3DGS (popping) and STP (see Fig.~\ref{fig:overlap_and_sorting}). Discontinuities and projection errors are most notable when Gaussians are large, which explains the advantage of the 3DGS Marcher with fewer primitives: Since there are fewer Gaussians to model the full scene, they also remain larger on average. Note that the more fine-granular ordering of STP yields a slight advantage over 3DGS. As the number of Gaussians increases, so does the ability to model fine scene details for all methods. With more Gaussians, the ability to consider overlap becomes insignificant, as Gaussians become so small that overlap is mostly avoided. In addition, 3DGS and STP have distinct advantages: first, in preserving opacity, they can trivially produce fine structures (anisotropic Gaussians) that appear solid from all sides. Second, discontinuities in the model via the enforced discrete primitive ordering can now be exploited to reproduce high-frequency detail and view-dependent effects.
Lastly, the computation of the volumetric integral becomes prone to numerical imprecision as the number of samples rises. Thus, the effectiveness of 3DGS is strongly tied to the use of larger primitive counts, which its fast rendering pointedly facilitates.

\begin{figure*}
	\centering
	\begin{subfigure}[t]{0.12\linewidth}
		\includegraphics[trim=0cm 0cm 0.3cm 0cm, clip,width=\linewidth]{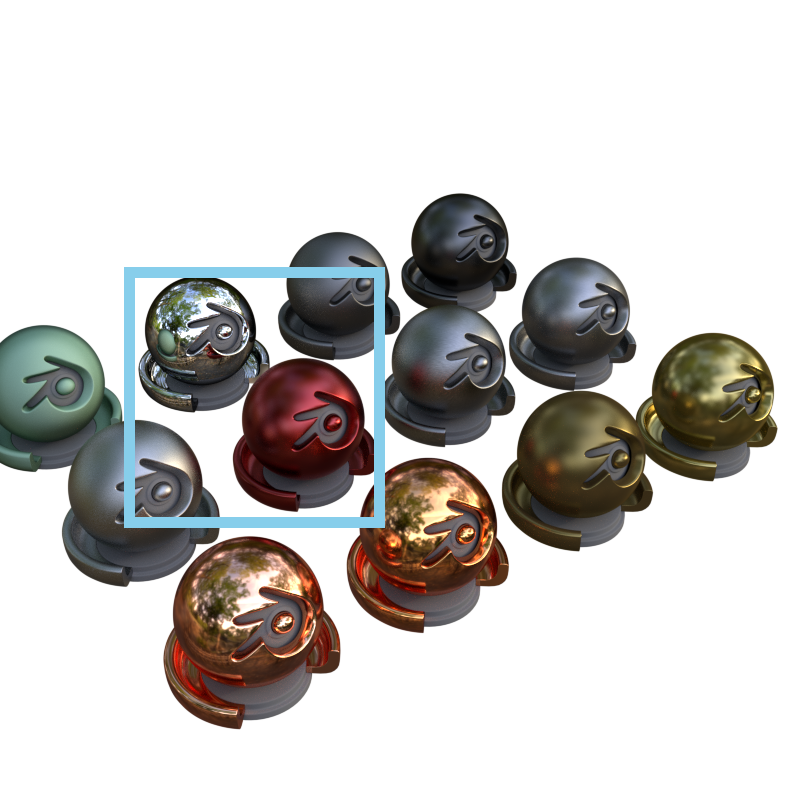}
		\caption{GT}
	\end{subfigure}%
	\hfill
	\begin{subfigure}[t]{0.12\linewidth}
		\includegraphics[trim=0cm 0cm 0.3cm 0cm, clip,width=\linewidth]{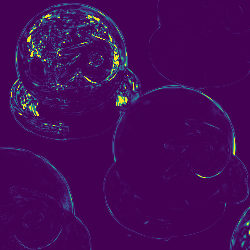}
		\caption{3DGS}
	\end{subfigure}%
	\hfill
	\begin{subfigure}[t]{0.12\linewidth}
		\includegraphics[trim=0cm 0cm 0.3cm 0cm, clip,width=\linewidth]{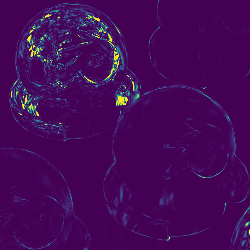}
		\caption{OTS Marcher}
	\end{subfigure}%
	\begin{subfigure}[t]{0.12\linewidth}
		\includegraphics[trim=0cm 0cm 0.3cm 0cm, clip,width=\linewidth]{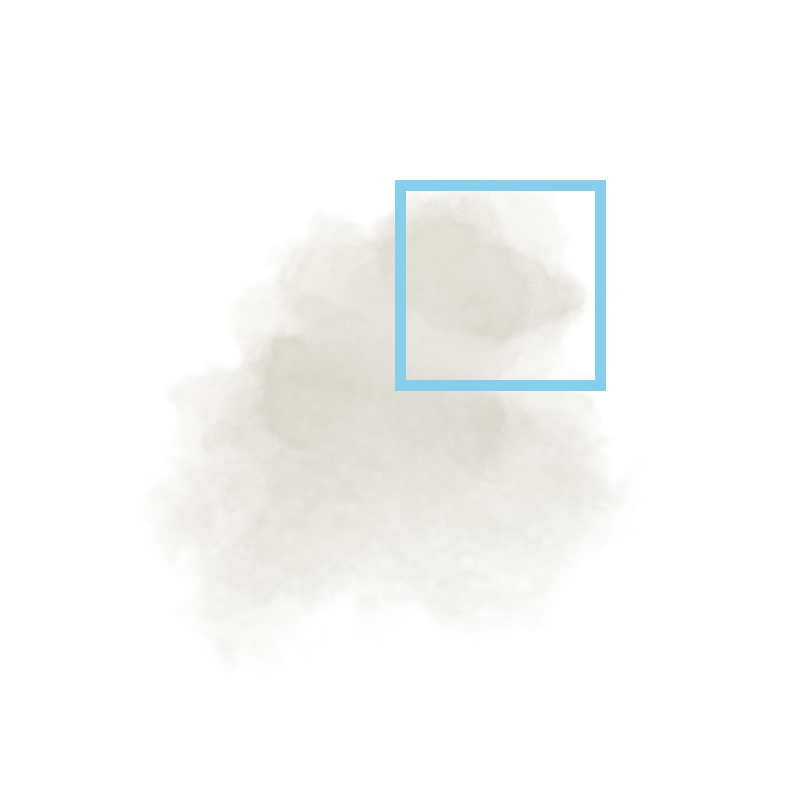}
		\caption{Overview}
	\end{subfigure}%
	\hfill
	\begin{subfigure}[t]{0.12\linewidth}
		\includegraphics[width=\linewidth]{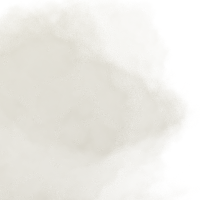}
		\caption{GT}
	\end{subfigure}%
	\hfill
	\begin{subfigure}[t]{0.12\linewidth}
		\includegraphics[width=\linewidth]{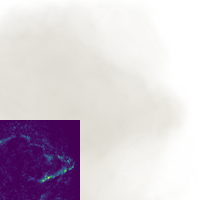}
		\caption{3DGS}
	\end{subfigure}%
	\hfill
	\begin{subfigure}[t]{0.12\linewidth}
		\includegraphics[width=\linewidth]{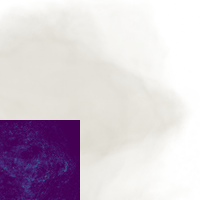}
		\caption{OTS Marcher}
	\end{subfigure}%
	\caption{(a--c)  \textsc{Materials} with 1M Gaussians:. Especially at grazing angles (object silhouettes), EWA-based approaches capture reflections more accurately, especially on object silhouettes, leading to lower squared error. (d--g) \textsc{WDAS Cloud} 1 scene with 1M Gaussians: Similar to \emph{Materials}, silhouettes are better captured with the more physically-inspired model. Squared error shown as insets.}
	\label{fig:example_materials}
\end{figure*}

An exception to this general trend can be observed in the \textsc{Materials} scene, where error metrics consistently favor the EWA-based methods OTS, OTS+SAtn, and OTS Marcher (see Table \ref{tab:results_materials_psnr}): 
Here, the image error is governed by reflections, especially at grazing angles on reflective objects (see Fig.~\ref{fig:example_materials}, left). This can be explained by the ability of EWA-based methods to produce Gaussians that are clearly visible at one angle (silhouettes), but see-through at others (frontal view) to model reflective Fresnel-like effects.


\newcommand{\first}[1]{\cellcolor{red!30}#1}

\newcommand{\second}[1]{\cellcolor{orange!30}#1}

\newcommand{\third}[1]{\cellcolor{yellow!30}#1}

\begin{table}
		\caption{PSNR error metrics (db) for the \textsc{Materials} scene.}
  \scalebox{0.89}
	{
		\setlength{\tabcolsep}{3pt} 
		\begin{tabular}{lr|r|r|r|r|r}
		\toprule
		& \multicolumn{6}{c}{Number of Gaussians} \\
		Algorithm & 4 000 & 12 000 & 36 000 & 100 000 & 330 000 & 1 000 000 \\
		\midrule
		3DGS & 22.48 & 24.42 & 26.23 & 27.74 & 28.81 & 29.54 \\
		3DGS + STP & 22.46 & 24.50 & 26.34 & 27.84 & 28.85 & 29.54 \\
		OTS & \second{23.65} & \second{25.62} & \first{27.28} & \third{28.51} & \third{29.43} & \second{29.95} \\
		OTS + SAtn & \third{23.59} & \third{25.61} & \second{27.26} & \first{28.59} & \second{29.44} & \first{29.96} \\
		3DGS marcher & 23.39 & 25.52 & 27.08 & 28.21 & 29.09 & 29.66 \\
		OTS marcher & \first{23.68} & \first{25.68} & \third{27.24} & \second{28.58} & \first{29.45} & \third{29.94} \\
		\bottomrule
		\end{tabular}
	}

	\label{tab:results_materials_psnr}
\end{table}


\paragraph*{3DGS-Based vs. EWA-Based Rendering}
Our test results show that for a small number of Gaussians, the EWA-based splatting solutions---OTS and OTS+SAtn---perform better than 3DGS-based splatting, closely matching the advantage of the slower ray-marching methods.
This aligns with the above trends: enforcing volumetric rendering properties yields better results with fewer Gaussians.
However, there is no discernible difference between OTS and OTS+SAtn, suggesting that self-attenuation is negligible for reconstruction quality.
As before, increasing the number of Gaussians eventually gives 3DGS-based splatting an advantage: It can more easily produce fine anisotropic structures (e.g., stripes, poles, branches) with identical opacity from all directions, which is particularly useful when modeling solid objects.
Accordingly, the volumetric, low-density \textsc{WDAS Cloud 1} scene is an exception to this trend: here, the more principled, EWA-based variants show an advantage regardless of model size (Fig.~\ref{fig:example_materials}, right).

We observe similar results when comparing the 3DGS Marcher and OTS Marcher: while the latter performs slightly better for small models, it is eventually surpassed as the model size grows. These results seem to confirm that 3DGS-based rendering---i.e., preserving a Gaussian's amplitude across views---is more effective in large models than conventional volumetric integration.




\revision{

\subsection{Training and Rendering Performance}
Our ray marching implementations are intended to provide guidelines or achievable quality with correct volume rendering and are not optimized, thus training a single scene can take several hours. All splatting methods require similar computation time (10--20 minutes for training), however, we assess more nuanced differences to identify potential tradeoffs and benefits in particular settings: OTS trains the fastest consistently, with a 3\% \emph{speedup} compared to 3DGS. For small models (4k), OTS+SAtn optimizes 3\% slower than 3DGS on average; 
Its overhead gradually rises as models become larger and plateaus at 10\% for 1M Gaussians.
During rendering, OTS and OTS+SAtn are, on average, 28\% and 43\% slower, respectively, across all scenes.
No trend with regard to number of Gaussians is visible.
}

\revision{
\section{Limitations and Future Work}
Our evaluation investigated quality and performance on small-scale scenes because they explicitly do not require \emph{densification}. 
Our assessment considers multiple parameters, (opacity vs. density, overlap, and self-attenuation) whose impact we want to isolate and quantify. Densification introduces complexity that aggravates objective conclusions, e.g., 3DGS densification is governed by gradient magnitude, and thus can yield different numbers of Gaussians for techniques with different gradient flow. However, we provide tentative results for larger scenes in our supplemental material. 

We also did not explore specialized use cases (relighting, sparse captures, dynamic scenes) where our new variants could potentially exceed previous work; This analysis is beyond the scope of this paper. However, the increased accuracy in the \textsc{Materials} scene suggests that EWA-based splatting helps reconstruct extreme material properties and thus might benefit inverse rendering use cases.
}

\section{Discussion and Conclusion}

Our in-depth analysis of Gaussian-primitive rendering approaches allowed us to clarify the distinction between using volumetric extinction (EWA-based) and ``opacity'' (3DGS-based) rendering in clear mathematical terms, which had caused significant confusion in the past. We also clearly identified several other approximations used in 3DGS. 
To allow careful experimental analysis of the effect of each approximation, we defined a set of algorithms which progressively replaced each approximation used in 3DGS with the principled volumetric rendering approach.

A key insight of this analysis is that accurate integration of Gaussians (e.g., using ray marching) provides advantages for a low number of Gaussians.
However, our results also show that, in practice, the advantage of these slow, principled methods in small models is closely matched by our EWA-based splatting method (OTS), preserving the rendering and optimization speed of the original 3DGS.

On the other hand, our experiments show that using the simpler splatters and ``opacity''-based models provides better results as the number of Gaussians increases.
We hypothesize that this behavior is due to the simpler approach providing a more opportunistic optimization landscape, exploiting, e.g., discontinuities to model finer details.
Thus, rather than using a complex, principled image formation, 3DGS relies on using larger models, paired with an inexpensive extinction function and a more approximate visibility resolution, which turns out to be most effective with more primitives.

Our analysis provides the mathematical foundation for understanding how 3DGS relates to volume rendering, offering insight into the effectiveness of the algorithm. We hope that our analysis will lead to further theoretical and practical developments, e.g., allowing interoperability between NeRF and 3DGS representations.

\section{Acknowledgments}
G. Drettakis was supported by the ERC Advanced Grant FUNGRAPH (788065, \url{https://project.inria.fr/fungraph}), and acknowledges support from NVIDIA and Adobe.

\bibliographystyle{eg-alpha-doi} 
\bibliography{article}       


\end{document}


\maketitle

\section{Proof for Equation used for Self-Attenuation}
\label{ap:proof_volumetric_equation_single_gaussian_closed_form}

\revision{
Given the computation:
\begin{align}
	I(\pixel) = c_0(\ray)\int_{-\infty}^{\infty} \mathcal{G}_3^n(\ray(t), 0) e^{-\int_{-\infty}^t \mathcal{G}_3^n(\ray(\tau), 0)\diff \tau} \diff t,
\end{align}
where $\G^n$ are normalised Gaussians, $f_0$ is a 2D extinction function:
\begin{align}
	f_0(\vr{p}) = \G_2^n(\pixel, w_0, \vr{\mu}'_0, \vr{\Sigma}'_0) = \int_{-\infty}^{\infty} \G_3^n(\ray(t), w_0, \vr{\mu}_0, \vr{\Sigma}_0)\diff t,
\end{align}
and $\ray$ is the ray going through pixel $\pixel$.
\\
\\
1. \textbf{Define the inner integral}:
\[
H(t) = \int_{-\infty}^t \mathcal{G}_3^n(\ray(\tau), 0) d\tau
\]
2. \textbf{Differentiate \( H(t) \)}.
By the Fundamental Theorem of Calculus:
\[
\frac{dH(t)}{dt} = \mathcal{G}_3^n(\ray(\tau), 0)
\]
3. \textbf{Rewrite the integral}.
Substitute \( \mathcal{G}_3^n(\ray(\tau), 0) dt \) with \( dH(t) \):
\[
\int_{-\infty}^{\infty} \mathcal{G}_3^n(\ray(\tau), 0) e^{-H(t)} dt = \int_{-\infty}^{\infty} e^{-H(t)} dH(t)
\]
4. \textbf{Integral evaluation}:
\[
\int_{-\infty}^{\infty} e^{-H(t)} dH(t)
\]
Given that \( H(t) \) for the Gaussian distribution is bounded and converges from $0$ to \( f_0(\pixel) \) as \( t \) goes from \(-\infty\) to \( \infty \), we can write:
\[
\int_{0}^{f_0(\pixel)} e^{-u} du
\]
5. \textbf{Evaluate the integral}:
\[
\int_{0}^{f_0(\pixel)} e^{-u} du = \left[ -e^{-u} \right]_{0}^{f_0(\pixel)} = -e^{-f_0(\pixel)} + e^{0} = 1 - e^{-f_0(\pixel)}
\]
Thus, the integral evaluates to \( 1 - e^{-f_0(\pixel)} \).
}

\revision{

\section{Implementation Details}

For initialization, we use the same policy that the original 3DGS codebase employed for training Nerf-Synthetic: here, the unit cube around the scene origin was initialized with 100k Gaussians with random positions and uniform opacity, i.e, all Gaussians were initialized to the same opacity / OTS weight.

We initialize the setup with 4k/12k/36k, etc., Gaussians, according to the experiment. We also compute a single value for all Gaussians to be used, however, ours is derived via the power function: Due to their different image formation solutions, we use $\frac{2}{N^{0.35}}$ and $\frac{2}{N^{0.55}}$ for 3DGS and OTS-based methods, respectively, where $N$ is the total number of Gaussians. 

\section{Results on Real-World Scenes}
To provide tentative evaluations on real-world datasets, we propose an exploration of larger scenes with the following experiment: we adapt the densification variables in our most principled splatting technique, OTS+SAtn, to match 3DGS as closely as possible and stop densification after reaching 100k/500k/1m/5m Gaussians. While this procedure is not as controlled as our main setup, it suffices to eliminate egregious discrepancies in primitive count, while still making use of densification.

Evaluating the outcomes on the MipNeRF360 dataset, we did not find significant trends for quality: across all tested sizes and scenes, 3DGS and OTS+SAtn differ by at most 4\% PSNR. Average training times were slightly lower (2\%) for OTS+SAtn, across all sizes. Similarly, rendering times are balanced, although with comparably high variance (OTS+SAtn 10\% faster in \textsc{Room}, 15\% slower in \textsc{Bonsai}). This suggests that the approximations used in the 3DGS visibility model also do not impact its effectiveness and robustness in larger, real-world scenarios.
}

\section{Details of the Backward Pass for Self-Attenuation Based Splatting}
\label{ap:self_attenuation_details}
3DGS implements the backward pass exactly.
That is, in forward mode the Gaussians are blended front-to-back, and in backward mode the gradient is computed back-to-front.
For this to work, the index of the last Gaussian and the remaining opacity is stored in forward mode.
Gaussians retrieved via this index, and remaining opacity are then used in the backward pass to recover all intermediate values.
In order to keep the numerical error in check, the forward mode blending is stopped \emph{before} the remaining opacity becomes too small.
However, this approach causes heavy artefacts if self-attenuation is used.
The reason is, that the linear falloff of the Taylor approximation is replaced with a true exponential falloff.
Therefore, we compute the gradient in the backward pass with a front-to-back strategy.
This works, because volumetric rendering can be performed either front-to-back or back-to-front, hence we can also compute the gradient in either direction.

\section{Details of the Ray Marching Algorithm}
\label{ap:ray_marching_details}
\paragraph*{Batching.}
For computational efficiency, we divide the work into batches of e.g. 128 bins.
Within each batch, the mixture is iterated twice as described in the main text.
The first batch starts at $t=0$ and ends at an position adapted to the data.
The next batch is started at the end-border of the last bin.
This repeats until we run out of Gaussians or accumulated opacity reaches a high enough threshold.

\paragraph*{Adaptive binning.}
This process starts by creating a fixed-size \emph{density section buffer}, where a sections are defined by start, end, and density.
Then, we iterate over all Gaussians to fill the buffer.
Initially, each Gaussian produces one sampling section by taking its position $\pm$ 3 standard deviations for start and end, and computing the density such that a pre-defined number of bins fit in this space.
We then add this section to the density section buffer.
When adding a section, existing and new sections are combined in such a way, that the resulting sections are in order and without overlap.
If sections do overlap during construction, the section with higher sampling density wins in the overlap area.
Accordingly, combining 2 sections can create up to 3 new sections.
Sections that fall out of the buffer are deferred to the next batch.

Once the buffer is filled, the bins are created by traversing the sections such that the size of each bin is inversely proportional to the density.

\paragraph*{Blending}
We then iterate over Gaussians again to compute integrals for each bin.
This results in 4 values per bin: red, green, blue, and integrated extinction (opacity).
Following Eq.~10 of the paper, colours are multiplied with per Gaussian integrated extinction.
Finally, the bin evaluations are blended together using
\begin{align}
	I(\vr{x}) = \sum_{i=0}^\infty \gamma_i \prod_{j=0}^{i-1} e^{- \rho_j} + c_b \prod_{j=0}^\infty e^{-\rho_j},
\end{align}
where the sum and products iterate over bins, $\gamma$ is one of the colours and $\rho$ the integrated extinction (opacity).
$\prod_{j=0}^{i-1} e^{- \rho_j}$ is the transmittance factor.
It is cached between iteration such that the product is computed only once.

\paragraph*{Gradient computation.}
The gradient was implemented manually using a similar front-to-back strategy as described in Sec. \ref{ap:self_attenuation_details}.
Fundamentally, there are two methods to compute a gradient for numerical integration methods: attached and detached \cite{zeltner2021monte}.
An attached gradient takes the numerical integration procedure into account.
In Monte Carlo integration that is sampling probabilities, or in our case the border positions.
A detached gradient, on the other hand, assumes that the numerical method can be ignored.
It therefore makes sampling or border positions constant with respect to the differentiation.
We implemented both variants and found, that the detached variant performs significantly better.
All gradient computations were verified by testing against a numerical variant.

\bibliographystyle{eg-alpha-doi} 
\bibliography{article}      

\onecolumn

\section{Aggregated Numerical Results}
\label{ap:avg_results}
In the the tables we show averages for various error metrics over all, the nerf, and the volumetric scenes.
Full numerical results, and error plots for each scene are further below.

\subsection{PSNR}
\begin{table}[h!]
\centering
\caption{Average PSNR results for all scenes.}
\label{tab:psnr_avg}
\begin{tabular}{lrrrrrr}
\toprule
& \multicolumn{6}{c}{Number of Gaussians} \\
Algorithm & 4000 & 12000 & 36000 & 108000 & 324000 & 972000 \\

\midrule
3DGS            & 31.63 & 33.22 & 34.73 & 36.05 & 37.02 & 37.58 \\
3DGS + STP      & 31.69 & 33.30 & 34.81 & 36.15 & 37.10 & 37.62 \\
OTS             & 32.17 & 33.66 & 34.83 & 35.78 & 36.52 & 36.93 \\
OTS + SAtn      & 32.24 & 33.72 & 34.90 & 35.81 & 36.51 & 36.93 \\
3DGS marcher    & 32.23 & 33.77 & 35.08 & 36.11 & 36.79 & 37.14 \\
OTS marcher     & 32.28 & 33.77 & 34.94 & 35.85 & 36.45 & 36.77 \\
\bottomrule
\end{tabular}
\end{table}

\begin{figure}[h!]
	\centering
	\includegraphics[width=0.7\linewidth]{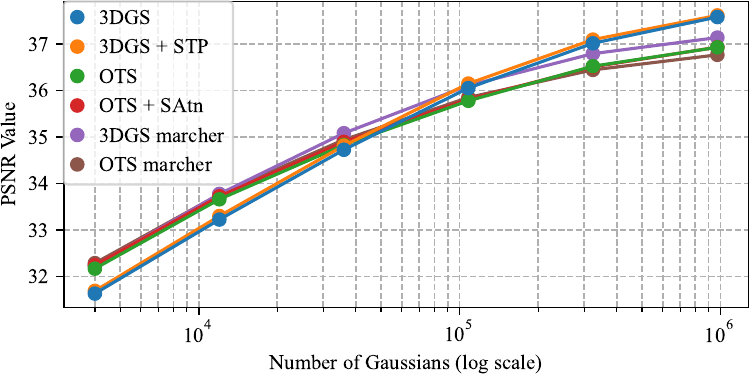}
	\caption{PSNR average over all scenes}
	\label{fig:ap_psnr_avg_all_plot}
\end{figure}
\pagebreak

\begin{table}[h!]
\centering
\caption{Average PSNR results for volumetric scenes}
\label{tab:psnr_avg_interest}
\begin{tabular}{lrrrrrr}
\toprule
& \multicolumn{6}{c}{Number of Gaussians} \\
Algorithm & 4000 & 12000 & 36000 & 108000 & 324000 & 972000 \\

\midrule
3DGS & 39.37 & 40.78 & 42.15 & 43.29 & 44.01 & 44.33 \\
3DGS + STP & 39.44 & 40.85 & 42.23 & 43.43 & 44.12 & 44.43 \\
OTS & 39.59 & 40.98 & 42.06 & 42.91 & 43.43 & 43.71 \\
OTS + SAtn & 39.66 & 41.04 & 42.15 & 42.93 & 43.38 & 43.71 \\
3DGS marcher & 39.51 & 41.03 & 42.31 & 43.15 & 43.59 & 43.66 \\
OTS marcher & 39.66 & 41.04 & 42.20 & 42.98 & 43.34 & 43.48 \\
\bottomrule
\end{tabular}
\end{table}

\begin{figure}[h!]
	\centering
	\includegraphics[width=0.7\linewidth]{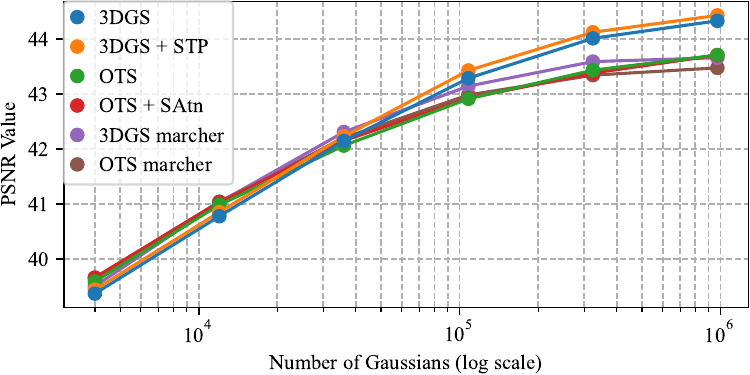}
	\caption{PSNR average over volumetric scenes}
	\label{fig:ap_psnr_avg_vol_plot}
\end{figure}
\pagebreak

\begin{table}[h!]
\centering
\caption{Average PSNR results for Nerf scenes}
\label{tab:psnr_avg_others}
\begin{tabular}{lrrrrrr}
\toprule
& \multicolumn{6}{c}{Number of Gaussians} \\
Algorithm & 4000 & 12000 & 36000 & 108000 & 324000 & 972000 \\

\midrule
3DGS & 25.82 & 27.56 & 29.16 & 30.63 & 31.77 & 32.52 \\
3DGS + STP & 25.87 & 27.63 & 29.25 & 30.69 & 31.83 & 32.51 \\
OTS & 26.60 & 28.16 & 29.40 & 30.43 & 31.34 & 31.85 \\
OTS + SAtn & 26.67 & 28.24 & 29.45 & 30.47 & 31.36 & 31.85 \\
3DGS marcher & 26.77 & 28.33 & 29.66 & 30.83 & 31.70 & 32.25 \\
OTS marcher & 26.76 & 28.31 & 29.50 & 30.50 & 31.27 & 31.74 \\
\bottomrule
\end{tabular}
\end{table}

\begin{figure}[h!]
	\centering
	\includegraphics[width=0.7\linewidth]{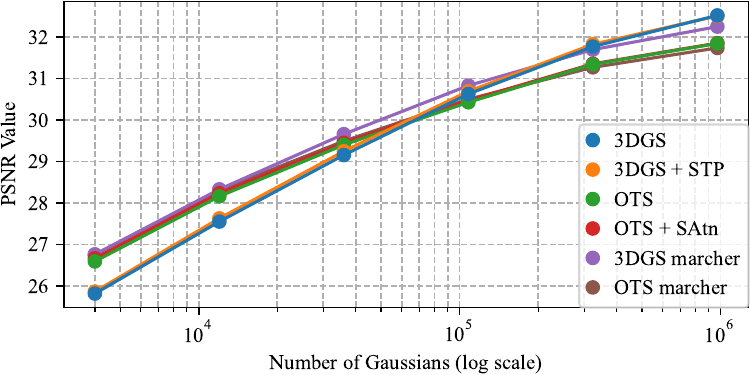}
	\caption{PSNR average over Nerf scenes}
	\label{fig:ap_psnr_avg_nerf_plot}
\end{figure}
\pagebreak

\subsection{SSIM}
\begin{table}[h!]
\centering
\caption{Average SSIM results for all scenes.}
\label{tab:ssim_avg}
\begin{tabular}{lrrrrrr}
\toprule
& \multicolumn{6}{c}{Number of Gaussians} \\
Algorithm & 4000 & 12000 & 36000 & 108000 & 324000 & 972000 \\

\midrule
3DGS            & 0.9320 & 0.9471 & 0.9596 & 0.9684 & 0.9736 & 0.9760 \\
3DGS + STP      & 0.9322 & 0.9476 & 0.9601 & 0.9688 & 0.9738 & 0.9760 \\
OTS             & 0.9394 & 0.9524 & 0.9619 & 0.9686 & 0.9729 & 0.9747 \\
OTS + SAtn      & 0.9394 & 0.9526 & 0.9621 & 0.9688 & 0.9730 & 0.9748 \\
3DGS marcher    & 0.9401 & 0.9536 & 0.9635 & 0.9701 & 0.9739 & 0.9756 \\
OTS marcher     & 0.9402 & 0.9536 & 0.9629 & 0.9694 & 0.9732 & 0.9749 \\
\bottomrule
\end{tabular}
\end{table}

\begin{figure}[h!]
	\centering
	\includegraphics[width=0.7\linewidth]{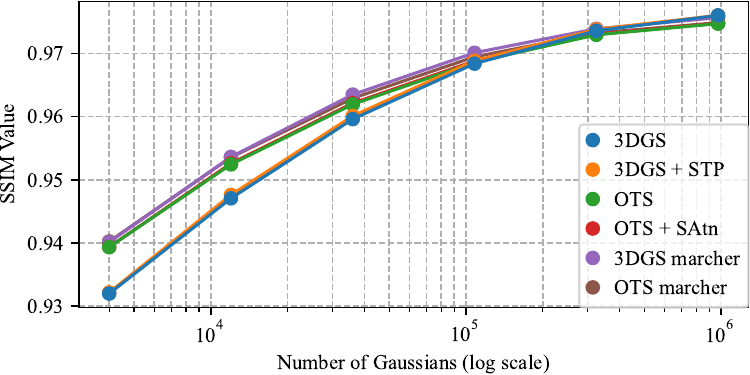}
	\caption{SSIM average over all scenes}
	\label{fig:ap_ssim_avg_all_plot}
\end{figure}
\pagebreak

\begin{table}[h!]
\centering
\caption{Average SSIM results for volumetric scenes}
\label{tab:ssim_avg_interest}
\begin{tabular}{lrrrrrr}
\toprule
& \multicolumn{6}{c}{Number of Gaussians} \\
Algorithm & 4000 & 12000 & 36000 & 108000 & 324000 & 972000 \\

\midrule
3DGS & 0.9733 & 0.9801 & 0.9853 & 0.9882 & 0.9896 & 0.9900 \\
3DGS + STP & 0.9733 & 0.9801 & 0.9852 & 0.9883 & 0.9897 & 0.9900 \\
OTS & 0.9754 & 0.9821 & 0.9862 & 0.9884 & 0.9893 & 0.9895 \\
OTS + SAtn & 0.9756 & 0.9823 & 0.9863 & 0.9885 & 0.9894 & 0.9895 \\
3DGS marcher & 0.9746 & 0.9816 & 0.9860 & 0.9883 & 0.9892 & 0.9893 \\
OTS marcher & 0.9755 & 0.9821 & 0.9863 & 0.9885 & 0.9893 & 0.9894 \\
\bottomrule
\end{tabular}
\end{table}

\begin{figure}[h!]
	\centering
	\includegraphics[width=0.7\linewidth]{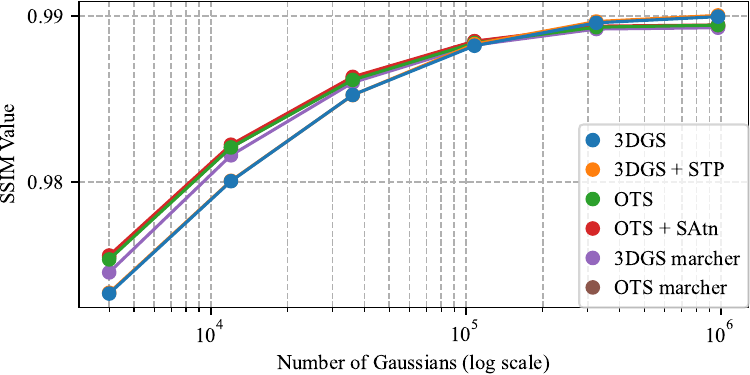}
	\caption{SSIM average over volumetric scenes}
	\label{fig:ap_ssim_avg_vol_plot}
\end{figure}
\pagebreak

\begin{table}[h!]
\centering
\caption{Average SSIM results for Nerf scenes}
\label{tab:ssim_avg_others}
\begin{tabular}{lrrrrrr}
\toprule
& \multicolumn{6}{c}{Number of Gaussians} \\
Algorithm & 4000 & 12000 & 36000 & 108000 & 324000 & 972000 \\

\midrule
3DGS & 0.9010 & 0.9223 & 0.9404 & 0.9535 & 0.9616 & 0.9655 \\
3DGS + STP & 0.9013 & 0.9233 & 0.9412 & 0.9542 & 0.9619 & 0.9655 \\
OTS & 0.9124 & 0.9302 & 0.9438 & 0.9538 & 0.9606 & 0.9636 \\
OTS + SAtn & 0.9122 & 0.9304 & 0.9440 & 0.9540 & 0.9607 & 0.9637 \\
3DGS marcher & 0.9143 & 0.9326 & 0.9466 & 0.9565 & 0.9624 & 0.9654 \\
OTS marcher & 0.9138 & 0.9322 & 0.9454 & 0.9552 & 0.9611 & 0.9640 \\
\bottomrule
\end{tabular}
\end{table}

\begin{figure}[h!]
	\centering
	\includegraphics[width=0.7\linewidth]{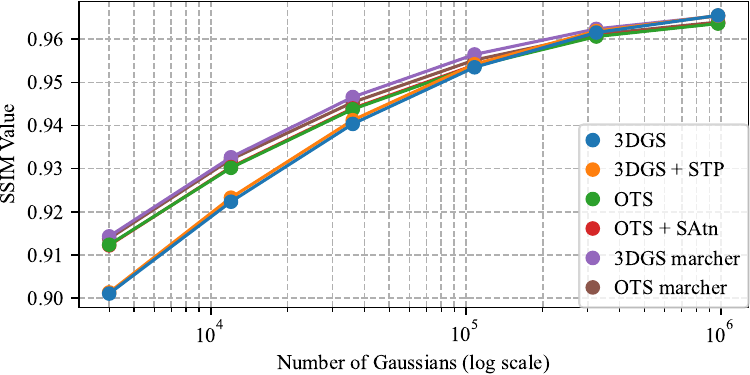}
	\caption{SSIM average over Nerf scenes}
	\label{fig:ap_ssim_avg_nerf_plot}
\end{figure}
\pagebreak

\subsection{LPIPS}
\begin{table}[h!]
\centering
\caption{Average LPIPS results for all scenes.}
\label{tab:lpips_avg}
\begin{tabular}{lrrrrrr}
\toprule
& \multicolumn{6}{c}{Number of Gaussians} \\
Algorithm & 4000 & 12000 & 36000 & 108000 & 324000 & 972000 \\

\midrule
3DGS            & 0.1231 & 0.1007 & 0.0800 & 0.0632 & 0.0523 & 0.0467 \\
3DGS + STP      & 0.1222 & 0.0995 & 0.0790 & 0.0625 & 0.0520 & 0.0467 \\
OTS             & 0.1115 & 0.0906 & 0.0742 & 0.0611 & 0.0517 & 0.0471 \\
OTS + SAtn      & 0.1122 & 0.0913 & 0.0744 & 0.0611 & 0.0518 & 0.0471 \\
3DGS marcher    & 0.1115 & 0.0908 & 0.0736 & 0.0606 & 0.0521 & 0.0477 \\
OTS marcher     & 0.1116 & 0.0908 & 0.0743 & 0.0610 & 0.0523 & 0.0481 \\
\bottomrule
\end{tabular}
\end{table}

\begin{figure}[h!]
	\centering
	\includegraphics[width=0.7\linewidth]{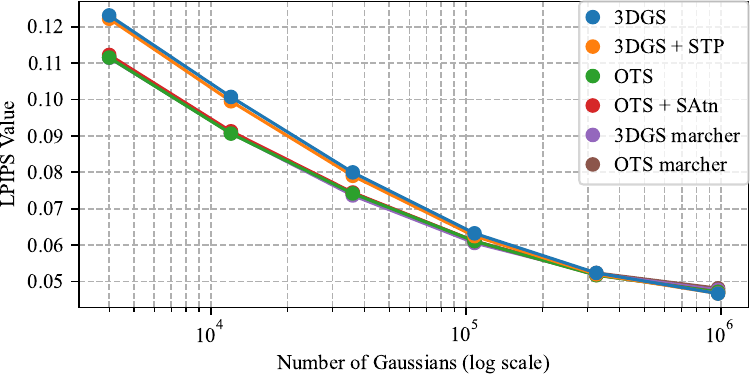}
	\caption{LPIPS average over all scenes}
	\label{fig:ap_lpips_avg_all_plot}
\end{figure}
\pagebreak

\begin{table}[h!]
\centering
\caption{Average LPIPS results for volumetric scenes}
\label{tab:lpips_avg_interest}
\begin{tabular}{lrrrrrr}
\toprule
& \multicolumn{6}{c}{Number of Gaussians} \\
Algorithm & 4000 & 12000 & 36000 & 108000 & 324000 & 972000 \\

\midrule
3DGS & 0.1200 & 0.1023 & 0.0853 & 0.0721 & 0.0641 & 0.0600 \\
3DGS + STP & 0.1192 & 0.1015 & 0.0849 & 0.0720 & 0.0641 & 0.0602 \\
OTS & 0.1146 & 0.0954 & 0.0803 & 0.0685 & 0.0613 & 0.0574 \\
OTS + SAtn & 0.1146 & 0.0955 & 0.0800 & 0.0683 & 0.0611 & 0.0573 \\
3DGS marcher & 0.1173 & 0.0986 & 0.0832 & 0.0719 & 0.0647 & 0.0609 \\
OTS marcher & 0.1148 & 0.0959 & 0.0809 & 0.0691 & 0.0623 & 0.0586 \\
\bottomrule
\end{tabular}
\end{table}

\begin{figure}[h!]
	\centering
	\includegraphics[width=0.7\linewidth]{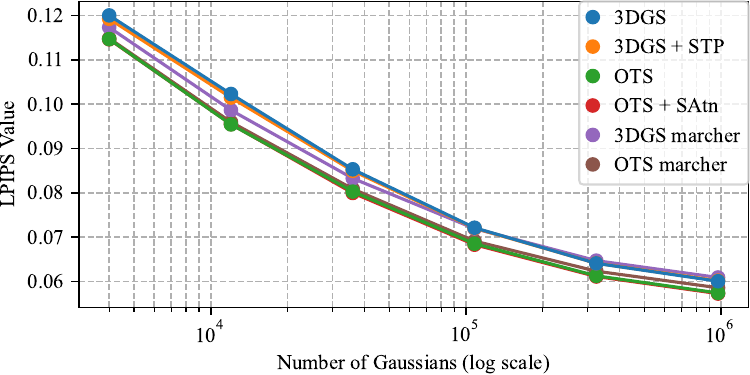}
	\caption{LPIPS average over volumetric scenes}
	\label{fig:ap_lpips_avg_vol_plot}
\end{figure}
\pagebreak

\begin{table}[h!]
\centering
\caption{Average LPIPS results for Nerf scenes}
\label{tab:lpips_avg_others}
\begin{tabular}{lrrrrrr}
\toprule
& \multicolumn{6}{c}{Number of Gaussians} \\
Algorithm & 4000 & 12000 & 36000 & 108000 & 324000 & 972000 \\

\midrule
3DGS & 0.1254 & 0.0995 & 0.0760 & 0.0566 & 0.0435 & 0.0366 \\
3DGS + STP & 0.1244 & 0.0980 & 0.0746 & 0.0553 & 0.0429 & 0.0366 \\
OTS & 0.1092 & 0.0871 & 0.0696 & 0.0555 & 0.0446 & 0.0394 \\
OTS + SAtn & 0.1105 & 0.0881 & 0.0703 & 0.0557 & 0.0448 & 0.0396 \\
3DGS marcher & 0.1071 & 0.0850 & 0.0664 & 0.0521 & 0.0426 & 0.0378 \\
OTS marcher & 0.1091 & 0.0869 & 0.0694 & 0.0549 & 0.0449 & 0.0403 \\
\bottomrule
\end{tabular}
\end{table}

\begin{figure}[h!]
	\centering
	\includegraphics[width=0.7\linewidth]{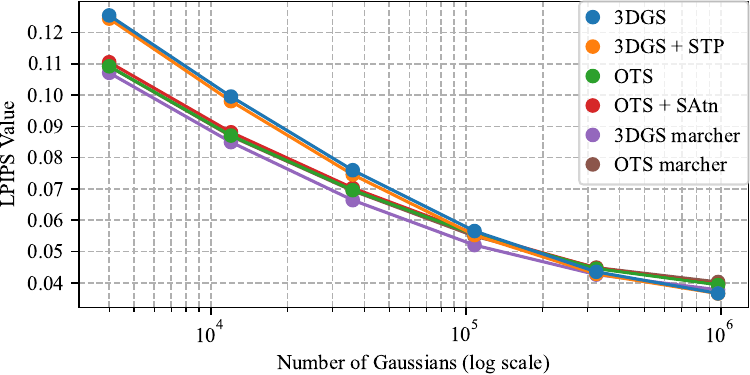}
	\caption{LPIPS average over Nerf scenes}
	\label{fig:ap_lpips_avg_nerf_plot}
\end{figure}
\pagebreak

\section{Full Numerical Results}
Here we show the full numerical results, and per-scene error plots.
\label{ap:full_results}
\subsection{PSNR}
\begin{longtable}[H]{llrrrrrr}
\toprule
& & \multicolumn{6}{c}{Number of Gaussians} \\
Scene & Algorithm & 4000 & 12000 & 36000 & 108000 & 324000 & 972000 \\
\midrule \endhead
Burning Ficus & 3DGS & 29.61 & 32.35 & 35.29 & 37.97 & 39.85 & 40.57 \\
 & 3DGS + STP & 29.75 & 32.34 & 35.30 & 38.18 & 40.01 & 40.64 \\
 & OTS & 30.07 & 32.96 & 34.66 & 35.64 & 36.20 & 36.43 \\
 & OTS + SAtn & 30.36 & 33.06 & 34.85 & 35.74 & 36.24 & 36.43 \\
 & 3DGS marcher & 29.99 & 33.28 & 35.48 & 36.93 & 37.79 & 38.11 \\
 & OTS marcher & 30.15 & 32.80 & 34.59 & 35.55 & 36.05 & 36.20 \\
Chair & 3DGS & 27.79 & 29.16 & 30.41 & 31.70 & 32.89 & 33.96 \\
 & 3DGS + STP & 27.68 & 29.11 & 30.37 & 31.58 & 32.95 & 33.99 \\
 & OTS & 28.32 & 29.74 & 30.79 & 31.91 & 33.14 & 33.97 \\
 & OTS + SAtn & 28.38 & 29.87 & 30.86 & 31.93 & 33.13 & 33.92 \\
 & 3DGS marcher & 28.23 & 29.18 & 30.25 & 31.90 & 33.01 & 33.96 \\
 & OTS marcher & 28.44 & 29.90 & 30.80 & 31.94 & 33.00 & 33.83 \\
Coloured Wdas & 3DGS & 37.19 & 38.13 & 39.02 & 39.65 & 39.99 & 40.17 \\
 & 3DGS + STP & 37.16 & 38.15 & 39.00 & 39.63 & 40.01 & 40.19 \\
 & OTS & 37.46 & 38.18 & 38.67 & 39.10 & 39.43 & 39.79 \\
 & OTS + SAtn & 37.48 & 38.22 & 38.67 & 39.09 & 39.43 & 39.78 \\
 & 3DGS marcher & 37.18 & 38.02 & 38.80 & 39.34 & 39.72 & 39.94 \\
 & OTS marcher & 37.32 & 37.92 & 38.35 & 38.85 & 39.19 & 39.56 \\
Drums & 3DGS & 21.72 & 23.32 & 24.45 & 25.22 & 25.68 & 25.98 \\
 & 3DGS + STP & 21.80 & 23.47 & 24.49 & 25.24 & 25.67 & 25.91 \\
 & OTS & 22.13 & 23.78 & 24.63 & 25.16 & 25.51 & 25.69 \\
 & OTS + SAtn & 22.18 & 23.77 & 24.66 & 25.17 & 25.52 & 25.69 \\
 & 3DGS marcher & 22.53 & 23.98 & 24.89 & 25.42 & 25.75 & 25.95 \\
 & OTS marcher & 22.26 & 23.83 & 24.69 & 25.17 & 25.49 & 25.64 \\
Explosion & 3DGS & 32.79 & 36.26 & 39.81 & 42.67 & 44.47 & 45.49 \\
 & 3DGS + STP & 32.92 & 36.41 & 39.96 & 42.75 & 44.48 & 45.43 \\
 & OTS & 33.29 & 36.92 & 40.46 & 43.36 & 44.96 & 45.62 \\
 & OTS + SAtn & 33.33 & 37.01 & 40.64 & 43.38 & 44.93 & 45.62 \\
 & 3DGS marcher & 33.43 & 37.10 & 40.88 & 43.46 & 44.52 & 44.49 \\
 & OTS marcher & 33.65 & 37.60 & 41.54 & 44.13 & 44.84 & 44.70 \\
Ficus & 3DGS & 25.78 & 28.30 & 30.93 & 33.39 & 35.12 & 35.86 \\
 & 3DGS + STP & 25.73 & 28.39 & 31.14 & 33.64 & 35.37 & 36.02 \\
 & OTS & 26.49 & 28.61 & 30.21 & 31.11 & 31.67 & 31.94 \\
 & OTS + SAtn & 26.57 & 28.84 & 30.46 & 31.26 & 31.74 & 31.95 \\
 & 3DGS marcher & 26.95 & 29.71 & 31.69 & 32.95 & 33.72 & 34.05 \\
 & OTS marcher & 26.58 & 28.71 & 30.36 & 31.27 & 31.71 & 31.87 \\
Hotdog & 3DGS & 30.96 & 32.91 & 34.44 & 35.56 & 36.32 & 36.86 \\
 & 3DGS + STP & 31.20 & 33.10 & 34.63 & 35.66 & 36.33 & 36.77 \\
 & OTS & 31.71 & 33.42 & 34.65 & 35.42 & 36.00 & 36.40 \\
 & OTS + SAtn & 32.04 & 33.58 & 34.73 & 35.46 & 36.04 & 36.42 \\
 & 3DGS marcher & 31.87 & 33.43 & 34.67 & 35.49 & 36.02 & 36.28 \\
 & OTS marcher & 32.18 & 33.64 & 34.55 & 35.28 & 35.59 & 35.77 \\
Lego & 3DGS & 24.59 & 26.55 & 28.45 & 30.63 & 32.69 & 34.14 \\
 & 3DGS + STP & 24.81 & 26.69 & 28.72 & 30.72 & 32.67 & 34.03 \\
 & OTS & 25.57 & 27.22 & 28.96 & 30.81 & 32.59 & 33.65 \\
 & OTS + SAtn & 25.54 & 27.30 & 28.97 & 30.76 & 32.57 & 33.67 \\
 & 3DGS marcher & 25.72 & 27.45 & 29.31 & 31.22 & 33.02 & 34.24 \\
 & OTS marcher & 25.61 & 27.38 & 29.17 & 30.79 & 32.51 & 33.53 \\
Materials & 3DGS & 22.48 & 24.42 & 26.23 & 27.74 & 28.81 & 29.54 \\
 & 3DGS + STP & 22.46 & 24.50 & 26.34 & 27.84 & 28.85 & 29.54 \\
 & OTS & 23.65 & 25.62 & 27.28 & 28.51 & 29.43 & 29.95 \\
 & OTS + SAtn & 23.59 & 25.61 & 27.26 & 28.59 & 29.44 & 29.96 \\
 & 3DGS marcher & 23.39 & 25.52 & 27.08 & 28.21 & 29.09 & 29.66 \\
 & OTS marcher & 23.68 & 25.68 & 27.24 & 28.58 & 29.45 & 29.94 \\
Mic & 3DGS & 28.47 & 29.93 & 31.31 & 32.64 & 33.64 & 34.26 \\
 & 3DGS + STP & 28.51 & 30.03 & 31.38 & 32.74 & 33.78 & 34.27 \\
 & OTS & 29.45 & 30.47 & 31.35 & 32.26 & 33.43 & 33.90 \\
 & OTS + SAtn & 29.62 & 30.54 & 31.34 & 32.35 & 33.45 & 33.95 \\
 & 3DGS marcher & 29.70 & 30.61 & 31.65 & 32.90 & 33.73 & 34.27 \\
 & OTS marcher & 29.65 & 30.62 & 31.41 & 32.38 & 33.38 & 33.92 \\
Ship & 3DGS & 24.78 & 25.87 & 27.06 & 28.15 & 29.01 & 29.58 \\
 & 3DGS + STP & 24.76 & 25.78 & 26.92 & 28.12 & 28.99 & 29.53 \\
 & OTS & 25.46 & 26.46 & 27.36 & 28.25 & 28.97 & 29.28 \\
 & OTS + SAtn & 25.45 & 26.41 & 27.33 & 28.27 & 28.96 & 29.26 \\
 & 3DGS marcher & 25.74 & 26.78 & 27.75 & 28.57 & 29.24 & 29.61 \\
 & OTS marcher & 25.66 & 26.71 & 27.74 & 28.60 & 29.03 & 29.41 \\
Wdas Cloud 1 & 3DGS & 48.87 & 49.05 & 49.11 & 49.17 & 49.17 & 49.23 \\
 & 3DGS + STP & 48.87 & 49.11 & 49.28 & 49.38 & 49.36 & 49.37 \\
 & OTS & 48.82 & 49.04 & 49.36 & 49.59 & 49.71 & 49.99 \\
 & OTS + SAtn & 48.96 & 49.12 & 49.43 & 49.51 & 49.70 & 49.95 \\
 & 3DGS marcher & 48.88 & 48.91 & 49.17 & 49.14 & 49.17 & 49.22 \\
 & OTS marcher & 49.03 & 49.31 & 49.51 & 49.68 & 50.00 & 50.17 \\
Wdas Cloud 2 & 3DGS & 45.29 & 45.57 & 45.80 & 46.08 & 46.20 & 46.14 \\
 & 3DGS + STP & 45.32 & 45.61 & 45.90 & 46.23 & 46.29 & 46.26 \\
 & OTS & 45.36 & 45.60 & 45.77 & 45.89 & 46.00 & 46.09 \\
 & OTS + SAtn & 45.36 & 45.63 & 45.79 & 45.90 & 45.93 & 46.06 \\
 & 3DGS marcher & 45.03 & 45.50 & 45.74 & 45.90 & 46.03 & 45.88 \\
 & OTS marcher & 45.42 & 45.68 & 45.86 & 46.00 & 46.13 & 46.15 \\
Wdas Cloud 3 & 3DGS & 42.48 & 43.32 & 43.86 & 44.20 & 44.40 & 44.39 \\
 & 3DGS + STP & 42.65 & 43.50 & 43.95 & 44.40 & 44.59 & 44.69 \\
 & OTS & 42.57 & 43.22 & 43.46 & 43.91 & 44.30 & 44.34 \\
 & OTS + SAtn & 42.49 & 43.17 & 43.54 & 43.93 & 44.06 & 44.42 \\
 & 3DGS marcher & 42.53 & 43.36 & 43.80 & 44.10 & 44.30 & 44.31 \\
 & OTS marcher & 42.36 & 42.95 & 43.38 & 43.70 & 43.85 & 44.08 \\
\bottomrule
\caption{PSNR results for all scenes}
\end{longtable}

\begin{figure}[h!]
\centering
\includegraphics[width=0.8\textwidth]{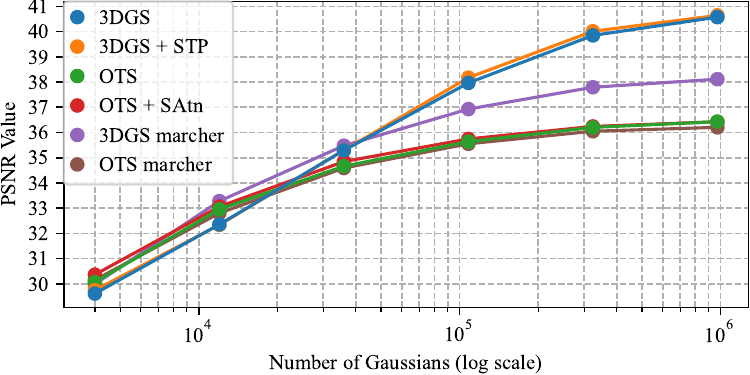}
\caption{PSNR results for Burning Ficus}
\label{fig:ap_psnr_Burning Ficus_plot}
\end{figure}

\begin{figure}[h!]
\centering
\includegraphics[width=0.8\textwidth]{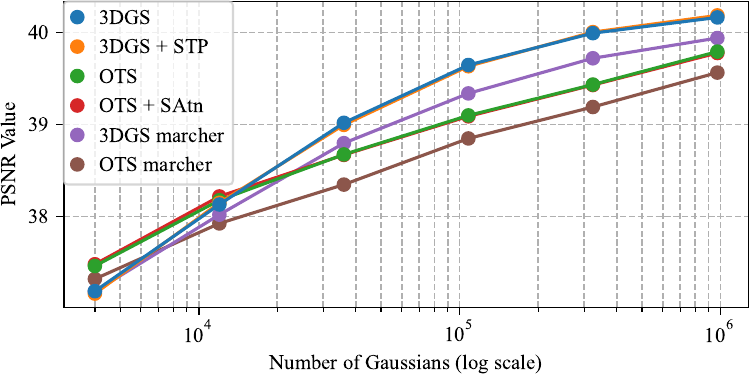}
\caption{PSNR results for Coloured Wdas}
\label{fig:ap_psnr_Coloured Wdas_plot}
\end{figure}

\begin{figure}[h!]
\centering
\includegraphics[width=0.8\textwidth]{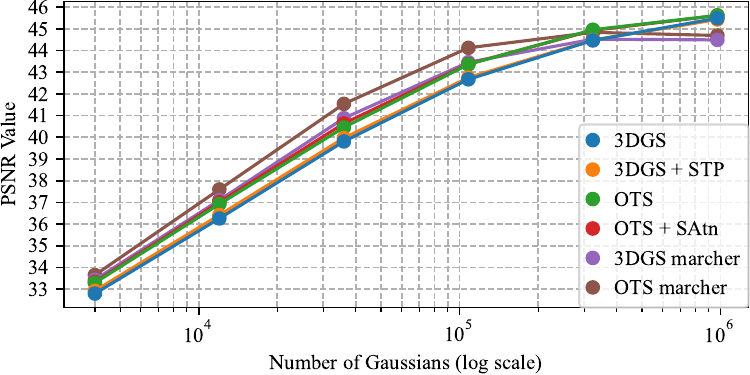}
\caption{PSNR results for Explosion}
\label{fig:ap_psnr_Explosion_plot}
\end{figure}

\begin{figure}[h!]
\centering
\includegraphics[width=0.8\textwidth]{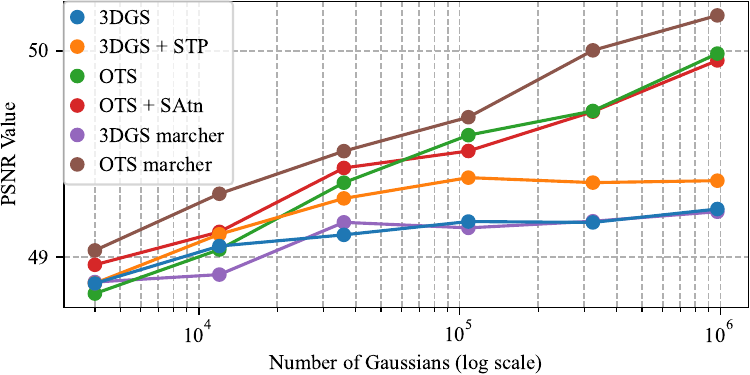}
\caption{PSNR results for Wdas Cloud 1}
\label{fig:ap_psnr_Wdas Cloud 1_plot}
\end{figure}

\begin{figure}[h!]
\centering
\includegraphics[width=0.8\textwidth]{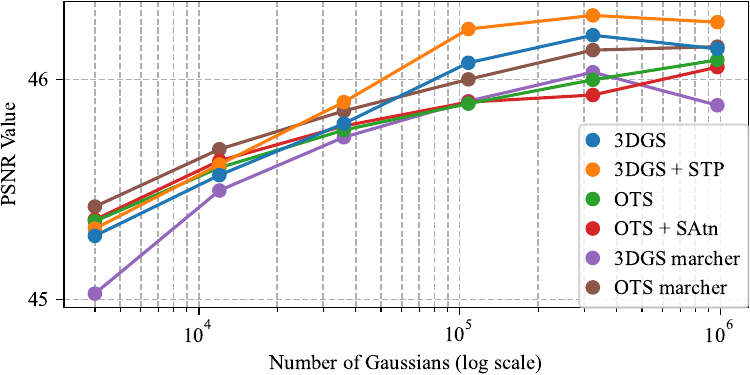}
\caption{PSNR results for Wdas Cloud 2}
\label{fig:ap_psnr_Wdas Cloud 2_plot}
\end{figure}

\begin{figure}[h!]
\centering
\includegraphics[width=0.8\textwidth]{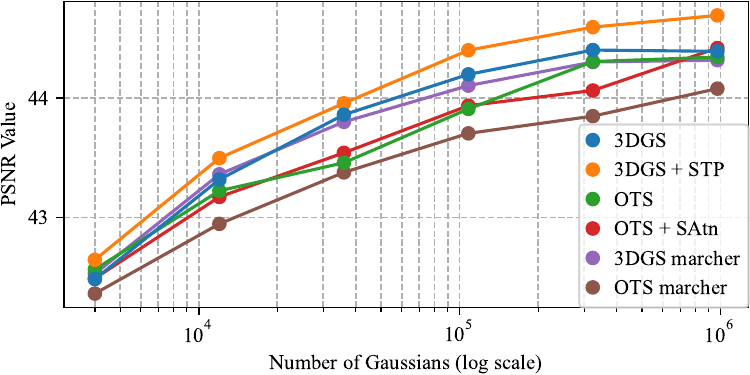}
\caption{PSNR results for Wdas Cloud 3}
\label{fig:ap_psnr_Wdas Cloud 3_plot}
\end{figure}

\begin{figure}[h!]
\centering
\includegraphics[width=0.8\textwidth]{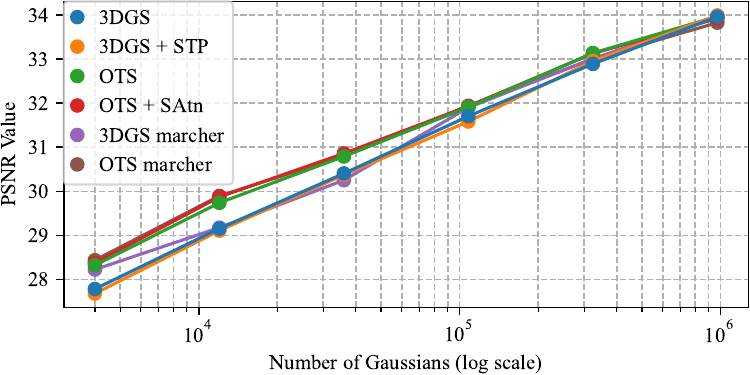}
\caption{PSNR results for Chair}
\label{fig:ap_psnr_Chair_plot}
\end{figure}

\begin{figure}[h!]
\centering
\includegraphics[width=0.8\textwidth]{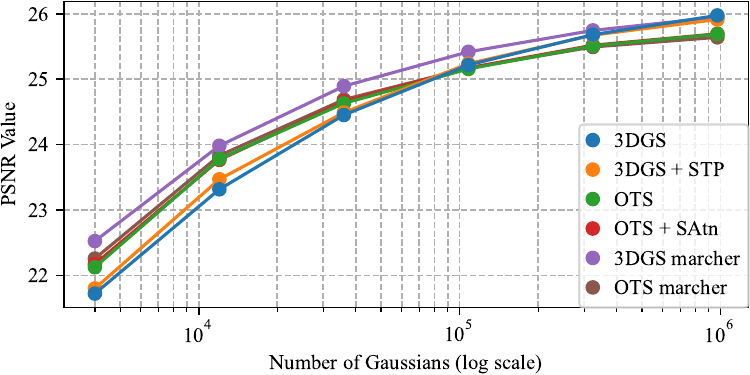}
\caption{PSNR results for Drums}
\label{fig:ap_psnr_Drums_plot}
\end{figure}

\begin{figure}[h!]
\centering
\includegraphics[width=0.8\textwidth]{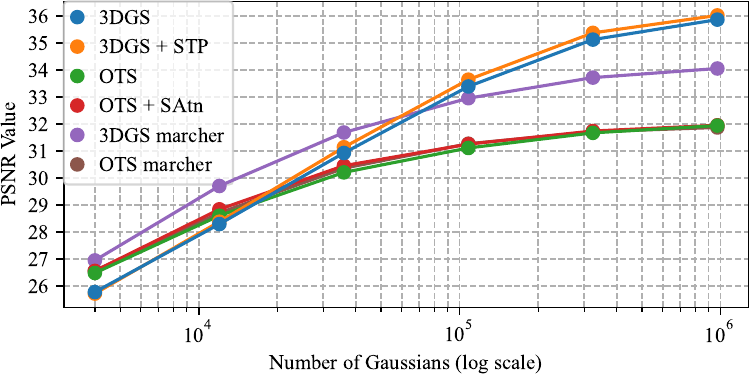}
\caption{PSNR results for Ficus}
\label{fig:ap_psnr_Ficus_plot}
\end{figure}

\begin{figure}[h!]
\centering
\includegraphics[width=0.8\textwidth]{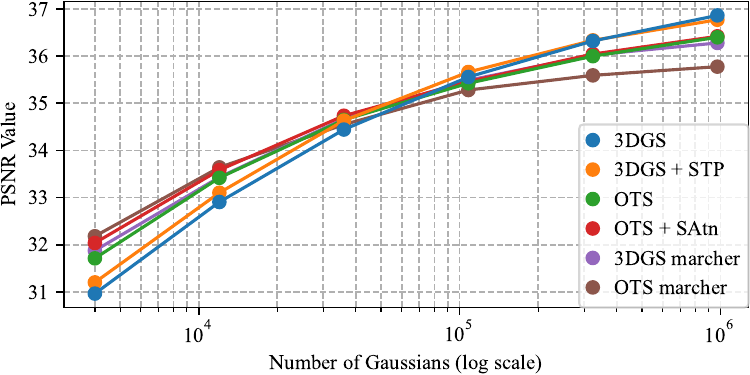}
\caption{PSNR results for Hotdog}
\label{fig:ap_psnr_Hotdog_plot}
\end{figure}

\begin{figure}[h!]
\centering
\includegraphics[width=0.8\textwidth]{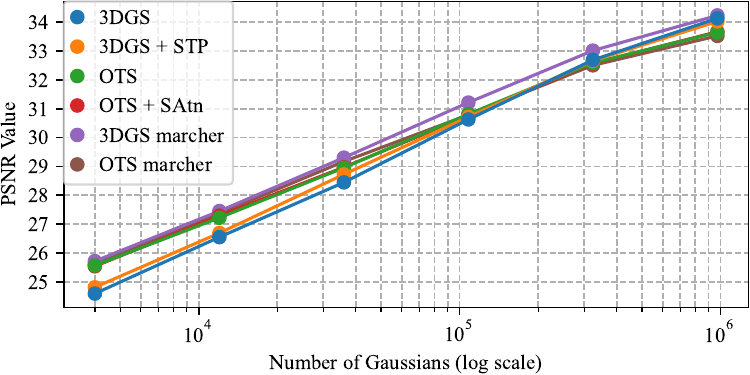}
\caption{PSNR results for Lego}
\label{fig:ap_psnr_Lego_plot}
\end{figure}

\begin{figure}[h!]
\centering
\includegraphics[width=0.8\textwidth]{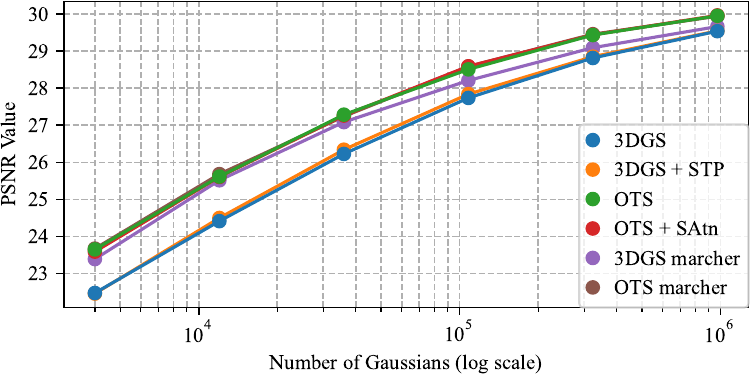}
\caption{PSNR results for Materials}
\label{fig:ap_psnr_Materials_plot}
\end{figure}

\begin{figure}[h!]
\centering
\includegraphics[width=0.8\textwidth]{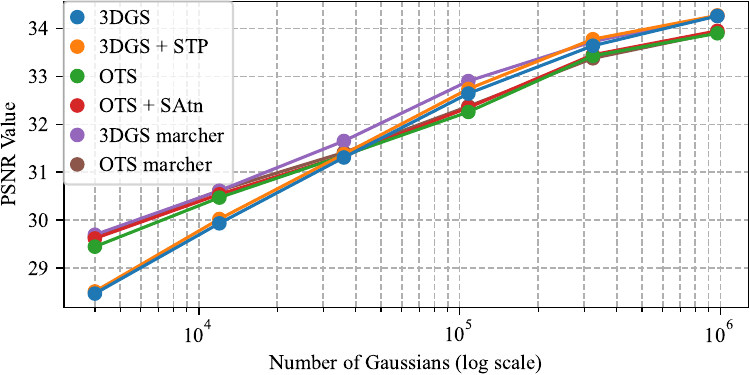}
\caption{PSNR results for Mic}
\label{fig:ap_psnr_Mic_plot}
\end{figure}

\begin{figure}[h!]
\centering
\includegraphics[width=0.8\textwidth]{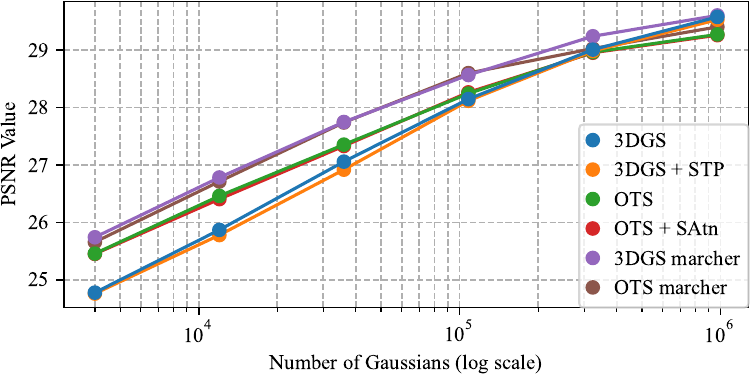}
\caption{PSNR results for Ship}
\label{fig:ap_psnr_Ship_plot}
\end{figure}

\subsection{SSIM}
\begin{longtable}[H]{llrrrrrr}
\toprule
& & \multicolumn{6}{c}{Number of Gaussians} \\
Scene & Algorithm & 4000 & 12000 & 36000 & 108000 & 324000 & 972000 \\
\midrule \endhead
Burning Ficus & 3DGS & 0.9431 & 0.9610 & 0.9761 & 0.9851 & 0.9895 & 0.9908 \\
 & 3DGS + STP & 0.9436 & 0.9613 & 0.9762 & 0.9857 & 0.9899 & 0.9909 \\
 & OTS & 0.9506 & 0.9674 & 0.9769 & 0.9821 & 0.9848 & 0.9855 \\
 & OTS + SAtn & 0.9512 & 0.9678 & 0.9776 & 0.9826 & 0.9851 & 0.9856 \\
 & 3DGS marcher & 0.9477 & 0.9674 & 0.9789 & 0.9851 & 0.9880 & 0.9885 \\
 & OTS marcher & 0.9496 & 0.9660 & 0.9766 & 0.9823 & 0.9848 & 0.9852 \\
Chair & 3DGS & 0.9198 & 0.9381 & 0.9547 & 0.9686 & 0.9774 & 0.9829 \\
 & 3DGS + STP & 0.9186 & 0.9398 & 0.9553 & 0.9687 & 0.9779 & 0.9831 \\
 & OTS & 0.9268 & 0.9462 & 0.9579 & 0.9694 & 0.9782 & 0.9829 \\
 & OTS + SAtn & 0.9270 & 0.9462 & 0.9581 & 0.9694 & 0.9780 & 0.9827 \\
 & 3DGS marcher & 0.9276 & 0.9438 & 0.9575 & 0.9694 & 0.9775 & 0.9826 \\
 & OTS marcher & 0.9290 & 0.9472 & 0.9586 & 0.9699 & 0.9781 & 0.9825 \\
Coloured Wdas & 3DGS & 0.9788 & 0.9810 & 0.9833 & 0.9849 & 0.9859 & 0.9862 \\
 & 3DGS + STP & 0.9786 & 0.9810 & 0.9832 & 0.9849 & 0.9860 & 0.9862 \\
 & OTS & 0.9800 & 0.9822 & 0.9841 & 0.9856 & 0.9865 & 0.9866 \\
 & OTS + SAtn & 0.9800 & 0.9824 & 0.9841 & 0.9857 & 0.9865 & 0.9866 \\
 & 3DGS marcher & 0.9787 & 0.9807 & 0.9827 & 0.9842 & 0.9852 & 0.9857 \\
 & OTS marcher & 0.9795 & 0.9815 & 0.9830 & 0.9848 & 0.9860 & 0.9865 \\
Drums & 3DGS & 0.8872 & 0.9108 & 0.9292 & 0.9410 & 0.9478 & 0.9512 \\
 & 3DGS + STP & 0.8878 & 0.9121 & 0.9298 & 0.9419 & 0.9483 & 0.9513 \\
 & OTS & 0.8992 & 0.9206 & 0.9332 & 0.9411 & 0.9463 & 0.9485 \\
 & OTS + SAtn & 0.8987 & 0.9200 & 0.9332 & 0.9414 & 0.9463 & 0.9486 \\
 & 3DGS marcher & 0.9029 & 0.9233 & 0.9369 & 0.9451 & 0.9494 & 0.9519 \\
 & OTS marcher & 0.9000 & 0.9211 & 0.9338 & 0.9418 & 0.9466 & 0.9486 \\
Explosion & 3DGS & 0.9515 & 0.9703 & 0.9829 & 0.9892 & 0.9918 & 0.9929 \\
 & 3DGS + STP & 0.9515 & 0.9701 & 0.9827 & 0.9890 & 0.9917 & 0.9929 \\
 & OTS & 0.9528 & 0.9724 & 0.9843 & 0.9901 & 0.9922 & 0.9929 \\
 & OTS + SAtn & 0.9535 & 0.9728 & 0.9846 & 0.9902 & 0.9923 & 0.9929 \\
 & 3DGS marcher & 0.9545 & 0.9732 & 0.9852 & 0.9905 & 0.9923 & 0.9925 \\
 & OTS marcher & 0.9549 & 0.9745 & 0.9863 & 0.9912 & 0.9926 & 0.9926 \\
Ficus & 3DGS & 0.9306 & 0.9534 & 0.9702 & 0.9816 & 0.9871 & 0.9890 \\
 & 3DGS + STP & 0.9309 & 0.9543 & 0.9715 & 0.9825 & 0.9878 & 0.9895 \\
 & OTS & 0.9442 & 0.9604 & 0.9705 & 0.9764 & 0.9799 & 0.9814 \\
 & OTS + SAtn & 0.9442 & 0.9613 & 0.9716 & 0.9770 & 0.9803 & 0.9815 \\
 & 3DGS marcher & 0.9463 & 0.9648 & 0.9758 & 0.9823 & 0.9855 & 0.9868 \\
 & OTS marcher & 0.9435 & 0.9608 & 0.9715 & 0.9778 & 0.9808 & 0.9818 \\
Hotdog & 3DGS & 0.9503 & 0.9610 & 0.9693 & 0.9754 & 0.9795 & 0.9818 \\
 & 3DGS + STP & 0.9512 & 0.9617 & 0.9698 & 0.9756 & 0.9792 & 0.9814 \\
 & OTS & 0.9545 & 0.9632 & 0.9703 & 0.9751 & 0.9786 & 0.9808 \\
 & OTS + SAtn & 0.9560 & 0.9641 & 0.9707 & 0.9753 & 0.9787 & 0.9809 \\
 & 3DGS marcher & 0.9564 & 0.9643 & 0.9708 & 0.9755 & 0.9790 & 0.9811 \\
 & OTS marcher & 0.9577 & 0.9654 & 0.9707 & 0.9752 & 0.9781 & 0.9799 \\
Lego & 3DGS & 0.8703 & 0.9029 & 0.9320 & 0.9558 & 0.9706 & 0.9778 \\
 & 3DGS + STP & 0.8739 & 0.9058 & 0.9355 & 0.9574 & 0.9711 & 0.9777 \\
 & OTS & 0.8839 & 0.9130 & 0.9373 & 0.9565 & 0.9695 & 0.9756 \\
 & OTS + SAtn & 0.8837 & 0.9129 & 0.9374 & 0.9561 & 0.9698 & 0.9759 \\
 & 3DGS marcher & 0.8898 & 0.9198 & 0.9454 & 0.9633 & 0.9741 & 0.9791 \\
 & OTS marcher & 0.8879 & 0.9182 & 0.9435 & 0.9607 & 0.9722 & 0.9770 \\
Materials & 3DGS & 0.8713 & 0.9019 & 0.9280 & 0.9448 & 0.9541 & 0.9589 \\
 & 3DGS + STP & 0.8705 & 0.9029 & 0.9287 & 0.9455 & 0.9545 & 0.9592 \\
 & OTS & 0.8913 & 0.9180 & 0.9380 & 0.9505 & 0.9581 & 0.9614 \\
 & OTS + SAtn & 0.8904 & 0.9180 & 0.9379 & 0.9513 & 0.9582 & 0.9615 \\
 & 3DGS marcher & 0.8887 & 0.9185 & 0.9377 & 0.9495 & 0.9562 & 0.9597 \\
 & OTS marcher & 0.8912 & 0.9193 & 0.9384 & 0.9519 & 0.9586 & 0.9616 \\
Mic & 3DGS & 0.9573 & 0.9687 & 0.9779 & 0.9839 & 0.9872 & 0.9886 \\
 & 3DGS + STP & 0.9578 & 0.9692 & 0.9784 & 0.9845 & 0.9877 & 0.9888 \\
 & OTS & 0.9636 & 0.9699 & 0.9775 & 0.9827 & 0.9868 & 0.9881 \\
 & OTS + SAtn & 0.9644 & 0.9696 & 0.9771 & 0.9830 & 0.9869 & 0.9882 \\
 & 3DGS marcher & 0.9650 & 0.9717 & 0.9802 & 0.9855 & 0.9879 & 0.9889 \\
 & OTS marcher & 0.9653 & 0.9708 & 0.9778 & 0.9839 & 0.9873 & 0.9886 \\
Ship & 3DGS & 0.8214 & 0.8419 & 0.8616 & 0.8769 & 0.8887 & 0.8939 \\
 & 3DGS + STP & 0.8199 & 0.8403 & 0.8609 & 0.8775 & 0.8886 & 0.8929 \\
 & OTS & 0.8353 & 0.8502 & 0.8653 & 0.8784 & 0.8876 & 0.8903 \\
 & OTS + SAtn & 0.8333 & 0.8507 & 0.8656 & 0.8787 & 0.8877 & 0.8906 \\
 & 3DGS marcher & 0.8377 & 0.8547 & 0.8685 & 0.8812 & 0.8895 & 0.8931 \\
 & OTS marcher & 0.8359 & 0.8545 & 0.8687 & 0.8802 & 0.8874 & 0.8917 \\
Wdas Cloud 1 & 3DGS & 0.9951 & 0.9951 & 0.9951 & 0.9951 & 0.9951 & 0.9950 \\
 & 3DGS + STP & 0.9951 & 0.9951 & 0.9951 & 0.9951 & 0.9951 & 0.9950 \\
 & OTS & 0.9953 & 0.9954 & 0.9954 & 0.9954 & 0.9954 & 0.9952 \\
 & OTS + SAtn & 0.9953 & 0.9954 & 0.9954 & 0.9954 & 0.9954 & 0.9952 \\
 & 3DGS marcher & 0.9951 & 0.9951 & 0.9951 & 0.9951 & 0.9950 & 0.9950 \\
 & OTS marcher & 0.9953 & 0.9954 & 0.9954 & 0.9954 & 0.9954 & 0.9953 \\
Wdas Cloud 2 & 3DGS & 0.9881 & 0.9885 & 0.9888 & 0.9890 & 0.9890 & 0.9889 \\
 & 3DGS + STP & 0.9882 & 0.9885 & 0.9888 & 0.9890 & 0.9890 & 0.9889 \\
 & OTS & 0.9888 & 0.9892 & 0.9895 & 0.9896 & 0.9896 & 0.9894 \\
 & OTS + SAtn & 0.9888 & 0.9892 & 0.9895 & 0.9896 & 0.9896 & 0.9894 \\
 & 3DGS marcher & 0.9882 & 0.9886 & 0.9888 & 0.9888 & 0.9887 & 0.9886 \\
 & OTS marcher & 0.9888 & 0.9892 & 0.9894 & 0.9896 & 0.9896 & 0.9895 \\
Wdas Cloud 3 & 3DGS & 0.9831 & 0.9845 & 0.9854 & 0.9861 & 0.9862 & 0.9860 \\
 & 3DGS + STP & 0.9832 & 0.9845 & 0.9854 & 0.9862 & 0.9864 & 0.9862 \\
 & OTS & 0.9847 & 0.9861 & 0.9868 & 0.9875 & 0.9875 & 0.9872 \\
 & OTS + SAtn & 0.9848 & 0.9861 & 0.9869 & 0.9875 & 0.9876 & 0.9872 \\
 & 3DGS marcher & 0.9833 & 0.9847 & 0.9854 & 0.9860 & 0.9860 & 0.9856 \\
 & OTS marcher & 0.9848 & 0.9860 & 0.9868 & 0.9875 & 0.9876 & 0.9872 \\
\bottomrule
\caption{SSIM results for all scenes}
\end{longtable}

\begin{figure}[h!]
\centering
\includegraphics[width=0.8\textwidth]{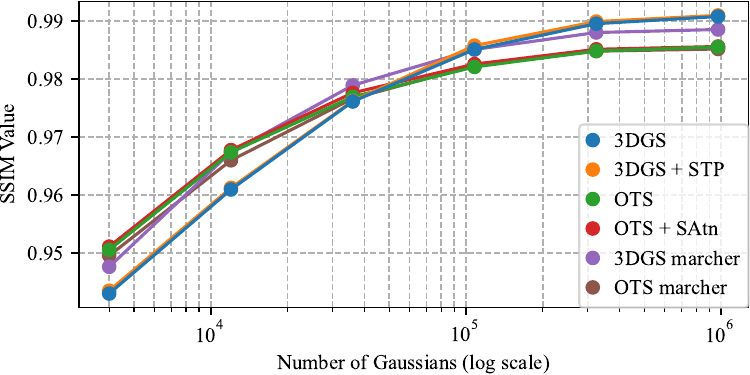}
\caption{SSIM results for Burning Ficus}
\label{fig:ap_ssim_Burning Ficus_plot}
\end{figure}

\begin{figure}[h!]
\centering
\includegraphics[width=0.8\textwidth]{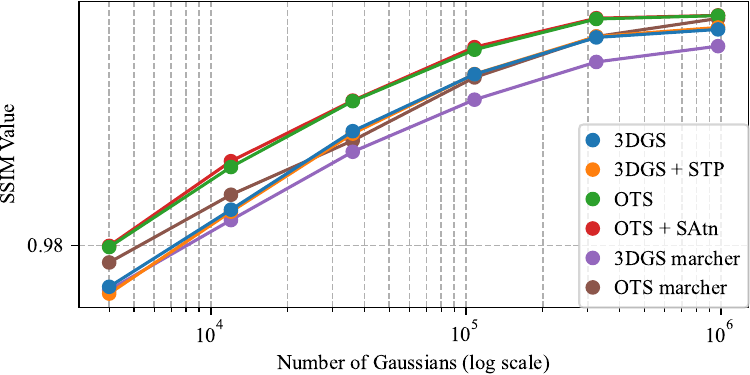}
\caption{SSIM results for Coloured Wdas}
\label{fig:ap_ssim_Coloured Wdas_plot}
\end{figure}

\begin{figure}[h!]
\centering
\includegraphics[width=0.8\textwidth]{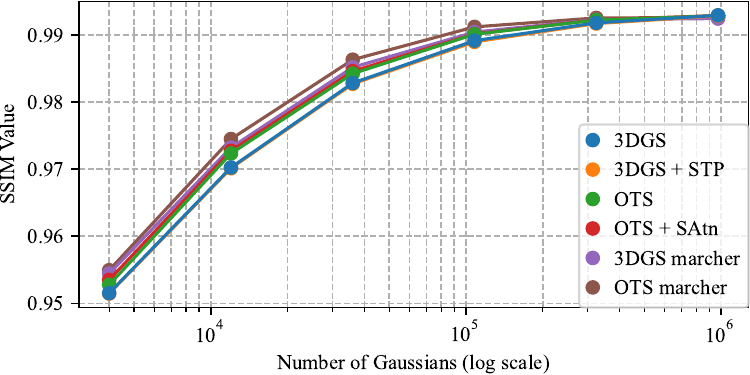}
\caption{SSIM results for Explosion}
\label{fig:ap_ssim_Explosion_plot}
\end{figure}

\begin{figure}[h!]
\centering
\includegraphics[width=0.8\textwidth]{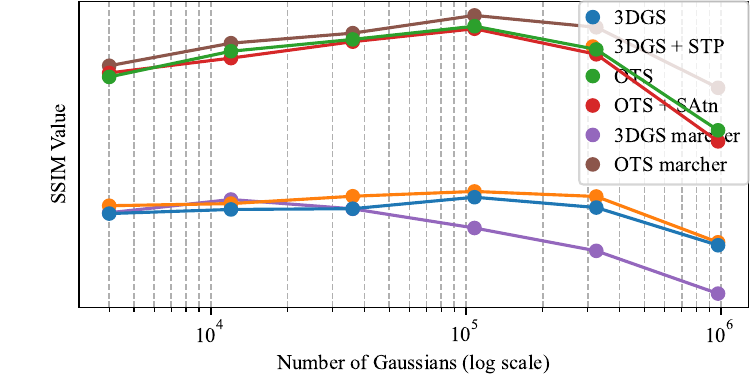}
\caption{SSIM results for Wdas Cloud 1}
\label{fig:ap_ssim_Wdas Cloud 1_plot}
\end{figure}

\begin{figure}[h!]
\centering
\includegraphics[width=0.8\textwidth]{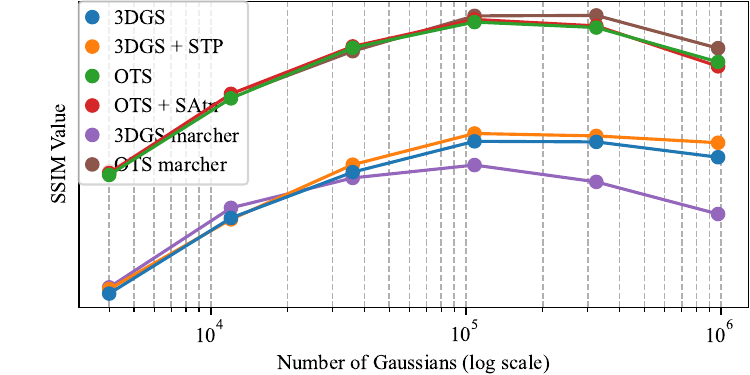}
\caption{SSIM results for Wdas Cloud 2}
\label{fig:ap_ssim_Wdas Cloud 2_plot}
\end{figure}

\begin{figure}[h!]
\centering
\includegraphics[width=0.8\textwidth]{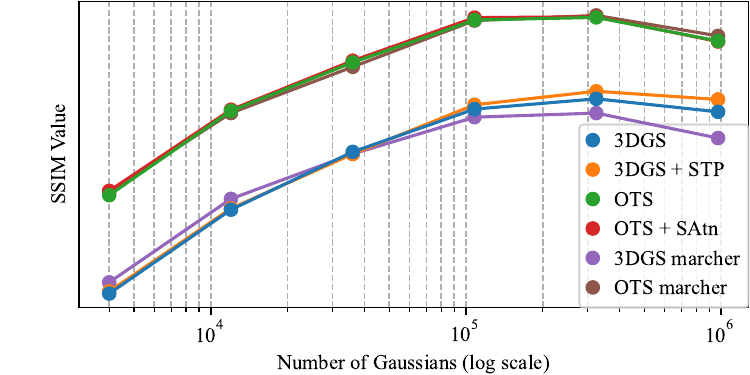}
\caption{SSIM results for Wdas Cloud 3}
\label{fig:ap_ssim_Wdas Cloud 3_plot}
\end{figure}

\begin{figure}[h!]
\centering
\includegraphics[width=0.8\textwidth]{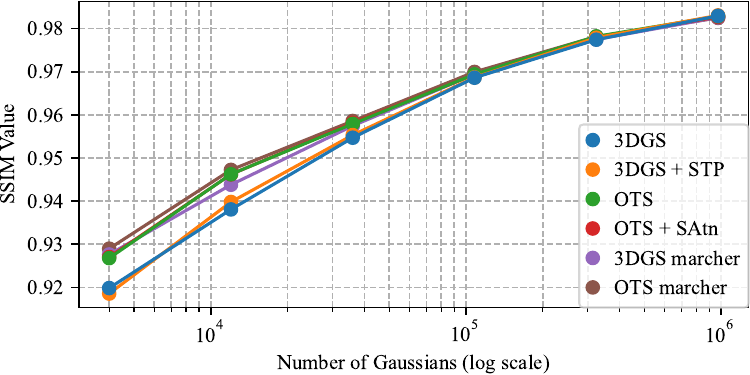}
\caption{SSIM results for Chair}
\label{fig:ap_ssim_Chair_plot}
\end{figure}

\begin{figure}[h!]
\centering
\includegraphics[width=0.8\textwidth]{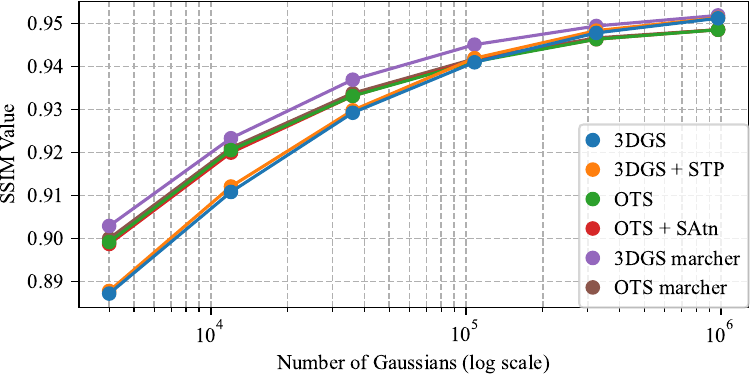}
\caption{SSIM results for Drums}
\label{fig:ap_ssim_Drums_plot}
\end{figure}

\begin{figure}[h!]
\centering
\includegraphics[width=0.8\textwidth]{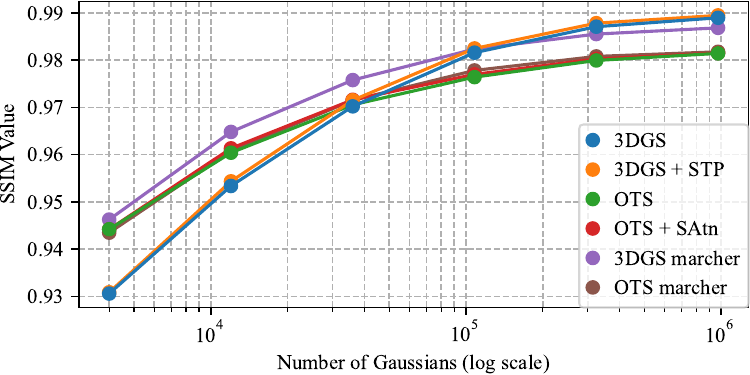}
\caption{SSIM results for Ficus}
\label{fig:ap_ssim_Ficus_plot}
\end{figure}

\begin{figure}[h!]
\centering
\includegraphics[width=0.8\textwidth]{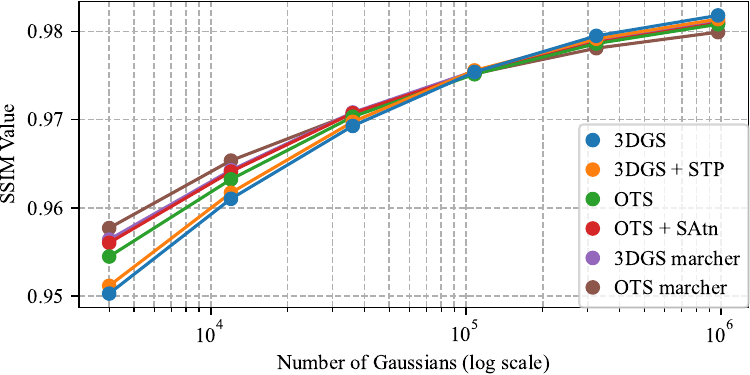}
\caption{SSIM results for Hotdog}
\label{fig:ap_ssim_Hotdog_plot}
\end{figure}

\begin{figure}[h!]
\centering
\includegraphics[width=0.8\textwidth]{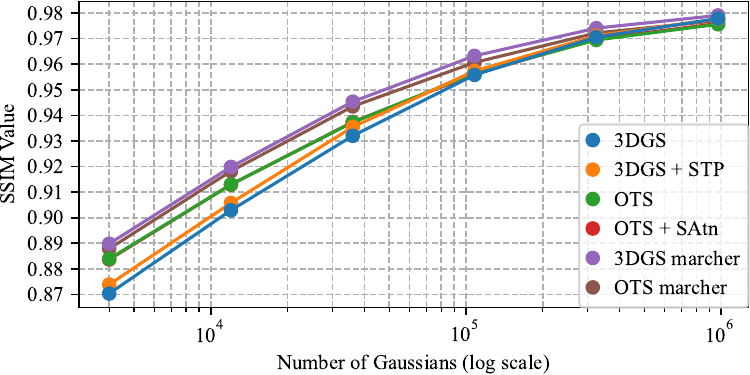}
\caption{SSIM results for Lego}
\label{fig:ap_ssim_Lego_plot}
\end{figure}

\begin{figure}[h!]
\centering
\includegraphics[width=0.8\textwidth]{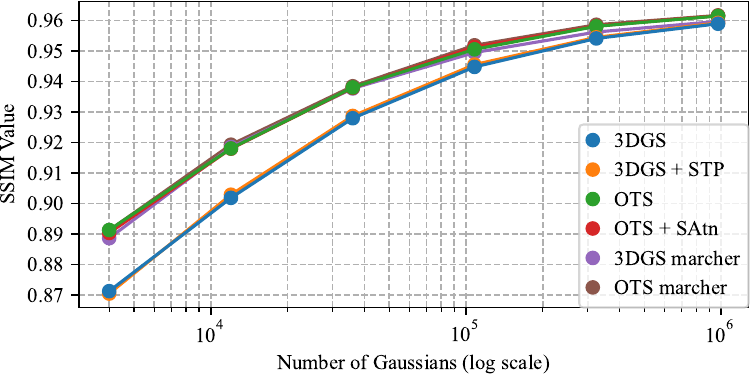}
\caption{SSIM results for Materials}
\label{fig:ap_ssim_Materials_plot}
\end{figure}

\begin{figure}[h!]
\centering
\includegraphics[width=0.8\textwidth]{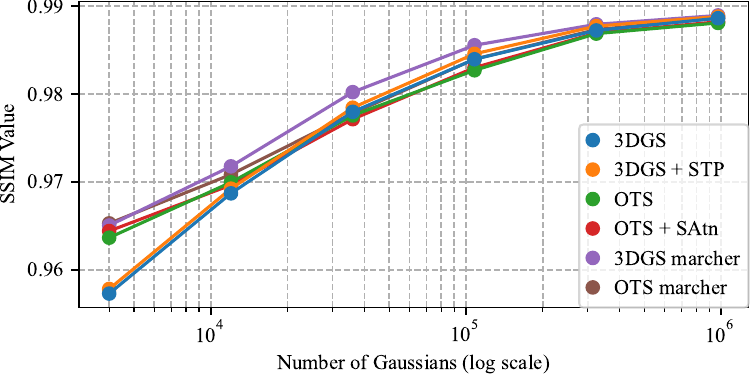}
\caption{SSIM results for Mic}
\label{fig:ap_ssim_Mic_plot}
\end{figure}

\begin{figure}[h!]
\centering
\includegraphics[width=0.8\textwidth]{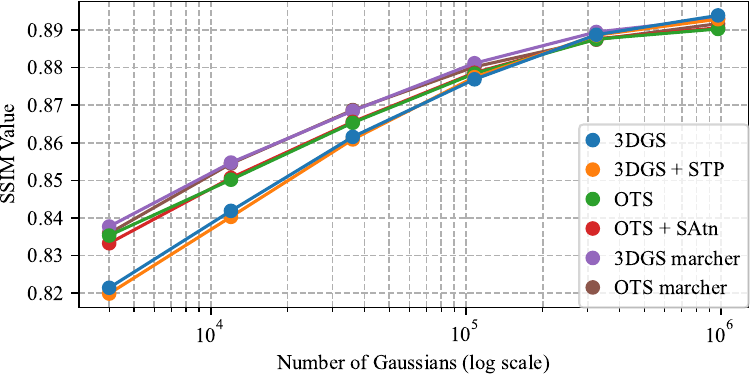}
\caption{SSIM results for Ship}
\label{fig:ap_ssim_Ship_plot}
\end{figure}

\subsection{LPIPS}
\begin{longtable}[H]{llrrrrrr}
\toprule
& & \multicolumn{6}{c}{Number of Gaussians} \\
Scene & Algorithm & 4000 & 12000 & 36000 & 108000 & 324000 & 972000 \\
\midrule \endhead
Burning Ficus & 3DGS & 0.1082 & 0.0805 & 0.0534 & 0.0336 & 0.0227 & 0.0187 \\
 & 3DGS + STP & 0.1052 & 0.0789 & 0.0522 & 0.0322 & 0.0222 & 0.0187 \\
 & OTS & 0.0920 & 0.0641 & 0.0456 & 0.0339 & 0.0274 & 0.0253 \\
 & OTS + SAtn & 0.0923 & 0.0654 & 0.0459 & 0.0341 & 0.0272 & 0.0253 \\
 & 3DGS marcher & 0.0954 & 0.0648 & 0.0443 & 0.0311 & 0.0238 & 0.0217 \\
 & OTS marcher & 0.0934 & 0.0669 & 0.0476 & 0.0347 & 0.0283 & 0.0268 \\
Chair & 3DGS & 0.0976 & 0.0780 & 0.0588 & 0.0407 & 0.0276 & 0.0195 \\
 & 3DGS + STP & 0.0980 & 0.0761 & 0.0583 & 0.0399 & 0.0265 & 0.0192 \\
 & OTS & 0.0860 & 0.0666 & 0.0530 & 0.0387 & 0.0252 & 0.0192 \\
 & OTS + SAtn & 0.0877 & 0.0681 & 0.0534 & 0.0384 & 0.0258 & 0.0196 \\
 & 3DGS marcher & 0.0861 & 0.0687 & 0.0542 & 0.0392 & 0.0285 & 0.0215 \\
 & OTS marcher & 0.0865 & 0.0671 & 0.0539 & 0.0387 & 0.0263 & 0.0206 \\
Coloured Wdas & 3DGS & 0.1184 & 0.1058 & 0.0900 & 0.0764 & 0.0662 & 0.0592 \\
 & 3DGS + STP & 0.1184 & 0.1058 & 0.0903 & 0.0769 & 0.0664 & 0.0592 \\
 & OTS & 0.1156 & 0.1008 & 0.0861 & 0.0731 & 0.0638 & 0.0572 \\
 & OTS + SAtn & 0.1152 & 0.1003 & 0.0860 & 0.0726 & 0.0637 & 0.0572 \\
 & 3DGS marcher & 0.1192 & 0.1078 & 0.0941 & 0.0813 & 0.0708 & 0.0628 \\
 & OTS marcher & 0.1179 & 0.1055 & 0.0929 & 0.0789 & 0.0685 & 0.0606 \\
Drums & 3DGS & 0.1386 & 0.1081 & 0.0841 & 0.0656 & 0.0533 & 0.0467 \\
 & 3DGS + STP & 0.1364 & 0.1056 & 0.0827 & 0.0633 & 0.0522 & 0.0466 \\
 & OTS & 0.1170 & 0.0895 & 0.0738 & 0.0630 & 0.0543 & 0.0506 \\
 & OTS + SAtn & 0.1204 & 0.0926 & 0.0764 & 0.0639 & 0.0548 & 0.0511 \\
 & 3DGS marcher & 0.1130 & 0.0885 & 0.0706 & 0.0576 & 0.0499 & 0.0452 \\
 & OTS marcher & 0.1192 & 0.0924 & 0.0761 & 0.0641 & 0.0551 & 0.0521 \\
Explosion & 3DGS & 0.1639 & 0.1148 & 0.0728 & 0.0476 & 0.0354 & 0.0298 \\
 & 3DGS + STP & 0.1614 & 0.1118 & 0.0709 & 0.0470 & 0.0352 & 0.0300 \\
 & OTS & 0.1613 & 0.1090 & 0.0684 & 0.0440 & 0.0333 & 0.0289 \\
 & OTS + SAtn & 0.1613 & 0.1085 & 0.0670 & 0.0435 & 0.0331 & 0.0287 \\
 & 3DGS marcher & 0.1596 & 0.1090 & 0.0668 & 0.0446 & 0.0352 & 0.0328 \\
 & OTS marcher & 0.1584 & 0.1038 & 0.0623 & 0.0410 & 0.0334 & 0.0316 \\
Ficus & 3DGS & 0.0740 & 0.0528 & 0.0341 & 0.0201 & 0.0134 & 0.0112 \\
 & 3DGS + STP & 0.0740 & 0.0517 & 0.0328 & 0.0193 & 0.0128 & 0.0109 \\
 & OTS & 0.0600 & 0.0423 & 0.0305 & 0.0242 & 0.0208 & 0.0203 \\
 & OTS + SAtn & 0.0598 & 0.0419 & 0.0303 & 0.0242 & 0.0209 & 0.0205 \\
 & 3DGS marcher & 0.0557 & 0.0378 & 0.0259 & 0.0188 & 0.0154 & 0.0145 \\
 & OTS marcher & 0.0598 & 0.0424 & 0.0308 & 0.0239 & 0.0210 & 0.0207 \\
Hotdog & 3DGS & 0.0840 & 0.0661 & 0.0521 & 0.0406 & 0.0316 & 0.0261 \\
 & 3DGS + STP & 0.0817 & 0.0641 & 0.0505 & 0.0396 & 0.0317 & 0.0266 \\
 & OTS & 0.0745 & 0.0607 & 0.0490 & 0.0398 & 0.0324 & 0.0274 \\
 & OTS + SAtn & 0.0730 & 0.0596 & 0.0486 & 0.0399 & 0.0323 & 0.0276 \\
 & 3DGS marcher & 0.0709 & 0.0588 & 0.0481 & 0.0398 & 0.0330 & 0.0291 \\
 & OTS marcher & 0.0705 & 0.0575 & 0.0487 & 0.0405 & 0.0339 & 0.0306 \\
Lego & 3DGS & 0.1674 & 0.1315 & 0.0953 & 0.0609 & 0.0372 & 0.0246 \\
 & 3DGS + STP & 0.1650 & 0.1303 & 0.0920 & 0.0601 & 0.0369 & 0.0251 \\
 & OTS & 0.1463 & 0.1142 & 0.0841 & 0.0573 & 0.0364 & 0.0264 \\
 & OTS + SAtn & 0.1494 & 0.1167 & 0.0855 & 0.0591 & 0.0367 & 0.0265 \\
 & 3DGS marcher & 0.1440 & 0.1100 & 0.0746 & 0.0488 & 0.0315 & 0.0243 \\
 & OTS marcher & 0.1489 & 0.1148 & 0.0809 & 0.0541 & 0.0340 & 0.0254 \\
Materials & 3DGS & 0.1463 & 0.1124 & 0.0832 & 0.0613 & 0.0476 & 0.0398 \\
 & 3DGS + STP & 0.1463 & 0.1105 & 0.0809 & 0.0589 & 0.0461 & 0.0388 \\
 & OTS & 0.1290 & 0.0965 & 0.0724 & 0.0557 & 0.0452 & 0.0398 \\
 & OTS + SAtn & 0.1294 & 0.0975 & 0.0730 & 0.0550 & 0.0450 & 0.0394 \\
 & 3DGS marcher & 0.1277 & 0.0944 & 0.0708 & 0.0547 & 0.0445 & 0.0390 \\
 & OTS marcher & 0.1276 & 0.0967 & 0.0738 & 0.0552 & 0.0454 & 0.0401 \\
Mic & 3DGS & 0.0560 & 0.0400 & 0.0252 & 0.0161 & 0.0112 & 0.0099 \\
 & 3DGS + STP & 0.0553 & 0.0387 & 0.0243 & 0.0148 & 0.0106 & 0.0096 \\
 & OTS & 0.0423 & 0.0325 & 0.0231 & 0.0159 & 0.0107 & 0.0101 \\
 & OTS + SAtn & 0.0414 & 0.0324 & 0.0235 & 0.0155 & 0.0106 & 0.0100 \\
 & 3DGS marcher & 0.0415 & 0.0307 & 0.0203 & 0.0132 & 0.0109 & 0.0107 \\
 & OTS marcher & 0.0406 & 0.0311 & 0.0224 & 0.0145 & 0.0104 & 0.0099 \\
Ship & 3DGS & 0.2395 & 0.2073 & 0.1750 & 0.1472 & 0.1263 & 0.1154 \\
 & 3DGS + STP & 0.2385 & 0.2072 & 0.1753 & 0.1463 & 0.1263 & 0.1160 \\
 & OTS & 0.2183 & 0.1941 & 0.1711 & 0.1495 & 0.1317 & 0.1214 \\
 & OTS + SAtn & 0.2227 & 0.1959 & 0.1718 & 0.1498 & 0.1319 & 0.1219 \\
 & 3DGS marcher & 0.2178 & 0.1907 & 0.1668 & 0.1446 & 0.1275 & 0.1179 \\
 & OTS marcher & 0.2202 & 0.1928 & 0.1690 & 0.1486 & 0.1329 & 0.1229 \\
Wdas Cloud 1 & 3DGS & 0.0741 & 0.0723 & 0.0703 & 0.0680 & 0.0670 & 0.0675 \\
 & 3DGS + STP & 0.0741 & 0.0720 & 0.0705 & 0.0681 & 0.0673 & 0.0675 \\
 & OTS & 0.0734 & 0.0709 & 0.0690 & 0.0651 & 0.0623 & 0.0611 \\
 & OTS + SAtn & 0.0733 & 0.0708 & 0.0689 & 0.0650 & 0.0623 & 0.0608 \\
 & 3DGS marcher & 0.0739 & 0.0719 & 0.0695 & 0.0667 & 0.0646 & 0.0647 \\
 & OTS marcher & 0.0734 & 0.0709 & 0.0692 & 0.0649 & 0.0624 & 0.0610 \\
Wdas Cloud 2 & 3DGS & 0.1211 & 0.1149 & 0.1082 & 0.1006 & 0.0960 & 0.0948 \\
 & 3DGS + STP & 0.1214 & 0.1150 & 0.1084 & 0.1012 & 0.0963 & 0.0952 \\
 & OTS & 0.1174 & 0.1106 & 0.1040 & 0.0975 & 0.0917 & 0.0888 \\
 & OTS + SAtn & 0.1173 & 0.1107 & 0.1039 & 0.0975 & 0.0915 & 0.0885 \\
 & 3DGS marcher & 0.1215 & 0.1144 & 0.1080 & 0.1007 & 0.0955 & 0.0928 \\
 & OTS marcher & 0.1176 & 0.1109 & 0.1044 & 0.0971 & 0.0914 & 0.0883 \\
Wdas Cloud 3 & 3DGS & 0.1343 & 0.1252 & 0.1171 & 0.1064 & 0.0971 & 0.0902 \\
 & 3DGS + STP & 0.1349 & 0.1253 & 0.1173 & 0.1067 & 0.0972 & 0.0904 \\
 & OTS & 0.1280 & 0.1170 & 0.1083 & 0.0974 & 0.0891 & 0.0832 \\
 & OTS + SAtn & 0.1282 & 0.1171 & 0.1079 & 0.0969 & 0.0889 & 0.0832 \\
 & 3DGS marcher & 0.1341 & 0.1239 & 0.1168 & 0.1072 & 0.0983 & 0.0905 \\
 & OTS marcher & 0.1282 & 0.1176 & 0.1087 & 0.0982 & 0.0898 & 0.0831 \\
\bottomrule
\caption{LPIPS results for all scenes}
\end{longtable}

\begin{figure}[h!]
\centering
\includegraphics[width=0.8\textwidth]{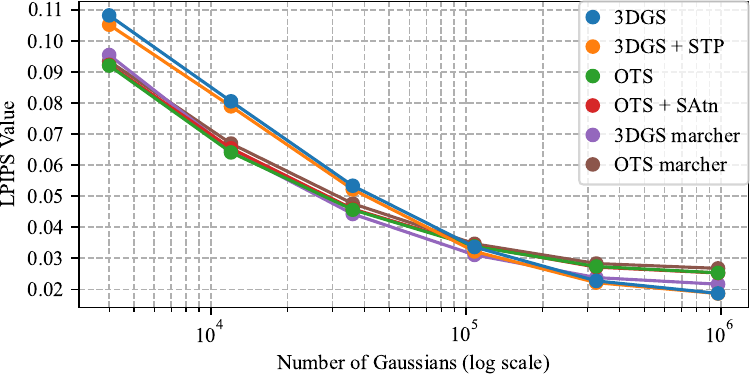}
\caption{LPIPS results for Burning Ficus}
\label{fig:ap_lpips_Burning Ficus_plot}
\end{figure}

\begin{figure}[h!]
\centering
\includegraphics[width=0.8\textwidth]{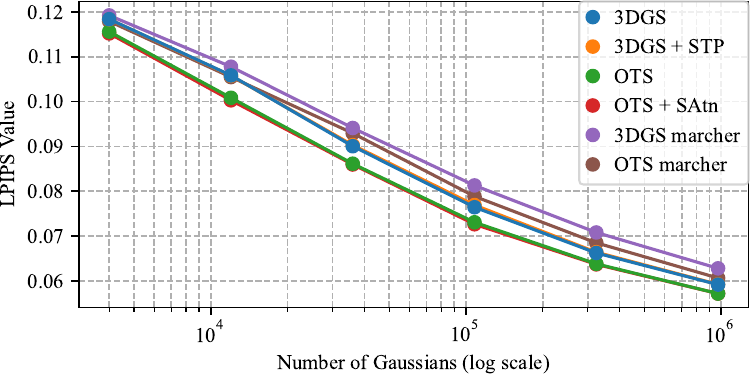}
\caption{LPIPS results for Coloured Wdas}
\label{fig:ap_lpips_Coloured Wdas_plot}
\end{figure}

\begin{figure}[h!]
\centering
\includegraphics[width=0.8\textwidth]{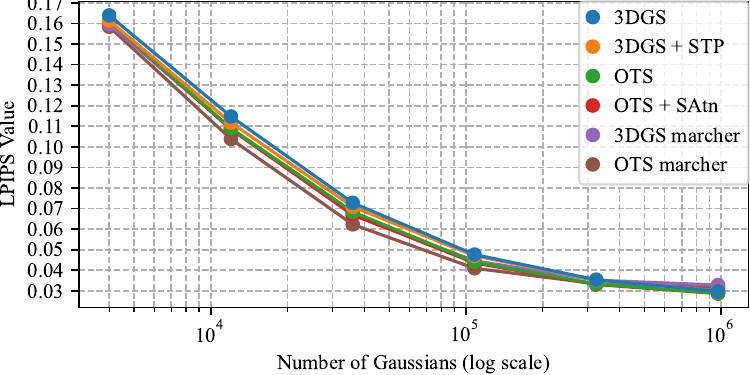}
\caption{LPIPS results for Explosion}
\label{fig:ap_lpips_Explosion_plot}
\end{figure}

\begin{figure}[h!]
\centering
\includegraphics[width=0.8\textwidth]{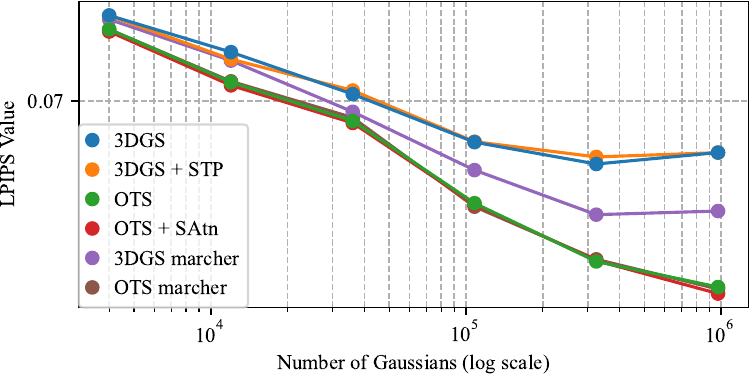}
\caption{LPIPS results for Wdas Cloud 1}
\label{fig:ap_lpips_Wdas Cloud 1_plot}
\end{figure}

\begin{figure}[h!]
\centering
\includegraphics[width=0.8\textwidth]{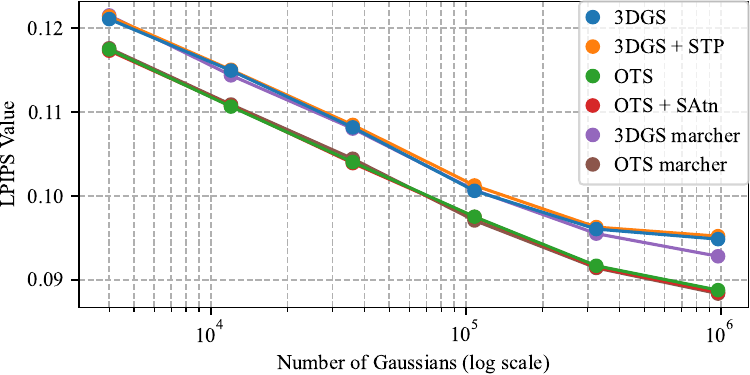}
\caption{LPIPS results for Wdas Cloud 2}
\label{fig:ap_lpips_Wdas Cloud 2_plot}
\end{figure}

\begin{figure}[h!]
\centering
\includegraphics[width=0.8\textwidth]{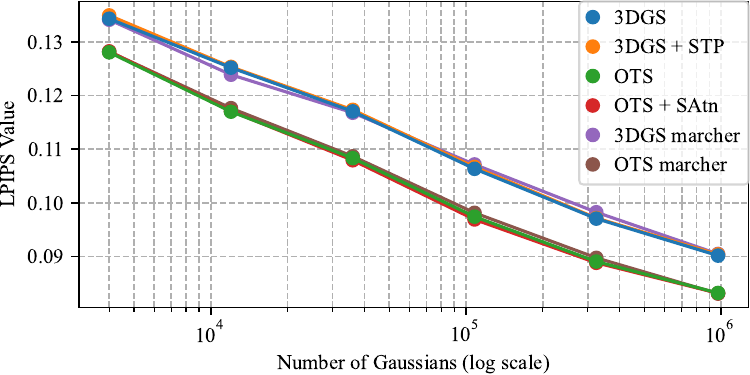}
\caption{LPIPS results for Wdas Cloud 3}
\label{fig:ap_lpips_Wdas Cloud 3_plot}
\end{figure}

\begin{figure}[h!]
\centering
\includegraphics[width=0.8\textwidth]{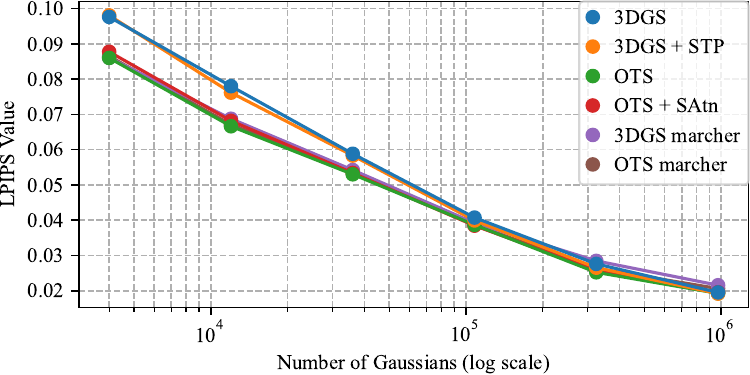}
\caption{LPIPS results for Chair}
\label{fig:ap_lpips_Chair_plot}
\end{figure}

\begin{figure}[h!]
\centering
\includegraphics[width=0.8\textwidth]{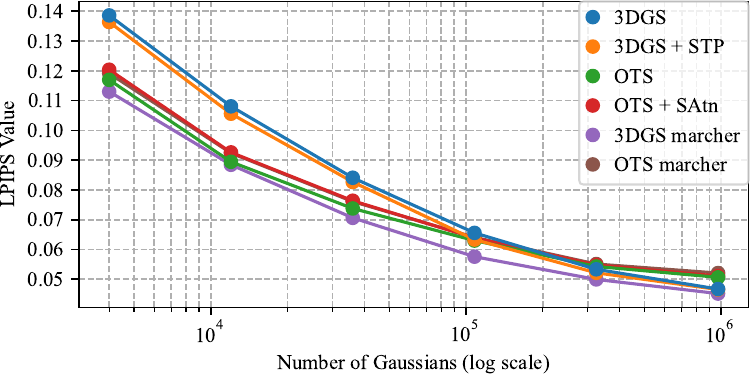}
\caption{LPIPS results for Drums}
\label{fig:ap_lpips_Drums_plot}
\end{figure}

\begin{figure}[h!]
\centering
\includegraphics[width=0.8\textwidth]{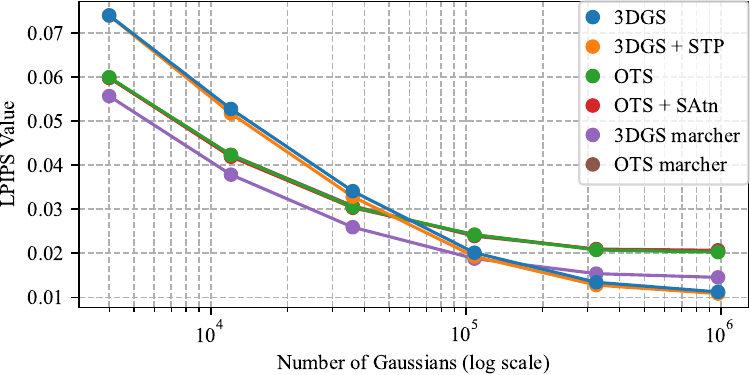}
\caption{LPIPS results for Ficus}
\label{fig:ap_lpips_Ficus_plot}
\end{figure}

\begin{figure}[h!]
\centering
\includegraphics[width=0.8\textwidth]{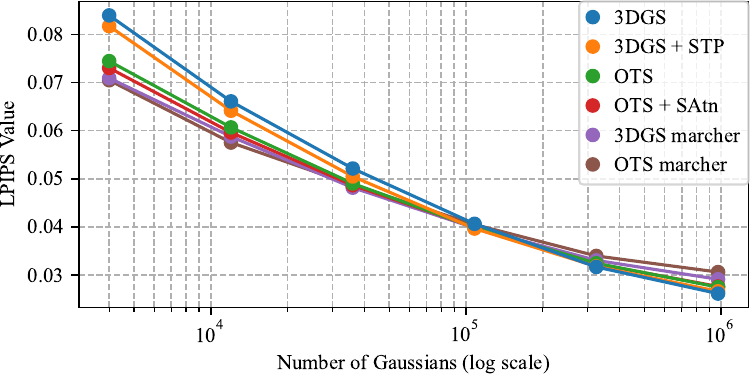}
\caption{LPIPS results for Hotdog}
\label{fig:ap_lpips_Hotdog_plot}
\end{figure}

\begin{figure}[h!]
\centering
\includegraphics[width=0.8\textwidth]{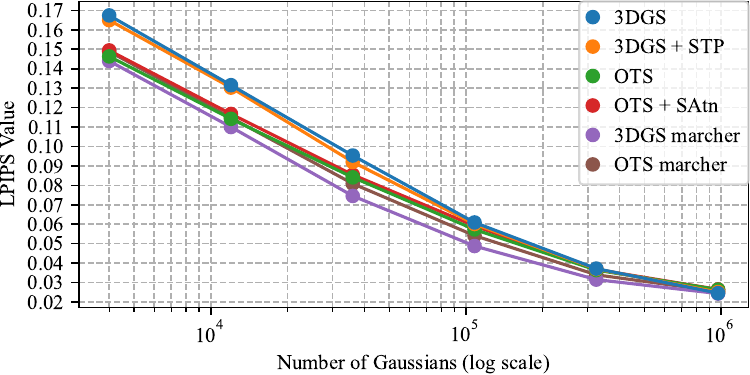}
\caption{LPIPS results for Lego}
\label{fig:ap_lpips_Lego_plot}
\end{figure}

\begin{figure}[h!]
\centering
\includegraphics[width=0.8\textwidth]{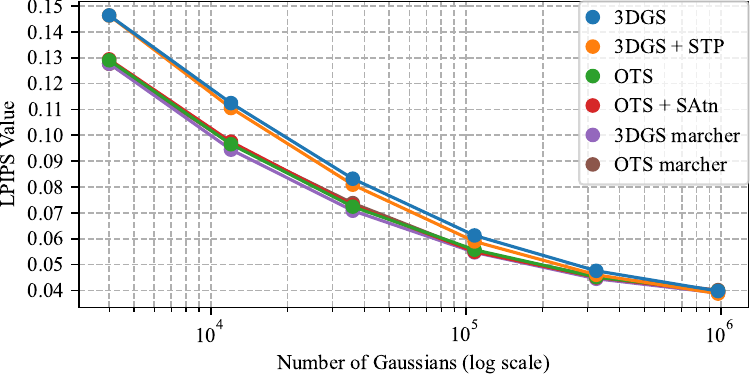}
\caption{LPIPS results for Materials}
\label{fig:ap_lpips_Materials_plot}
\end{figure}

\begin{figure}[h!]
\centering
\includegraphics[width=0.8\textwidth]{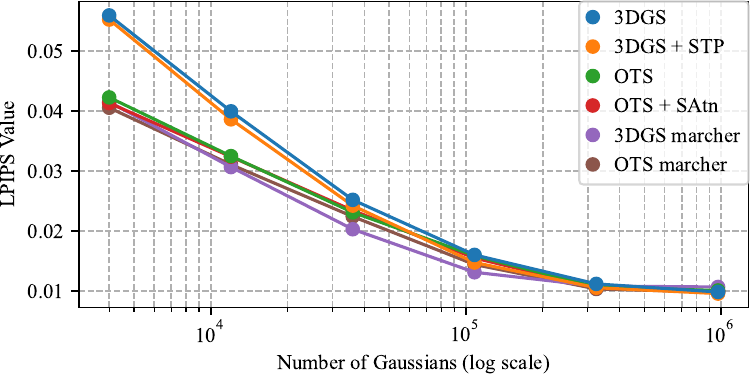}
\caption{LPIPS results for Mic}
\label{fig:ap_lpips_Mic_plot}
\end{figure}

\begin{figure}[h!]
\centering
\includegraphics[width=0.8\textwidth]{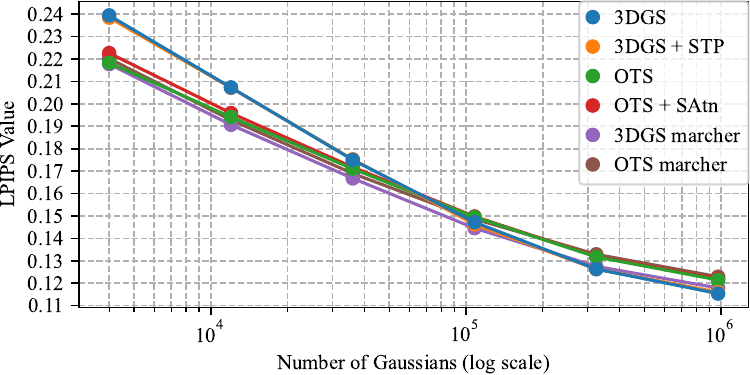}
\caption{LPIPS results for Ship}
\label{fig:ap_lpips_Ship_plot}
\end{figure}